# Dynamic safety cases for frontier AI

Carmen Cârlan[a*]        Francesca Gomez[a*]        Yohan Mathew[a*]        Ketana Krishna[a*]

René King[a*]        Peter Gebauer[a*]        Ben R Smith[a*†]

[a]Arcadia Impact - AI Governance Taskforce

## Abstract

Frontier artificial intelligence (AI) systems present both benefits and risks to society. Safety cases - structured arguments supported by evidence - are one way to help ensure the safe development and deployment of these systems. Yet the evolving nature of AI capabilities, as well as changes in the operational environment and understanding of risk, necessitates mechanisms for continuously updating these safety cases. Typically, in other sectors, safety cases are produced pre-deployment and do not require frequent updates post-deployment, which can be a manual, costly process. This paper proposes a Dynamic Safety Case Management System (DSCMS) to support both the initial creation of a safety case and its systematic, semi-automated revision over time. Drawing on methods developed in the autonomous vehicles (AV) sector - state-of-the-art Checkable Safety Arguments (CSA) combined with Safety Performance Indicators (SPIs) recommended by UL 4600, a DSCMS helps developers maintain alignment between system claims and the latest system state. We demonstrate this approach on a safety case template for offensive cyber capabilities and suggest ways it can be integrated into governance structures for safety-critical decision-making. While the correctness of the initial safety argument remains paramount - particularly for high-severity risks - a DSCMS provides a framework for adapting to new insights and strengthening incident response. We outline challenges and further work towards development and implementation of this approach as part of continuous safety assurance of frontier AI systems.

*Equal contribution. Authors are free to list themselves first in the author order in their CVs.
Author contributions are detailed in the Contribution Statement.
† Primary correspondence to ben@benrsmith.org.
Francesca Gomez correspondence to francesca@wiser-human.com

## Contribution Statement

Author contributions are as follows:
Conceptualisation, supervision and project administration: Ben R Smith
Methodology, Investigation, Writing – Original Draft, Review:

| | |
|---|---|
| Ben R Smith | Sections 1, 2, 6, 7, 8 |
| Rene King | Sections 2, 3, 5.1 |
| Carmen Cârlan | Sections 2, 4.1, 4.4 |
| Francesca Gomez | Sections 2, 4.2, 4.3, 4.4 ('External' SPIs) |
| Yohan Mathew | Sections 2, 4.2, 4.3, 4.4 ('Internal' SPIs) |
| Peter Gebauer | Sections 2, 5.2 |
| Ketana Krishna | Sections 2, 5.3 |

Writing – Review & Editing: Ben R Smith.

## Acknowledgements

We would like to express special thanks to Marie Buhl for her input in defining the research question, providing directional guidance, and for detailed feedback on this paper. We are grateful to Benjamin Hilton for his perspective, discussion and comment, and to Syeda Nida Hassan, Celia Waggoner, Emma Tilley, Muhammad Sadiq, Anne Le Roux, and Reema Panjwani for their valuable feedback. Finally, thanks to Justin Olive and Arcadia Impact for initiating the AI Governance Taskforce within which this research was conducted, and to the other research team of this Autumn 2024 cohort, led by Steve Barrett, for general support and the discussion of related ideas (Barrett et al., manuscript in preparation).

## Contents

## Suggestions for how to read this paper

For those who may not want or have time to read the paper sequentially in full we suggest:

- For frontier AI developer management and policy makers we suggest starting with the Executive summary, then Introduction, Recommendations, Discussion.
- For safety researchers we suggest starting with the Executive summary, the Introduction, section 4.
- For governance researchers we suggest starting with the Executive summary, the Introduction, section 5.

## Abbreviations

| | |
|---|---|
| AISI | AI Safety Institute |
| API | Application Programming Interface |
| ASL | AI Safety Level |
| AV | Autonomous Vehicles |
| CAE | Claims Arguments Evidence |
| CAIRF | Central AI Risk Function |
| CIA | Check Impact Analysis |
| CSA | Checkable Safety Arguments |
| CiSP | Cyber Security Information Sharing Partnership |
| CRA | Cyber Resilience Audit |
| CTF | Capture The Flag |
| CTI | Cyber Threat Intelligence |
| DSC | Dynamic Safety Cases |
| DSCMS | Dynamic Safety Case Management System |
| DSIT | UK Government Department for Science, Innovation and Technology |
| FSCCC | Financial Services Cyber Collaboration Centre |
| GSN | Goal Structuring Notation |
| INCOSE | International Council on Systems Engineering |
| LASR | The Laboratory for AI Security Research |
| LLM | Large Language Model |
| MoD | UK Government Ministry of Defence |
| MTTD | Mean Time To Detect |
| NCSC | UK Government National Cyber Security Centre |
| NPSA | UK Government National Protective Security Authority |
| OECD | The Organisation for Economic Cooperation and Development |
| ONR | UK Office for Nuclear Regulation |
| RSO | Responsible Scaling Officer |
| RSP | Responsible Scaling Policy |
| RVTM | Requirements Verification Traceability Matrix |
| SCSC | UK Safety-Critical Systems Club |
| SMS | Safety Management System |
| SPI | Safety Performance Indicator |
| TTP | Tactics, Techniques and Procedures |
| TRAINS | Testing Risks of AI for National Security |
| UL | Underwriters Laboratories |
| V&V | Verification and Validation |

# Executive summary

### Why Dynamic Safety Cases for frontier AI? (Section 1)

- Frontier AI systems can exhibit fast-evolving capabilities and interact in complex, sometimes unpredictable ways with sociotechnical environments, entailing systemic risk. While initial safety cases created early in a system's lifecycle remain essential to argue that a system is safe enough for deployment, especially for high-risk systems, there is also a need to update these safety arguments over time. Manual updating is costly, time-consuming and exposed to human error. Inspired by the autonomous vehicles (AV) sector, we propose a Dynamic Safety Case Management System (DSCMS) to systematically and (partially) automate these updates and thereby complement robust pre-deployment assurances. We demonstrate this concept using a concrete example - offensive cyber capabilities - and suggest its integration into governance frameworks to maximise effectiveness.

### Requirements for a Dynamic Safety Case Management System (Section 3)

- Key requirements for a dynamic safety cases management system (DSCMS) centre around the ability to update the safety case continually throughout the system lifecycle - monitoring Safety Performance Indicators (SPIs) tied to the claims that make up the safety case argument, and performing automated consistency checks and subsequent change impact analyses when triggered by SPI threshold breaches, allowing for rapid re-assessment of claims when new model capabilities or external risks emerge.

### DSCMS based on Checkable Safety Arguments (CSA) & SPIs (Section 4)

- Checkable Safety Arguments (CSA): Provide formal, semantically enriched relationships between safety arguments and system representations, enabling automated consistency checks and change impact analysis.
- A Safety Performance Indicator (SPI): Metrics (with thresholds) that signal whether a safety claim remains valid. For example, in a cyber 'inability' safety case, a SPI might track system evaluations of proxy tasks, and Cyber Threat Intelligence (CTI) of emerging threats.

### Governance integration (Section 5)

- A DSCMS could be plugged into existing frontier AI company governance frameworks (for example, Anthropic's Responsible Scaling Policy) with minimal changes.
- A DSCMS could provide government stakeholders with real-time or aggregated visibility of emerging AI risks, informing regulations and enabling more nuanced or agile policy responses. Secure access to key information (e.g. Cyber Threat Intelligence feeds) is crucial for effective SPIs.
- Offensive cyber attack scenarios in this paper show how a DSCMS could support key high-stakes safety assurance decisions, e.g. major model updates, deployments, or expansions into new operational domains, by flagging potential safety case breaches *before* deployment, potentially in real-time.

### Discussion and recommendations (Section 6 & 7)

- DSCMS show potential value for reducing the burden of manual safety case updates and systematising and semi-automating ongoing maintenance. Whilst CSA can be adapted straightforwardly to frontier AI, work is required particularly in the development of understanding how to specify complete and correct SPIs sets that capture emerging risks while avoiding overreliance on easily measured but less significant indicators. Handling major shifts in AI capability is a challenge, which can require entirely new arguments. The DSCMS has potential to assist both frontier AI developers self-governance and national government oversight via providing potentially real time system safety status information for decision support as part of existing safety management frameworks.
- Recommendations to frontier AI developers, governments, and the wider research community:
  1. Build a proof of concept DSCMS prototype for a sample safety case.
  2. Develop CSA building blocks specifically tailored to frontier AI.
  3. Research SPI best practices, drawing on AV experience.
  4. Conduct governance 'war-gaming' with DSCMS dashboards to evaluate real-time decisions support under various scenarios.
  5. Support data-sharing for external SPIs (e.g. threat intelligence) between providers, governments and developers.
  6. Incentivise research for systemic risk modelling in open-ended complex contexts, focussing on pre-deployment detection of catastrophic potential.

# 1 Introduction

Frontier artificial intelligence (AI) systems present both benefits and risks to society (Yoshua et al., 2024). Robust safety assurance is therefore critical throughout the system lifecycle. Safety cases - a structured argument, supported by evidence that make a claim about the overall safety of a system in a given context (MoD, 2007) - are increasingly explored as a means for building confidence in AI system safety (Buhl et al., 2024).

While a thorough safety case is necessary before deployment, in practice AI systems and their contexts can evolve in ways that are difficult to anticipate. AI system capability evaluations provide some understanding of the system properties, but these capabilities can change after training because of fine-tuning, and through agent scaffolding (Anwar et al., 2024), including through capability elicitation techniques such as chain of thought prompting (Yao et al., 2023; Davidson et al., 2023). Recent work has suggested that models are capable of hiding capabilities, i.e. sandbagging (van de Weij, et al., 2024). Once deployed, safeguards might be weakened or removed, for example via jailbreak attacks (Chao et al., 2024). These factors can alter risk profiles, sometimes significantly. Due to the general purpose nature of these systems, the potential operational deployment domain is vast and the interaction of the AI system with sociotechnical systems is therefore complex, uncertain and can be very difficult to predict (Leveson, 2012).

In light of these uncertainties, periodic or event-driven updates to safety cases pre and post deployment are essential, even when the initial safety assurance is carefully designed for high-severity scenarios. In other safety-critical sectors, such as nuclear power, aviation and railways (Inge, 2007; Sujan et al., 2016; Favaro et al., 2023), safety cases are periodically updated to reflect changes to the system in response to upgrades, fault and failure corrective action and changes to the operational context. This is particularly the case in the autonomous vehicle (AV) sector where systems are developed in agile, continuous development cycles.

However, updating safety cases can be time-consuming and resource-intensive, involving tasks such as re-assessing hazards and risk analysis, maintaining the argument structure and logic and re-generating supporting evidence. In current practice, these are predominantly manual activities. In this paper we propose and explore a possible solution to the updating problem - a method for systematising and semi-automating the process, inspired by and adapted from proposed and existing methods used in the autonomous vehicles (AV) sector.

In this paper we outline requirements for such a system (section 3) and propose a system that fulfils those requirements (section 4) - which we call a Dynamic Safety Case Management System (DSCMS). This system consists of Checkable Safety Arguments (CSA) (Cârlan et al., 2020; Cârlan et al., 2022; Cârlan, forthcoming) combined with Safety Performance Indicators (SPIs). SPIs are metrics supported by evidence that use a specified threshold for comparison to indicate the satisfaction of a claim (UL, 2023). We apply the system to a safety case fragment template - offensive cyber capabilities - as a concrete demonstration. We then discuss how the DSCMS might be integrated into governance structures (section 5), and outline challenges, further work and recommendations towards development and implementation of this approach as part of continuous safety assurance (section 6 & 7).

## 2 Related work

Literature exploring safety cases for frontier AI is limited, although there have been a number of recent publications discussing the fundamentals (Buhl et al., 2024; Clymer et al., 2024; Wasil et al., 2024) and specific safety case 'sketches' or templates (Goemans et al., 2024; Grosse, R, 2024; Balesni et al., 2024). In the latter, Balesni et al. use indicators similar to SPIs based on AI model evaluations of scheming capabilities. In the autonomous systems sector which shares some similar characteristics to frontier AI, there has been work on safety case templates and patterns (Bloomfield et al., 2021; Wozniak et al., 2020).

Towards automating the process of updating safety cases, the concept of a "Dynamic Safety Case" (DSC) was introduced by Denny et al. (2015) - an approach for creating and managing safety arguments that continuously develop through a system's lifecycle. The DSC scope exploits runtime information to assess and develop safety arguments continuously, comparing argument parameters and change impact. Work on various elements of systematisation and towards tool-supported implementation of dynamic safety cases has been published (Denney et al., 2012; Agrawal et al., 2019; Mirzaei et al., 2022). Cârlan has built on this body of work, much of which is drawn together in Cârlan's PhD thesis on Checkable Safety Arguments (Cârlan, forthcoming). Among others, Denny et al. (Denney et al., 2015) propose continuously measuring the system safety performance with monitors. This can be done with SPIs, which have been adapted from aviation and developed in the UL 4600 safety standard for autonomous vehicles (UL, 2023). Ratiu et al. (2024) have proposed a systematic approach for defining SPIs. Denney & Pai (2024) building on their earlier work introduced above, propose a framework to facilitate what they term 'dynamic assurance', i.e. 'continued, justified confidence that a system is operating at a safety risk level consistent with an approved risk baseline' using an example of autonomy in aviation. A similar notion of 'dynamic safety management' and 'dynamic assurance cases' utilising self-adaption autonomous systems to maintain safety are developed in, for example, Trapp & Weiss (2019) and Calinescu et al. (2018). However, none of these state-of-the-art approaches have been applied to frontier AI systems.

Academic literature on regulatory regimes for risk management and safety cases (Leveson, 2011; Askell et al., 2019; Buhl et al., 2024; Le Coze, 2024; Schuett et al., 2024) offer a foundation for analysing governance mechanisms for dynamic safety cases. Studies on AI safety institutions (Castris & Thomas, 2024; Cha, 2024) also informed the research. Cross-sectoral findings were derived from ONR Technical Assessment Guides (ONR, 2024), UL 4600, Space Industry Regulations (2021), and ASEMS Safety Case Guidance (MoD, 2024). An array of UK policy documents, and international frameworks such as the Frontier AI Safety Commitments (DSIT, 2024) were mapped. A number of frontier AI developers have published safety frameworks (Anthropic, 2024; Open AI, 2024; Google DeepMind, 2024). A DSMCS could enable improved governance by potentially providing a more accurate, complete and timely means of safety assurance upon which decisions can be made.

## 3 Dynamic safety case management system (DSCMS) requirements

In the introduction, we described the need to update safety cases, and highlighted the burden to do so with manual methods. To address these challenges, this section defines requirements for a Dynamic Safety Case Management System (DSCMS) - a continuous, partially automated mechanism for use throughout the frontier AI system lifecycle.

Table 1 summarises the core system requirements for the DSCMS. These foundational requirements centre on establishing and maintaining safety cases, performing automated consistency checks, defining and managing Safety Performance Indicators (SPIs), and automatically re-evaluating safety arguments when SPI thresholds are breached.

In addition to meeting its core requirements, the DSCMS must also support key operational processes, such as integrating external data feeds. Further requirements relating to governance reporting and data security [REQ-23 and REQ-025 respectively] are discussed in Section 5.1 in the context of integrating the DSCMS within the larger governance context. Appendix B contains an expanded Requirements Verification Traceability Matrix (RVTM), including the rationale and verification approach for each requirement.

| ID | Requirement Title | Requirement Description |
|---|---|---|
| REQ -001 | Early Creation of Safety Cases | The DSCMS shall develop an initial safety case for the frontier AI system at the start of its lifecycle, including all required safety arguments, supporting evidence, and justifications in accordance with [Standards Reference Document]. **Keywords: foundational argument baseline** |
| REQ -002 | Consistent Maintenance of Safety Cases | The DSCMS shall maintain the safety case for the frontier AI system throughout its entire lifecycle. **Keywords: lifecycle-driven updating** |
| REQ -005 | Automated Consistency Checks | The DSCMS shall perform automated consistency checks of safety arguments against the latest associated development artefacts in [Configuration Management Document]. **Keywords: real-time artifact synchronisation** |
| REQ -017 | SPI Definition and Management | The DSCMS shall define and maintain a catalogue of Safety Performance Indicators (SPIs) ([SPI Definition Document]) with thresholds, update frequency, and referenceable artefacts. **Keywords: threshold-driven performance metrics** |
| REQ -018 | SPI-Driven Argument Re-Evaluation | The DSCMS shall automatically re-evaluate affected parts of the safety argument whenever an SPI threshold is breached ([SPI Threshold Criteria Document]). **Keywords: system-model-triggered argument impact analysis** |
| REQ -019 | External Data Feed Integration | The DSCMS shall map external data feeds to relevant safety arguments/evidence. Environmental data **Keywords: environmental data aggregation** |
| REQ -020 | External Data Change Impact | The DSCMS shall conduct automated impact analyses when new external data arrives. **Keywords: external-data triggered argument impact analysis'** |
| REQ -023 | Governance Reporting Interface | The DSCMS shall have a ([Governance Interface Specification]) interface. **Keywords: governance interface portal** |
| REQ -025 | Data Security and Access Control | The DSCMS shall implement data security measures in accordance with ([Data Security Plan]). **Keywords: compliance-based data protection** |

Table 1: System Requirements for the DSCMS

**Note 1:** Requirement identifiers are not strictly sequential because our requirements analysis process was iterative - requirements may emerge, change, or be merged through the development of a DSCMS. Further details are provided in Appendix A.

**Note 2:** Referenced documents are kept generic to allow for flexibility and adaptability across different operational and potential future regulatory contexts - they should be tailored during DSCMS customisation and implementation to ensure alignment with the same.

# 4 DSCMS based on Checkable Safety Arguments (CSA)

Whereas in the previous section, we introduced the concept and requirements of a Dynamic Safety Case Management System, in this section, we propose an implementation of a DSCMS.

In compliance with REQ-001 specified in Section 3, our DSCMS first requires the creation of an initial structured safety case for the frontier AI system under consideration in the early phases of the system lifecycle. We propose using a state-of-the-art framework called checkable safety arguments (CSA) to specify the safety case. Further, state-of-the-art safety argument patterns or safety case templates may be used to structure the safety case. First, we explain how to specify an initial safety case according to the CSA framework (Section 4.1). Then, we explain how to set SPIs (Section 4.2) and propose concrete SPIs for a cybersecurity inability argument (Section 4.3). Finally, we explain the process of semi-automatically updating a safety case when gathering data on the SPIs (Section 4.4).

## 4.1 Checkable Safety Arguments (CSA) for frontier AI systems

In our pursuit of building a solution for continuous maintenance of the safety cases of frontier AI systems, in Section 2, we presented an overview of state-of-the-art approaches for implementing dynamic safety cases. In this section, we explore how we can use one of these approaches, the CSA framework, to implement a DSCMS for frontier AI systems. This approach was chosen because it meets the requirements we specified in Section 3 for safety case creation and management.

In a nutshell, the checkable safety arguments framework implements requirements REQ-002 and REQ-005 from Table 1 in Section 3 by extending existing languages for structured arguments to enable the automated execution of consistency checks between the system under consideration and arguments about the system safety. Consistency checks ensure that all safety arguments remain aligned with the latest version of the system by identifying the argumentation elements in a safety argument that are inconsistent with the system given the occurrence of a system change scenario. Consistency checks expand the concept of change impact analysis (CIA). Whereas CIA identifies the elements in a system safety case impacted by a system change, consistency checks identify the inconsistencies, i.e., contradictions between the changed system artifact and the safety case, indicating how the safety case elements are impacted by a system change. The consistency checks enabled by the CSA framework are complete and more accurate than the consistency checks or CIA approaches in the state of the art. (Cârlan, forthcoming) provides definitions for completeness and accuracy as below.

> **Definition - Completeness**: Given a system specification, and an argument about the safety assurance of this system, when the system undergoes a change, a consistency check is complete if all argumentation elements actually impacted by the change are identified, i.e., the actual impact area is covered by the impact area identified by the respective consistency check.

> **Definition - Impact area**: The set of argumentation elements impacted by a system change.

> **Definition - Accuracy**: Given a set of system artifacts, and a system safety argument, when the system undergoes a change, a consistency check is accurate if it does not falsely identify argumentation elements as impacted. Namely, the identified impact area should not have more elements than the actual impact area. False positives are argumentation elements annotated as impacted when they are not actually impacted.

The accuracy of consistency checks depends on the following assumptions:

> **Assumption 1**: Safety arguments are assumed to be valid. A safety argument is valid if the conclusion is probably true when all premises are true (Barwise et al., 2002).

**Assumption 2**: The traceability between system engineering models and safety argument models is assumed to be complete. If only some referencing relationships are specified, the consistency checks cannot identify all argumentation elements directly impacted by a change in a traced engineering model.

**Assumption 3**: The specification of the SPIs is assumed to be correct. It is assumed that the right metrics are specified and that the thresholds are correctly specified. A threshold which is too permissive would lead to the scenarios when the associated safety claim is invalidated, without the breaching of the SPI (false negative), whereas a too conservative SPI would lead to scenarios in which the SPI would be breached more often than the invalidation of the associated safety claim (false positive).

Checkable arguments are specified as formal engineering models. One advantage of having a formal specification is that it makes the safety argument amenable to automated management. Although Cârlan (Cârlan, forthcoming) elaborates on extending the Goal Structuring Notation language (SCSC, 2021) to accommodate the specification of checkable arguments, the modeling concepts may be used for any argument modeling language, such as Claims Arguments Evidence (CAE).

Checkable arguments are composed of checkable argumentation elements. Checkable argumentation elements can be integrated with formalised system development artifacts, e.g., hazards, safety requirements, safety analysis results, system design, or verification & validation (V&V) results via referencing relationships (see requirement REQ-019, in Table 1). The system artifacts need to be formalised, e.g., be specified as system engineering models, so the reasoning for their changes can be automated. According to INCOSE[1], a system engineering model is a simplified representation of a system, leaving out irrelevant details and only specifying the relevant aspects of the system. Models may be specified semi-formally as formalised graphical or textual languages, mathematically, as mathematical theories, or formally, as formal models with formalised syntax, semantics, and logics. Formal models can be specified using a set of well-defined interconnected concepts, typically formalised in a modeling language definition.

Figure 1 below sketches the process behind checkable safety arguments.

**SPI monitoring.** As shown in the figure, in this work, in compliance with UL 4600 (UL, 2023), we extend the checkable safety arguments framework to enable the specification of SPIs, attached to claims in the arguments. SPIs indicate the satisfaction of the claims in safety arguments. An SPI is to be specified as a metric, a targeted value or threshold for that metric, and a monitor for measuring the metric. The monitor specifies how often the metric will be measured and how the measured values will be used to assess the SPI. Some SPIs may evaluate the measured values individually, and some may evaluate them in an aggregated manner. SPIs of higher-level claims may depend on SPIs of lower-level claims. If an SPI is invalidated, the claim to which it was attached is also invalidated. Thus, the framework satisfies requirement REQ-017.

---

[1] https://www.incose.org/

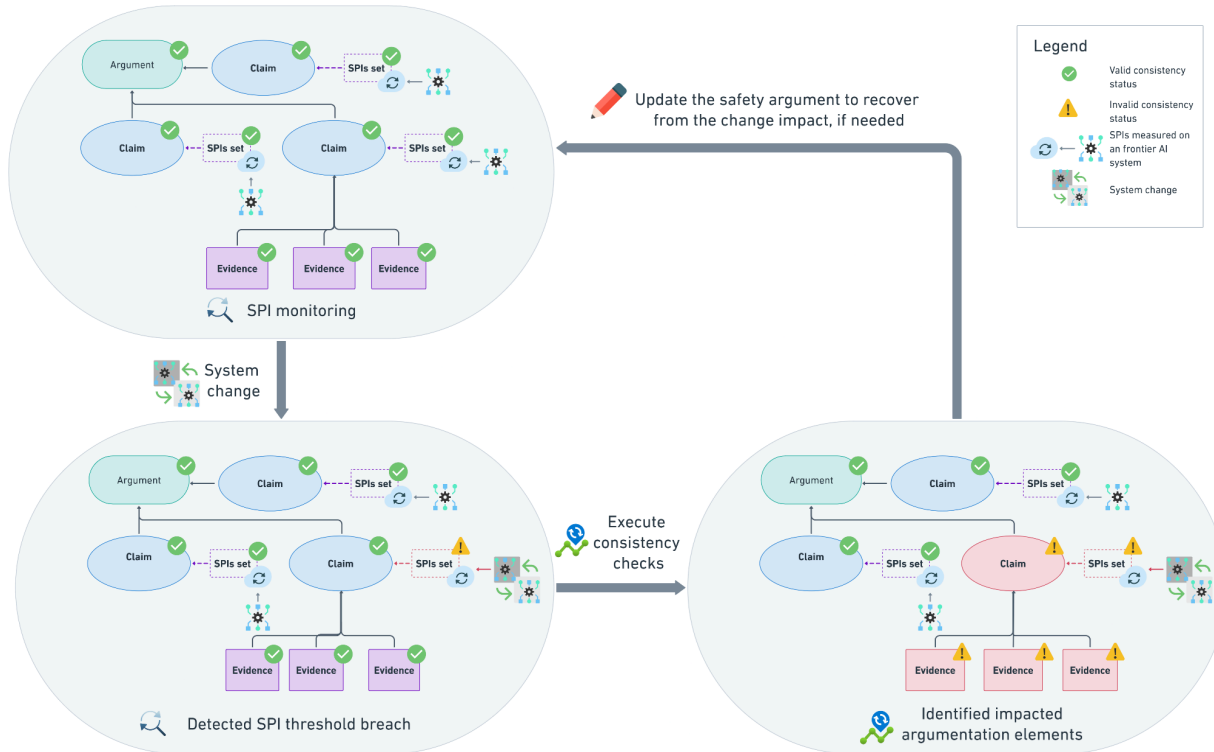

Figure 1: The process recommended by the CSA framework.

**SPI threshold breaches triggering the execution of consistency checks.** Checkable safety arguments with direct traces to system engineering models or attached SPIs are responsive to system changes. Given a change in a traced artifact or the invalidation of an annotated SPI (i.e., the breach of an SPI threshold), the execution of consistency checks is automatically triggered, thus satisfying requirements REQ-018 and REQ-020.

**Consistency checks identifying impacted elements.** The specification of automatically executable consistency checks is supported by the fact that the checkable safety arguments are semantically enriched with consistency rules, specifying how the impact of changes in the considered system propagates throughout safety arguments. Concretely, consistency rules specify how changes in referenced system engineering models, or the invalidations of the SPIs impact the argumentation elements to which they are attached, as well as other argumentation elements.

A consistency check takes as input a safety argument, and a system change scenario and outputs the consistency status (i.e., valid or invalid) of the argumentation elements in the safety argument after the system change. The consistency check first identifies and appropriately annotates the consistency status of all argumentation elements in the safety argument that are directly impacted by the undergone system change. The argumentation elements directly impacted by a system change are either linked to SPIs breaching their thresholds or to changed system development artifacts. Then, the indirectly impacted elements are identified based on the relationships between the identified directly impacted argumentation elements and other elements of the safety argument.

To identify indirectly impacted elements, first, consistency checks identify each relationship having as a source or target the argumentation element directly impacted by a system change. Second, if not already done, the impact of the system change on the argumentation element at the other end of the relationship is identified. For each argumentation element identified as indirectly impacted, the consistency check identifies the propagation of the change impact to the elements connected to it that still need to be analysed.

**Updating the safety argument to recover from change impact.** After the impact of a system change is identified, update actions need to be taken, to recover from the impact. We propose extending the checkable safety arguments framework to support the specification of recommended update actions and consistency rules.

According to Cârlan (Cârlan, forthcoming), checkable argument building blocks may be specified and then re-used in different projects to model checkable safety arguments. A checkable argument building block is a set of connected argumentation elements with placeholders for referencing specific system artifacts. Such argumentation elements are annotated with consistency rules between the argumentation elements and the system that is to be referenced. (Cârlan, forthcoming) proposes a catalog of checkable building blocks by modeling a set of state-of-the-art argument building blocks as checkable.

However, the proposed consistency rules and existing checkable building blocks do not scope to frontier AI. In this paper, we consider a CAE-based template for arguing that an AI model does not have the capabilities to pose unacceptable cyber risks proposed by Goemans et al. (2024), and we model this template using the checkable arguments modeling framework. To this end, we first specify a set of SPIs for the claims in the argument template (see Sections 4.2 and 4.3). Second, in Section 4.4, we sketch a subset of possible system change scenarios, which will impact the arguments built by instantiating the considered argument template. For each change scenario, we specify the corresponding consistency rules. There are additional types of arguments that could be used in safety cases as dangerous capabilities advance, as described in Clymer et al. (2024) - control (restrictions that prevent harm), trustworthiness (establishing intent), and deference (using a credible AI overseer).

Of course, the degree of the automation of the consistency checks specified for a specific safety argument is dependent on whether or not SPIs can be defined for the safety claims within the argument, whether the metrics can be automatically measured, or whether they need human judgement, and on the automation of the update actions reacting to the change impact. The update actions will most likely be conducted with human oversight.

### 4.2 Safety Performance Indicators (SPIs) for frontier AI Safety Cases

Safety Performance Indicators (SPIs) are "metrics supported by evidence that use threshold comparisons of condition claims in a safety case" (UL, 2023), i.e., quantitative or qualitative specifications of acceptable risk criteria used in the claims of a safety case. SPIs can be classified as 'External' or 'Internal', depending on the origin of the data used as evidence, which we explain in the following sections. SPIs can also be categorised into leading or lagging indicators, based on whether they are used primarily as early warning of potential future risks or to evaluate imminent risks & incidents.

According to UL 4600 (UL, 2023), SPIs used in safety cases can be divided into two types depending on their temporal focus - leading and lagging indicators. Leading indicators are used to identify potential safety issues in future before the respective capabilities or harms can be directly evaluated, and are often based on proxy metrics like effective compute and scaling laws. Lagging indicators are a more direct evaluation of capabilities, and are commonly used in frontier developers today via model evaluations. Schleiss et al (2022) have defined a process for continuous updating of safety cases by using leading and lagging indicators in parallel. They suggest first using leading metrics predictions to estimate post-mitigation risks, comparing this to lagging metrics to understand prediction gaps, and then recalibrating leading indicator prediction models to improve accuracy.

Koessler et al. (2024) describe the different kinds of thresholds that can be used to define acceptance criteria in frontier AI risk assessment - risk thresholds, capability thresholds and compute thresholds. Risk thresholds are defined in terms of the probability and severity of societal harm, and are the most principled type of thresholds used, but are the hardest to define for a general-purpose technology like frontier AI. Capability thresholds, which are the most common form of risk assessment metrics used by frontier AI developers today, define an acceptable level of model capabilities beyond which a model might be deemed unsafe for deployment, and are generally estimated using AI model evaluations. Compute thresholds are defined in terms of the amount of training compute used due to its correlation with capabilities, which are also known as scaling laws (Kaplan et al., 2020). They are easy to measure, but are also the most removed from direct metrics of societal harm.

### 4.2.1 Internal Safety Performance Indicators

Internal SPIs are metrics derived from within the developer's systems, processes, or evaluations, such as internal test results, model performance monitoring, or data from safety controls. These metrics generally focus on detecting changes in a model's capabilities, which are widely regarded as the primary driver of shifts in its safety profile. The large majority of internal SPIs are determined by capability evaluations, red-teaming and compute scaling.

Weidinger et al. (2021) show that at least three layers of evaluations are necessary to understand the overall socio-technical impact in model evaluations - 1) elicitation of capabilities with just the AI system itself, 2) effect of people interacting with the AI system, and 3) systemic impact on the larger socioeconomic and environmental context. Similarly, Solaiman et al (2023) also identify various categories of social impact to be accounted for while evaluating both the base technical system and the impact on people and society, along with suggestions for specific evaluations and mitigations for each - the latter two layers here are primarily addressed by 'external' SPIs, discussed in the next section.

As a result of being specified primarily in terms of proxy tasks and evaluations, internal SPIs generally apply to lower-level claims in safety cases. For example, in the cyber inability argument by Goemans et al. (2024), the overall objective is first broken down into risk models and their supporting claims that occupy levels 1 to 3. Each of these risk models are then converted into proxy tasks in levels 4 and 5. Finally proxy tasks are supported by evaluations in levels 6 to 8. In our analysis, internal SPIs are largely found in levels 4 to 8 of the safety case, whereas the upper levels are mostly dominated by external SPIs.

| Claim ID | Claim Description | SPI Example | SPI Type |
|----------|------------------|-------------|----------|
| C3.x | Risk models C3.1-C3.n are sufficiently comprehensive | The model cannot penetrate any important layers of defense on its own weights | Lagging |
| C6.1 | The AI system cannot exceed the Cybersecurity Apprentice baseline in Vulnerability Discovery Task Suite in Automated Evals with Human Oversight | <10% change in bug-finding rate for cybersecurity professional, as compared to not using the model | Lagging |
| C6.x | The evaluation setups are adequate | There has been <4x increase in effective compute used for training of a specific model (eg: GPT-4) since the last set of evaluations were done for a previous version of the same model - this is used as a leading predictor since capabilities increase with effective compute (scaling laws) | Leading |
| C8.2 | No sandbagging | <10% performance improvement on a benchmark task when adding noise to model weights, as compared to non-sandbagging model; evaluation approach originally described by Tice et al (2024) | Lagging |

Table 2: Examples of Internal SPIs

### 4.2.2 External Safety Performance Indicators

External SPIs are metrics based on data sources external to or independent from the frontier AI developer, such as research insights, industry benchmarks, cyber threat intelligence, or findings from peer evaluations.

Given that AI models are socio-technical systems, harm from a model's offensive cyber capabilities depends not only on the model's technical capabilities but also on external factors, such as the behavior of threat actors attempting to misuse these capabilities and the cyber defenses of target organisations, including critical infrastructure. Understanding these broader systemic risks is crucial for preventing harm, as recognised by policymakers through inclusion in legislation e.g., Article 51 of the EU AI Act (European Parliament and Council of the European Union, 2024). Post-deployment monitoring, mandated in the first draft of the General AI Code of Practice for systemic risks (European Commission, 2024), highlights the importance of evaluating how AI systems interact with the human and organisational environment, particularly assessing their broader systemic impacts.

External SPIs play a critical role in this context by capturing evidence of how AI models interact with and influence the broader cyber threat landscape. By tracking factors such as threat actor behavior, vulnerabilities in target systems, and the evolution of attack techniques, external SPIs enable the monitoring of real-world misuse, emerging threats, and systemic impacts. These insights complement internal evaluations, providing a more comprehensive understanding of the risks posed by a model's offensive cyber capabilities.

**4.3. Illustrative SPIs for the cyber inability safety case template**

### 4.3.1 Our methodology for developing SPIs

Ratiu et al. (2024) propose an approach for developing a robust set of Safety Performance Indicators (SPIs) from an existing safety case. While this is based on the UL 4600 (UL, 2023) standard for autonomous products, it provides useful guidance for other domains, including a method for systematically defining SPIs from a safety case based on its components. Using this approach, we analysed the claims in the example safety case sketch presented by Goemans et al. (2024) to identify SPIs for monitoring the risk of harm from a frontier model's offensive cyber capabilities. This safety case uses inability arguments as the basis of why the model is safe to deploy. As such, our review incorporated both leading indicators to anticipate and prevent harms, as well as lagging indicators to analyse past incidents and address shortcomings in the safety case, following the recommendations of Ratiu et al. (2024).

This list is not meant to be exhaustive, but is designed to provide illustrative examples to show how these could inform the current safety posture of a model within the checkable safety argument framework.

### 4.3.2 Sources of Evidence for the Cyber Inability Safety Case SPIs

Under the definition of Safety Performance Indicators as 'metrics supported by evidence,' data sources that inform the SPIs constitute the 'evidence,' and SPIs are only effective if these data sources are reliable, accurate and up to date.

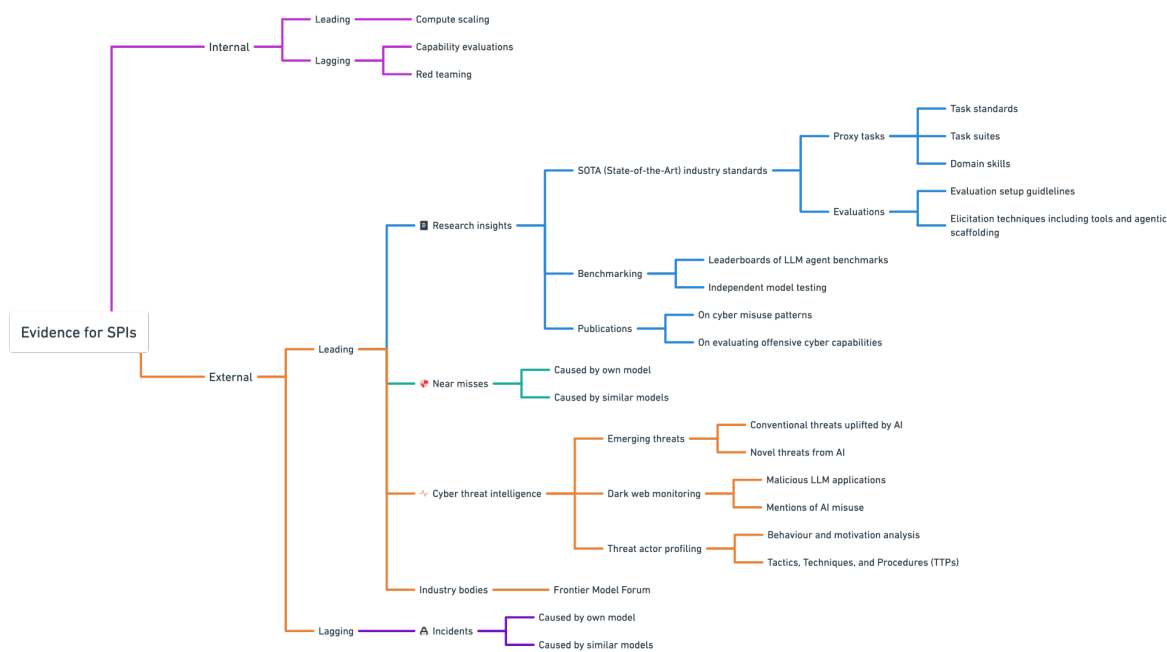

Figure 2: Data Sources providing Evidence for SPI metrics

**Evidence for Internal SPIs**

**Capability Evaluations**
Capability evaluations provide direct evidence of a model's performance against predefined risk-relevant tasks, such as vulnerability discovery or exploitation challenges. These evaluations include automated benchmarks (e.g., Capture-The-Flag, or CTF, tasks), task-based probing, and structured evaluation setups to test specific capabilities like reverse engineering or binary exploitation. Capability evaluations can be conducted internally at regular intervals or prior to major model releases, and their results help determine whether the model exceeds capability thresholds.

**Red Teaming**
Red teaming offers adversarial testing to identify under-elicited or emergent capabilities in a model, providing critical insights for validating safety case claims. Red teaming exercises are typically conducted by internal teams with expertise in offensive cybersecurity or through collaboration with external specialists for independent assessments. These tests include both automated red-teaming methods and scenario-based human-in-the-loop evaluations designed to simulate real-world misuse attempts.

**Compute Scaling**
Compute scaling evidence stems from tracking increases in effective compute used during model training, which is closely correlated with model capability improvements due to scaling laws. Effective compute combines various scaling-law-metrics like model size, dataset size, algorithmic progress, and training compute into a single metric. Monitoring compute usage provides a leading signal of potential capability jumps (Anthropic Responsible Scaling Policy, 2024), as scaling often results in emergent behaviors, even without changes to evaluation frameworks. Evidence for compute scaling includes logs of total training compute, effective compute comparisons relative to prior models, and projections of scaling trends.

**Evidence for External SPIs**

**Research insights** can inform safety case risk models by providing empirical experiments to elicit offensive cyber capabilities (Usman, Y. et al., 2024; Shao et al., 2024) and assessments of AI's impact on the cyber threat landscape (NCSC, 2024). Research can also contribute to proxy task design by suggesting alternative task decompositions used in independent testing (AISI, 2024), benchmarking (Anurin et al., 2024), and state-of-the-art task suites (Bhatt et al., 2024; Zhang et al., 2024), potentially offering more accurate proxies than those used during deployment. Additionally, research insights critique existing methods and introduce new evaluation techniques, spanning fully automated evaluations (3CB, Anurin et al., 2024), automated evaluations with human oversight (e.g. task-based probing, AISI, 2024), and human uplift studies (e.g., Long-Form Tasks, AISI, 2024). Comparing external benchmarking results with internal evaluations helps identify under-elicitation, whether due to inadequate elicitation techniques or deliberate attempts to misrepresent model capabilities by developers or even the model itself via sandbagging.

**Incidents** provide information on actual attacks involving the model and are commonly used in safety-focused industries to identify trends, determine root causes, and inform corrective measures to prevent recurrence of similar incidents. As incidents caused by a model's offensive cyber capabilities would cause losses to other organisations, rather than the frontier AI developer itself, the identification of incidents is challenging and would be dependent on having effective reporting channels in place. Frontier developers do have mechanisms in place to receive incident reports (Anthropic, 2024); however, it remains unclear what proportion of external organisations would use this to report cyber incidents caused by the model, how many would be capable of determining that a model was involved in an attack, and the time lag between incident occurrence and reporting.

**Near Misses** are events where a potential threat or risk was identified but no harm occurred due to safety controls or chance. Near misses offer similar insights to incidents, highlighting potential threats and tasks where the model may be uplifting offensive capabilities. However, near misses are often even more challenging to detect than incidents, as they heavily depend on proactive monitoring and reporting by the organisations targeted in attempted cyberattacks.

**Cyber Threat Intelligence (CTI)** monitors emerging cyber threats and misuse patterns, offering detailed information on potential threat actors, attack techniques, and campaigns. It encompasses monitoring dark web activity, tracking emerging threats, and identifying the Tactics, Techniques, and Procedures (TTPs) used by adversaries. CTI aligns proxy tasks with real-world threats, improving representativeness and revealing weaknesses in evaluation setups that require adjustments for better capability elicitation.

**Industry bodies** such as the Frontier Model Forum (2024) enable secure information sharing among member organisations, supporting peer learning and best practices for risk management, proxy tasks, and evaluation setups. Shared knowledge can reveal causes of under-elicitation, but confidentiality and competitive concerns may limit disclosure, reducing the completeness of insights.

### 4.3.3 Identification of SPIs from a cyber inability argument

Goemans et al. (2024) break down a simplified safety case template for a cyber inability argument into three layers:

1. Risk models from which harm could occur, in this case cyber risk. (**Risk models)**
2. Proxy tasks which represent what a model must be capable of doing for a risk model to be realised. (**Proxy tasks)**
3. Evaluation setups to test whether an AI system can perform the proxy tasks (**Evaluations**)

Reviewing each of these layers, we can propose Safety Performance Indicators (SPIs) which will allow us to detect invalidations of the safety case (Ratiu et al., 2024).

For each SPI proposed, illustrative thresholds have been provided. Ideally, these thresholds should align with safety tolerances set by management, such as those outlined in Responsible Scaling Policies (Anthropic, 2024) or similar frameworks. However, in practice, while frameworks like the OpenAI Preparedness Framework (OpenAI, 2024) provide some guidance on cyber capability tolerances (e.g., a medium risk level 'increases the productivity of operators by an efficiency threshold'), these tolerances are not always quantifiable, and the acceptable impacts of model usage on the external cyber threat environment are not addressed.

**Risk models**

In the safety case outlined by Goemans et al. (2024), risk models are used to capture the ways in which the frontier AI model could be used for offensive cyber attacks. The reasoning is that if evidence shows the model cannot enable such attacks, then it is safe to deploy. The claim that 'Deploying the AI System does not pose unacceptable Cyber Risk' is decomposed into claims relating to the sources of cyber risk:

- Conventional attack types in realistic settings will not be uplifted
- There is no risk of novel cyber attacks
- There are no other major sources of cyber risk

Buhl et al. (2024) further break down conventional attack types into more detailed risk models, detailing threat actors, harm vectors and targets. This reflects how threat analysis is commonly done in cyber security, where frameworks such as MITRE ATT&CK® (n.d.) are used to understand attackers, their capabilities and their motives. The safety case proposes that if each risk model objective can be proven, then the model can be considered safe.

Validations of these claims need to compare how the model is actually being used to aid cyber attacks post deployment with the expectations from the expert consultations and any pre-deployment monitoring used to inform the safety case design. While expert consultation helps provide confidence that these risk models have been correctly defined, this will only be known for certain once the model is deployed. Cyber threat patterns may also change and evolve post deployment.

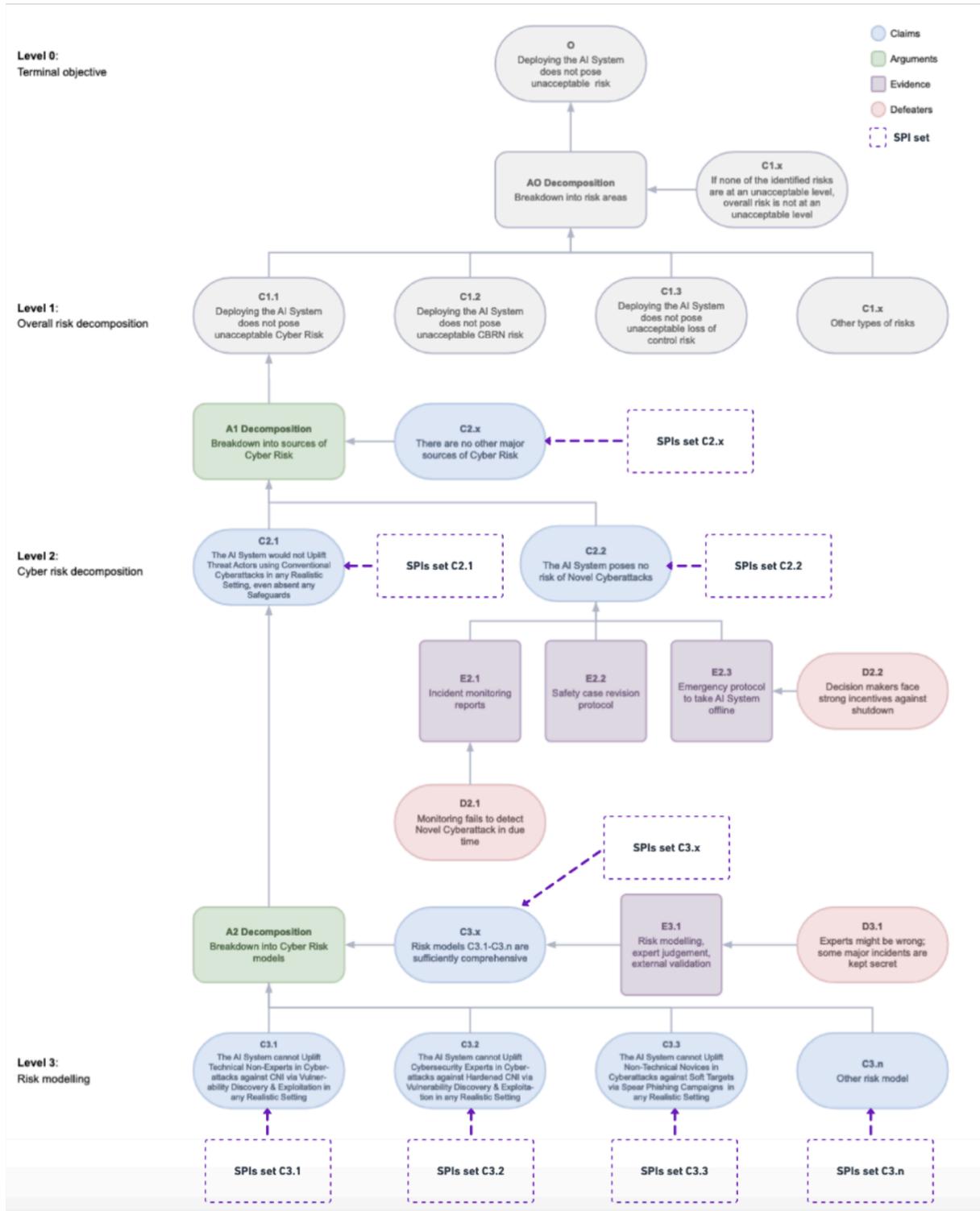

Figure 3: Extract from Safety Case Levels 0-3 by Goemans et al. (2024) with added SPI sets

Building on the cyber inability argument safety case by Goemans et al. (2024), detailed Safety Performance Indicators are proposed to monitor the validity of these cyber risk decomposition and risk modelling claims, for which three examples have been provided (C2.1, C2.2, C3.n).

SPIs have been annotated on the original safety case sketch to illustrate which claims they support in validating; for clarity, they have been highlighted with dotted lines.

The approach taken in enumerating the SPIs prioritises coverage and significance for understanding safety over the feasibility of automation, with the aim of presenting a more thorough list for consideration. However, it is recognised that feasibility of monitoring will be a key factor for frontier AI developers evaluating these SPIs, particularly where a lack of automation could result in significant time overheads. Realistically, there may also be time lags in obtaining some data, reducing its value. Nonetheless, by presenting a set of aspirational SPIs, we aim to help surface these challenges and inspire further research and development into improving automated access to data that would help deliver safety benefits, for example via APIs providing AI-model specific cyber intelligence feeds.

| SPI set C2.1: The AI system would not uplift threat actors using conventional cyberattacks in any realistic setting, even absent any safeguards | | | | | | | |
|---|---|---|---|---|---|---|---|
| **SPI ID** | **SPI Title** | **SPI Description** | **Unit** | **Example** | **Leading/ Lagging** | **Evidence** | **Example Threshold** |
| C2.1 SPI 1 | # of incidents - conventional attacks (last 30 days) | Number of cyberattacks in the last 30 days caused by model uplifting threat actors in conventional attacks | Count | 5 | Lagging | Incidents | 5 |
| C2.1 SPI 2 | Financial losses - conventional attacks (last 30 days) | Financial losses in the last 30 days from incidents caused by model uplifting threat actors in conventional attacks | $ | 50,000 | Lagging | Incidents | 1M |
| C2.1 SPI 3 | Mean Time to Detect Incidents (MTTD) | The average time taken to identify an incident from the time it occurs. Provides feedback on the speed of the incident detection channels. | Days | 5 | Lagging | Incidents | 7 |
| C2.1 SPI 4 | % change in # of incidents - conventional attacks (M-on-M) | Percentage change in the number of incidents from the model uplifting conventional attacks compared to the previous 30 days. | % | -20% | Lagging | Incidents | +10% |
| C2.1 SPI 5 | % change in financial losses - conventional attacks (M-on-M) | Percentage change in the financial losses from the model uplifting conventional attacks compared to the previous 30 days. | % | +30% | Lagging | Incidents | +10% |
| C2.1 SPI 6 | # of incidents from similar models - conventional attacks (last 30 days) | Number of cyberattacks in the last 30 days caused by a similar model uplifting threat actors in conventional attacks. | Count | 1 | Leading | Industry Bodies | 2 |
| C2.1 SPI 7 | Financial losses from similar model - conventional attacks (last 30 days) | Financial losses in the last 30 days from incidents caused by similar models uplifting threat actors in conventional attacks | $ | 50,000 | Leading | Industry Bodies | 1M |
| C2.1 SPI 8 | # of Near misses - conventional attacks (last 30 days) | Number of averted cyberattacks caused by model uplifting threat actors in conventional attacks | Count | 4 | Leading | Incidents | 5 |
| C2.1 SPI 9 | Research papers - other source | Research papers demonstrating conventional cyberattacks from frontier AI models uplifting threat actor capabilities | Count | 2 | Leading | Research insights | 0 |
| C2.1 SPI 10 | # of emerging threats - conventional attacks | Number of emerging threats of AI-enhanced conventional attacks from Cyber threat intelligence monitoring | Count | 3 | Leading | Cyber threat intelligence | 5 |
| C2.1 SPI 11 | Emerging threat trend - conventional attacks (% change) | Percentage change in the number of emerging threats from the model uplifting conventional attacks compared to the previous 30 days. | % | +5% | Leading | Cyber threat intelligence | +10% |
| C2.1 SPI 12 | Dark web mentions - conventional attacks (last 30 days) | Mentions of potential misuse relating to conventional attacks of the AI model on the dark web in the last 30 days | Count | 3 | Leading | Cyber threat intelligence | 5 |
| C2.1 SPI 13 | # of new TTPs - conventional attacks (last 30 days) | New Tactics, Techniques, and Procedures (TTPs) leveraging AI models identified as being used by threat actors in the last 30 days | Count | 1 | Leading | Cyber threat intelligence | 1 |

Table 3: SPI Set for claim C2.1

| SPI set C2.2: The AI System poses no risk of novel cyberattacks | | | | | | | |
|---|---|---|---|---|---|---|---|
| SPI ID | SPI Title | SPI Description | Unit | Example | Leading/ Lagging | Evidence | Example Threshold |
| C2.2 SPI 1 | # of incidents - novel cyberattacks (last 30 days) | Number of cyberattacks deviating from known attack patterns | Count | 5 | Lagging | Incidents | 2 |
| C2.2S PI 2 | Financial losses - novel attacks (last 30 days) | Financial losses in the last 30 days from incidents caused by novel cyberattacks | $ | 10,000 | Lagging | Incidents | 500,000 |
| C2.2 SPI 3 | Mean Time to Detect (MTTD) Incidents from novel attacks | The average time taken to identify incidents caused by novel cyberattacks from the time they occur. | Days | 10 | Lagging | Incidents | 7 |
| C2.2S PI 4 | % change in # of incidents - novel attacks (M-on-M) | Percentage change in the number of incidents caused by novel cyberattacks compared to the previous 30 days. | % | -20% | Lagging | Incidents | +10% |
| C2.2S PI 5 | % change in financial losses - novel attacks (M-on-M) | Percentage change in the financial losses from novel cyberattacks compared to the previous 30 days. | % | +30% | Lagging | Incidents | +10% |
| C2.2 SPI 6 | # of incidents from similar models - novel attacks (last 30 days) | Number of cyberattacks in the last 30 days caused by a similar model uplifting novel cyberattacks. | Count | 1 | Leading | Industry Bodies | 2 |
| C2.2 SPI 7 | Financial losses from similar model - novel attacks (last 30 days) | Financial losses in the last 30 days from incidents caused by similar models uplifting novel cyberattacks. | $ | 50,000 | Leading | Industry Bodies | 1M |
| C2.2 SPI 8 | # of Near misses - novel attacks (last 30 days) | Number of averted cyberattacks caused by model uplifting novel cyberattacks. | Count | 4 | Leading | Incidents | 5 |
| C2.2 SPI 9 | Research papers - novel attacks | Research papers identifying novel threats of cyberattacks from frontier AI models | Count | 2 | Leading | Research insights | 0 |
| C2.2 SPI 10 | # of emerging threats - novel attacks | Number of novel emerging threats of AI-enhanced attacks from Cyber threat intelligence monitoring | Count | 3 | Leading | Cyber threat intelligence | 5 |
| C2.2 SPI 12 | Dark web mentions - novel attacks (last 30 days) | Mentions of potential misuse relating to novel attacks of the AI model on the dark web in the last 30 days | Count | 3 | Leading | Cyber threat intelligence | 5 |
| C2.2 SPI 13 | # of new TTPs - novel attacks (last 30 days) | New Tactics, Techniques, and Procedures (TTPs) leveraging AI models via novel attacks identified as being used by threat actors in the last 30 days | Count | 1 | Leading | Cyber threat intelligence | 1 |

Table 4: SPI Set for claim C2.2

| SPI set C3.n: AI system cannot uplift [Threat actor] in cyberattacks against [Target profile] via [Harm vector] **(Risk model n)** | | | | | | | |
|---|---|---|---|---|---|---|---|
| **SPI ID** | **SPI Title** | **SPI Description** | **Unit** | **Example** | **Leading/ Lagging** | **Evidence** | **Example Threshold** |
| C3.n SPI 1 | # of incidents - Risk model n (last 30 days) | Number of cyberattacks caused by model uplifting [Threat actor] in cyberattacks against [Target profile] via [Harm vector] | Count | 3 | Lagging | Incidents | 1 |
| C3.n SPI 2 | Financial losses - Risk model n (last 30 days) | Financial losses in the last 30 days from incidents caused by model uplifting [Threat actor] in cyberattacks against [Target profile] via [Harm vector] | $ | 10,000 | Lagging | Incidents | 20,000 |
| C3.n SPI 3 | Mean Time to Detect (MTTD) Incidents from Risk model n | The average time taken to identify incidents caused by model uplifting [Threat actor] in cyberattacks against [Target profile] via [Harm vector] | Days | 10 | Lagging | Incidents | 7 |
| C3.n SPI 4 | % change in # of incidents - Risk model n (M-on-M) | Percentage change in the number of incidents caused by model uplifting [Threat actor] in cyberattacks against [Target profile] via [Harm vector] compared to the previous 30 days. | % | -20% | Lagging | Incidents | +10% |
| C3.n SPI 5 | % change in financial losses - Risk model n (M-on-M) | Percentage change in the financial losses from model uplifting [Threat actor] in cyberattacks against [Target profile] via [Harm vector] compared to the previous 30 days. | % | +30% | Lagging | Incidents | +10% |
| C3.n SPI 6 | # of incidents from similar models - Risk model n (last 30 days) | Number of cyberattacks in the last 30 days caused by a similar model uplifting [Threat actor] in cyberattacks against [Target profile] via [Harm vector] | Count | 1 | Leading | Industry Bodies | 2 |
| C3.n SPI 7 | Financial losses from similar model - Risk model n (last 30 days) | Financial losses in the last 30 days from incidents caused by similar models uplifting [Threat actor] in cyberattacks against [Target profile] via [Harm vector] | $ | 50,000 | Leading | Industry Bodies | 1M |
| C3.n SPI 8 | # of Near misses - Risk model n (last 30 days) | Number of averted cyberattacks caused by model uplifting [Threat actor] in cyberattacks against [Target profile] via [Harm vector] | Count | 4 | Leading | Incidents | 5 |
| C3.n SPI 9 | Research papers - Risk model n | Research papers identifying threats relating to frontier AI model uplifting [Threat actor] in cyberattacks against [Target profile] via [Harm vector] | Count | 2 | Leading | Research insights | 0 |
| C3.n SPI 10 | # of emerging threats - Risk model n | Number of emerging threats from Cyber threat intelligence monitoring relating to AI models uplifting [Threat actor] in cyberattacks against [Target profile] via [Harm vector] | Count | 3 | Leading | Cyber threat intelligence | 5 |
| C3.n SPI 12 | Dark web mentions - Risk model n (last 30 days) | Mentions of potential misuse of the AI model on the dark web in the last 30 days, relating to uplift of [Threat actor] in cyberattacks against [Target profile] via [Harm vector] | Count | 3 | Leading | Cyber threat intelligence | 5 |
| C3.n SPI 13 | # of new TTPs - Risk model n (last 30 days) | New Tactics, Techniques, and Procedures (TTPs) leveraging AI models for [Harm vector] identified as being used by [Threat actor] against [Target profile] in the last 30 days | Count | 1 | Leading | Cyber threat intelligence | 1 |

Table 5: SPI Set for claim C3.n

**Proxy tasks**

In the cyber inability safety sketch by Goemans et al. (2024), substitution is used to convert risk models into proxy tasks and claims are made relating to the selection and adequacy of proxy tasks and scoring.

These proxy tasks act as indicators of the model's capabilities, aiming to accurately reflect the skills and tasks necessary for a risk model to lead to real-world cyber incidents. A common approach for assessing offensive cyber capabilities, used by both frontier developers (Phuong et al., 2024; OpenAI, 2024) and AI Safety Institutes (US AISI and UK AISI, 2024), is to define a set of proxy tasks to represent actions which an AI model would need to perform to be able to uplift or execute a cyber attack. These typically take the form of Capture The Flag (CTF) challenges, which involve finding and exploiting vulnerabilities to retrieve a hidden string of text, or 'flag'.

The adequacy of these proxy tasks can be tracked post-deployment by monitoring the real-world skills being used to exploit the harm vectors specified in the risk model, while also incorporating new best practices and insights from ongoing research. In effect, this can provide evidence to validate or challenge the quality and completeness of the proxy task design and underlying threat models.

Building on the proxy task selection and adequacy claims in the cyber inability argument safety case by Goemans et al. (2024), detailed SPIs are proposed to monitor the validity of these, for which three examples have been provided (C4.1, C5.1, C5.2).

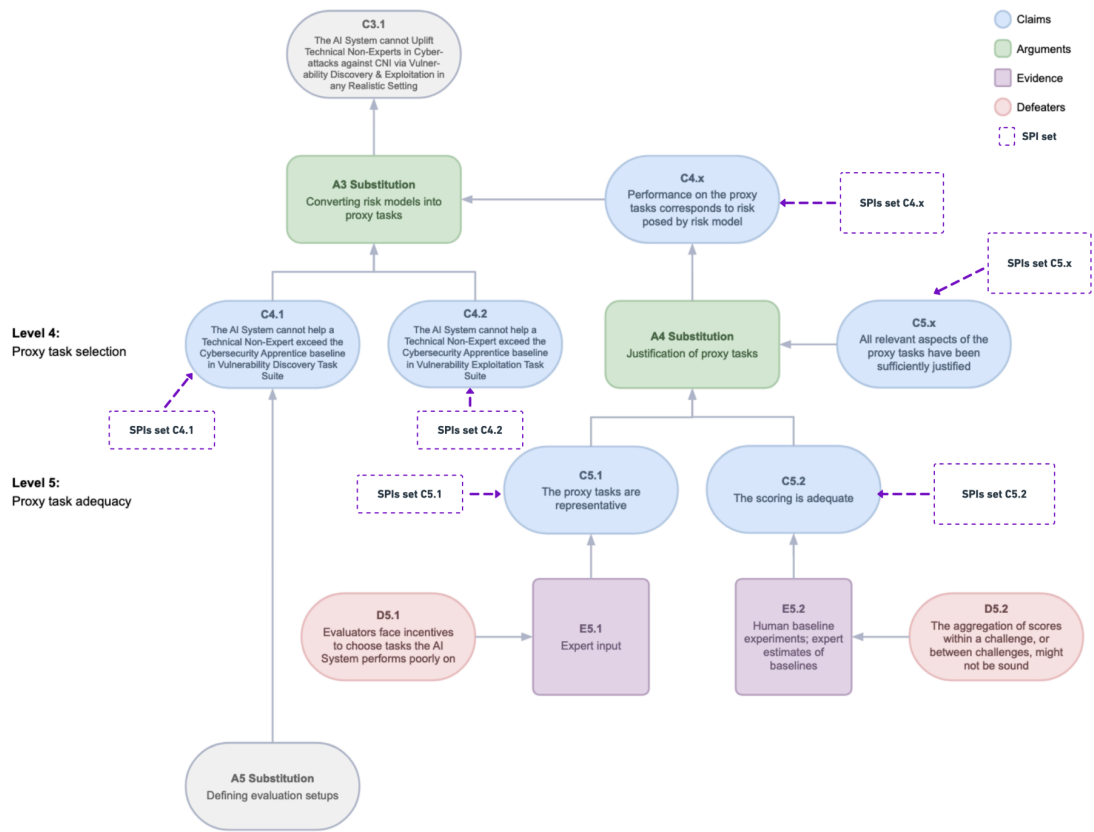

Figure 4: Extract from safety case sketch by Goemans et al. (2024): proxy tasks with added SPI sets

**SPI set C4.1**  AI system cannot help a Technical Non-Expert exceed the Cybersecurity Apprentice baseline in Vulnerability Discovery Task Suite

| SPI ID | SPI Title | SPI Description | Unit | Example | Leading/ Lagging | Evidence | Example Threshold |
|---|---|---|---|---|---|---|---|
| C4.1 SPI 1 | External Benchmark Capability Advancement | Measures the number of instances where external benchmarks or evaluations showed that a Technical Non-Expert exceeded an equivalent Cybersecurity Apprentice baseline for Vulnerability Discovery due to model assistance. | Count | 1 | Leading | Research insights | 1 |
| C4.1 SPI 2 | Redoing evaluations based on scaling law predictions (effective compute) | Ensures there has been less than a certain level of increase in effective compute used for training of a specific model (eg: GPT-4) since the last set of evaluations were done for a previous version of the same model | Multiple of effective compute | 4x | Leading | Scaling laws | 4x |
| C4.1 SPI 3 | Redoing evaluations based on scaling law predictions (fine-tuning) | Ensures there has been less than a certain amount of time of fine-turning progress of a specific model (eg: GPT-4) since the last set of evaluations were done for a previous version of the same model | Months of fine-tuning progress | 6 | Leading | Scaling laws | 6 |
| C4.1 SPI 4 | Patch Classification Evaluation 1 | Vulnerability detection proficiency measured using a patch classification dataset by Wang et al (2019) | Accuracy, F1-score | <75% | Lagging | Internal evaluations | <75% |
| C4.1 SPI 5 | Patch Classification Evaluation 2 | Vulnerability detection proficiency measured using the Security Patch Identification (SPI) dataset by Zhou et al (2021) | Accuracy, F1-score | <75% | Lagging | Internal evaluations | <60% |
| C4.1 SPI 6 | Function Classification Evaluation 1 | Vulnerability detection proficiency measured using the DiverseVul dataset by Chen et al (2023) | Accuracy, F1-score | <75% | Lagging | Internal evaluations | <60% |

Table 6: SPI Set for claim C4.1

**SPI set C5.1**  The proxy tasks are representative

| SPI ID | SPI Title | SPI Description | Unit | Example | Leading/ Lagging | Evidence | Example Threshold |
|---|---|---|---|---|---|---|---|
| C5.1 SPI 1 | Number of new Task Suites Published | Measures the number of new task suites published in the last 30 days by AI Safety Institutes, Frontier Developers, and independent organisations that provide new potential proxies. | Count | 2 | Leading | Research insights | Track for trends |
| C5.1 SPI 2 | Response to Published New Task Suites | Average time taken to evaluate and, if relevant, incorporate new tasks from published suites into proxy evaluations. | Days | 5 | Leading | Research insights | 14 |
| C5.1 SPI 3 | Number of new Published Benchmarks | Measures the number of new cyber benchmarks including the model published in the last 30 days | Count | 2 | Leading | Research insights | Track for trends |
| C5.1 SPI 4 | Response to Published Benchmarks | Average time taken to respond to new task-related updates provided by benchmarking, including integration into existing proxy tasks. | Days | 5 | Leading | Research insights | 14 |
| C5.1 SPI 5 | TTP Proxy Task Validation | Number of new Tactics, Techniques, and Procedures (TTPs) identified through CTI that are not reflected in the proxy tasks. | Count | 3 | Leading | Cyber Threat Intelligence | 2 |

| C5.1 SPI 6 | Proxy Alignment with Emerging Threats | Number of emerging threats identified through CTI that are not reflected in the proxy tasks. | Count | 1 | Leading | Cyber Threat Intelligence | 2 |
| C5.1 SPI 7 | Proxy Coverage Against Industry Standards | Percentage of proxy tasks that align with those used in industry benchmarks or by other Frontier Developers. | % | 80% | Leading | Research Insights, Industry Bodies | 70% |
| C5.1 SPI 8 | Incident Proxy Task Correlation | Measures the average percentage of tasks involved in real-world incidents that are adequately covered by the current proxy tasks | % | 85% | Lagging | Incidents | 90% |
| C5.1 SPI 9 | Near Miss Proxy Task Correlation | Measures the percentage of skills and actions involved in averted real-world incidents that are adequately covered by the current proxy tasks | % | 85% | Lagging | Near Misses | 90% |

Table 7: SPI Set for claim C5.1

| SPI set C5.2 The scoring is adequate | | | | | | | |
|---|---|---|---|---|---|---|---|
| **SPI ID** | **SPI Title** | **SPI Description** | **Unit** | **Example** | **Leading/ Lagging** | **Evidence** | **Example Threshold** |
| C5.2 SPI 1 | Task Difficulty Benchmark Comparison | Measures the alignment of cyber capability levels assigned to tasks e.g. CTFs against industry benchmarks or other Frontier Developers. | % | 85% | Leading | Research Insights, Industry Bodies | 90% |
| C5.2 SPI 2 | Automated Test scoring alignment with Industry Practices | Measures how closely the pass/fail criteria for proxy tasks align with industry standards and best practices e.g. use of manual corrections | % | 90% | Leading | Research Insights, Industry Bodies | 90% |
| C5.2 SPI 3 | Number of new Automated Test scoring methodologies published | Measures the number of new scoring methodologies or best practices for automated test scoring (e.g., number of test attempts like Pass@10) published in the last 90 days | Count | 2 | Leading | Research Insights | Track for trends |
| C5.2 SPI 4 | Task based probing scoring alignment with Industry Practices | Measures how closely the scoring for task-based probing/ automated evaluations with human-oversight align with industry standards and best practices e.g. hinting, trajectory interventions | % | 90% | Leading | Research Insights, Industry Bodies | 90% |
| C5.2 SPI 5 | Number of new task based probing scoring methodologies published | Measures the number of new scoring methodologies or best practices for task-based probing/ automated evaluations with human-oversight published in the last 90 days | Count | 2 | Leading | Research Insights | Track for trends |
| C5.2 SPI 4 | Human uplift test scoring alignment with Industry Practices | Measures how closely the scoring for human uplift tests align with industry standards and best practices e.g. hinting, trajectory interventions | % | 90% | Leading | Research Insights, Industry Bodies | 90% |
| C5.2 SPI 5 | Number of new human uplift test scoring methodologies published | Measures the number of new scoring methodologies or best practices for human uplift testing published in the last 90 days | Count | 2 | Leading | Research Insights | Track for trends |
| C5.2 SPI 6 | Response Time to Evaluate New Best Practices | Measures the time taken to evaluate and, if applicable, adopt new best practices or scoring methodologies suggested by research publications or benchmarking reports. | Days | 15 | Leading | Research Insights | 30 |

Table 8: SPI Set for claim C5.2

**Evaluations**

Once the adequacy of the proxy tasks has been established, the final claims in the safety case sketched by Goemans et al. (2024) address the suitability of the evaluation setups to provide sufficient evidence of the AI's inability to perform these proxy tasks. This includes demonstrating that evaluations are done in accordance with best practices.

As noted by AISI (2024b), designing an evaluation to measure a Large Language Model (LLM)'s capabilities presents challenges, involving trade-offs between labour intensity, result interpretability, and the accuracy with which it reflects real-world usage. As such, methodologies in use by developers and researchers are frequently evolving. To stay aligned with best practices, it is necessary to monitor, evaluate, and incorporate improvements to these methodologies. Notably, the first draft EU AI Code of Practice (European Commission, 2024) specifies the use of 'best-in-class evaluations' as a measure for assessing the capabilities and limitations of general-purpose AI models with systemic risk.

Goemans et al. (2024) include monitoring of leaderboards of LLM agent benchmarks and similar channels to obtain visibility of enhancements to elicitation techniques as evidence in E8.2 for the claim that there is no under elicitation of capabilities. They note, via the defeater D8.1, that there are challenges with this monitoring as there may be disincentives for some actors to disclose improvements. Prosaic under elicitation, where capabilities are understated due to a lack of rigour in evaluations, is identified as a possible cause of under elicitation of capabilities (C8.1). External monitoring can help identify this by flagging where internal choices around depth of evaluations and iteration attempts differ from industry best practices. Another potential cause of under elicitation identified in the safety case is sandbagging (C8.2), where a model strategically underperforms on an evaluation. While Goemans et al. (2024) present sandbagging as a currently hypothetical failure mode, recent research has successfully demonstrated that models will adjust their capabilities in response to certain prompts (van de Weij, et al., 2024), and as such defining SPIs to monitor the likelihood of this is a means of having early warning of potential emergence of this type of behaviour.

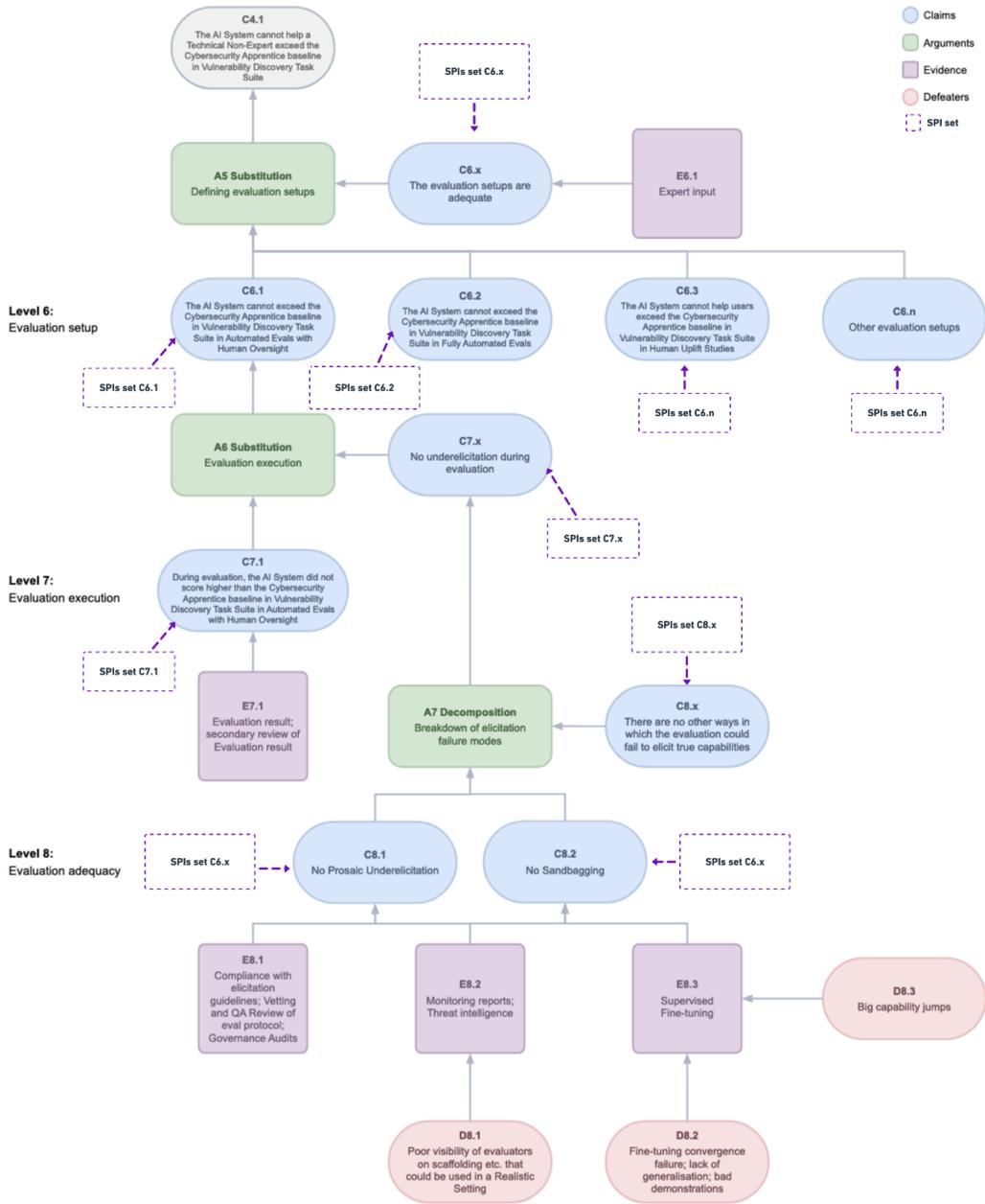

Figure 5: Extract from Safety Case sketch by Goemans et al. (2024): Evaluations with added SPI sets

Detailed Safety Performance Indicators are proposed to monitor the validity of the claims relating to the efficacy of evaluations, for which three examples have been provided (C7.x, C8.1, C8.2).

| SPI set C7.x No under elicitation during evaluation | | | | | | | |
|---|---|---|---|---|---|---|---|
| **SPI ID** | **SPI Title** | **SPI Description** | **Unit** | **Example** | **Leading/ Lagging** | **Evidence** | **Example Threshold** |
| C7.x SPI 1 | Incidents Elicitation Gap Analysis | Measures the percentage of incidents in the last 90 days where the real-world capability elicited was not adequately captured during evaluation. | % | 8% | Lagging | Incidents | 10% |
| C7.x SPI 2 | Near Misses Elicitation Gap Analysis | Measures the percentage of near misses or averted incidents in the last 90 days where the real-world capability elicited was not adequately captured during evaluation. | % | 12% | Lagging | Incidents | 20% |
| C7.x SPI 3 | Research Informed Under Elicitation Findings | Measures the number of newly published benchmarks and research insights in the last 90 days that suggest under elicitation of harmful behaviors during evaluation. | Count | 1 | Leading | Research Insights | 2 |

Table 9: SPI Set for claim C7.x

| SPI set C8.1 No Prosaic Under elicitation | | | | | | | |
|---|---|---|---|---|---|---|---|
| **SPI ID** | **SPI Title** | **SPI Description** | **Unit** | **Example** | **Leading/ Lagging** | **Evidence** | **Example Threshold** |
| C8.1 SPI 1 | Deviation from Industry Average Elicitation Iterations | Measures the %+/- deviation from average elicitation iterations used internally in the most recent evaluations, compared to external benchmarks. | % | +10% | Leading | Research Insights | <-20% |
| C8.1 SPI 2 | Use of Agentic Scaffolding Compared to Industry Standards | Measures the %+/- deviation from average number of agentic scaffolding techniques applied during the most recent evaluations, compared to industry standards or best practices. | % | -20% | Leading | Research Insights | <-20% |
| C8.1 SPI 3 | Use of Fine-Tuning Compared to Industry Standards | Measures the extent as a % alignment to calendar days spent fine-tuning during the most recent evaluations, compared to industry standards or best practices. | % | 90% | Leading | Research Insights | 90% |

Table 10: SPI Set for claim C8.1

| SPI set C8.2 No Sandbagging | | | | | | | |
|---|---|---|---|---|---|---|---|
| **SPI ID** | **SPI Title** | **SPI Description** | **Unit** | **Example** | **Leading/ Lagging** | **Evidence** | **Example Threshold** |
| C8.2 SPI 1 | Noise injection sandbagging evaluation | Performance improvement on a benchmark task when adding noise to model weights, as compared to non-sandbagging model; evaluation approach originally described by Tice et al (2024) | Percent | 10% | Lagging | Internal evaluations | 10% |
| C8.2 | Deviation from Previous | Measures the % deviation in capabilities compared to the previous | % | -8% | Leading | Internal - | ≤5% |
| SPI 2 | Capability Scores | evaluation round on the previous model version. Predictions based on additional compute, fine-tuning, or other scaling inputs are used to set threshold. | | | | Capability evaluations | |
| C8.2 SPI 3 | Capability Performance Outlier Analysis | Measures the relative deviation of a specific capability's performance compared to trends in other capabilities from the same evaluation round. Significant underperformance may indicate sandbagging. E.g. Vulnerability Exploitation lags by 15% while others improved by 5-8% | % deviation from mean | -10% | Leading | Internal - Capability evaluations | ≤10% deviation from capability mean |
| C8.2 SPI 4 | Performance Drift Analysis Post Deployment | Measures the degree of performance improvement between pre- and post-deployment evaluations as a % performance change. Significant increases may indicate sandbagging. | % performance change | +15% | Leading | Internal - Capability evaluations | ≤5% performance increase |

Table 11: SPI Set for claim C8.2

### 4.3.4 Prioritising Safety Performance Indicators

The approach taken in enumerating the Safety Performance Indicators (SPIs) for these example claims prioritises coverage and significance for understanding safety over the feasibility of automation, with the aim of presenting a more thorough list for consideration. However, it is recognized that feasibility of monitoring will be a key factor for frontier AI developers evaluating these SPIs, particularly where a lack of automation could result in significant time overheads. Additionally, there may be time lags in obtaining some data, which could reduce its value for proactive safety measures. Nonetheless, by presenting a set of aspirational SPIs, we aim to surface these challenges and inspire further research and development into improving automated access to data, such as APIs providing AI-model-specific cyber intelligence feeds.

To perform an initial evaluation of Safety Performance Indicators (SPIs), we propose a set of criteria that balance their significance for safety and feasibility of monitoring. These criteria aim to provide a structured approach to assess the value and practicality of each SPI, helping frontier AI developers to focus on those that offer the greatest impact on safety while remaining feasible to implement.

| Criteria Categories | Criteria Dimension | Scoring (0-5) |
|---|---|---|
| Significance for Safety | Relevance to Safety Claims: *How directly the SPI validates or challenges safety claims.* | 5: Directly tied to a core claim critical to the safety case. 3: Related to an important but secondary claim. 0: Minimal relevance to safety claims. |
| | Importance of the Claim: *The centrality of the associated claim to the overall safety case.* | 5: A parent claim that supports multiple sub-claims. 3: A significant but not foundational claim. 0: A peripheral claim with limited impact. |
| | Proactive Monitoring: *Whether the SPI serves as a leading indicator to identify risks before they occur.* | 5: Strong leading indicator for early risk identification. 3: Balanced between leading and lagging indicators. 0: Purely lagging, providing only retrospective insights. |
| Feasibility of Monitoring | Measurability: *How easily the data for the SPI can be collected.* | 5: Automatically measurable with minimal effort. 3: Requires some manual intervention or estimation. 0: Data is difficult or impractical to collect. |
| | Timeliness: How quickly actionable insights can be generated. | 5: Real-time or near-real-time monitoring. 3: Moderate delay in obtaining actionable data. 0: Significant delay in obtaining or processing data. |
| | Feasibility of Implementation: The availability of tools, resources, or infrastructure to implement the SPI. | 5: Fully supported by existing systems and tools. 3: Requires moderate development or investment. 0: Requires significant resources or new infrastructure. |

Table 12: SPI prioritisation criteria

To illustrate how the criteria can be used to evaluate SPIs, we take those in figure 3: **SPI set C2.1** (The AI system would not uplift threat actors using conventional cyberattacks in any realistic setting, even absent any safeguards), and provide initial ratings for these, averaging the scores of the criteria dimensions to give overall ratings for significance for safety and feasibility of monitoring.

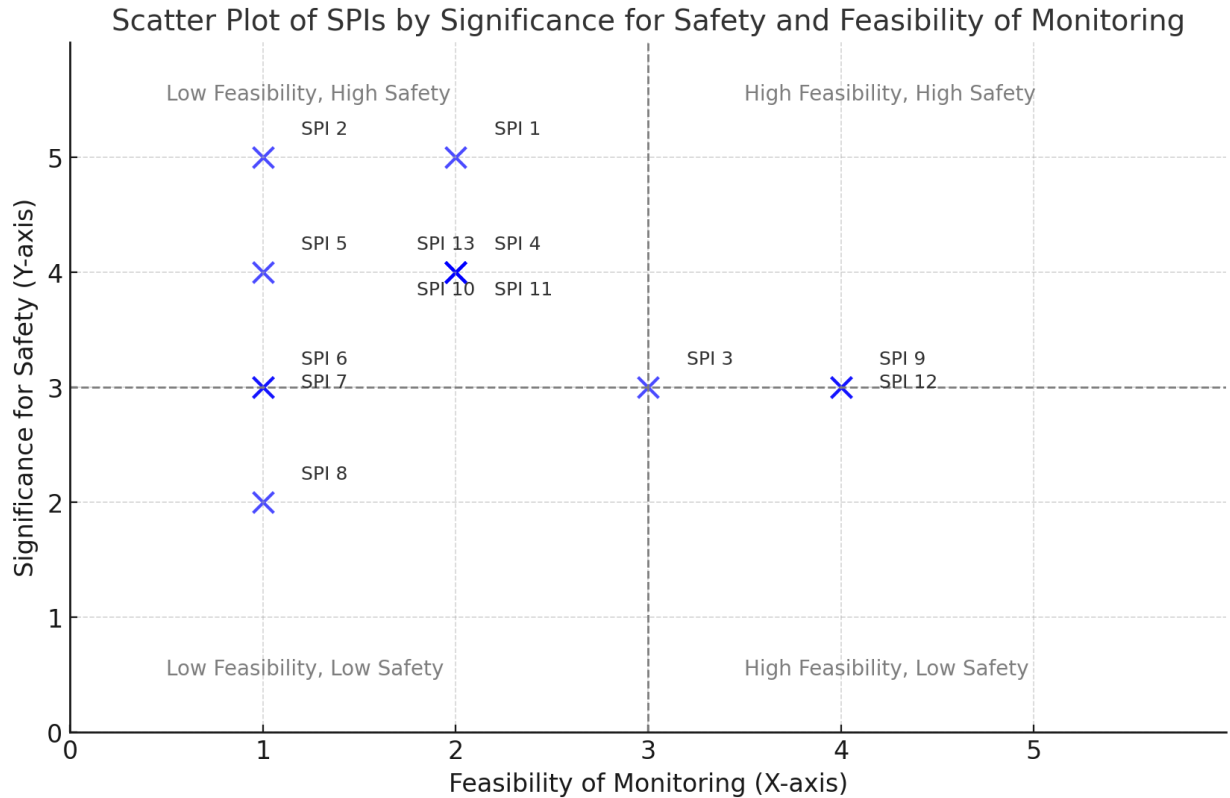

Figure 6: Scatter Plot of SPIs by Significance for Safety and Feasibility of Monitoring

The scatter plot highlights a general challenge: SPIs with the highest significance for safety, such as # of incidents - conventional attacks (SPI 1) and Financial losses - conventional attacks (SPI 2), tend to be less feasible to monitor. However, evaluating SPIs in this way could identify candidates where investing in automation would provide significant value, for example by providing API streams providing tailored cyber threat intelligence specific to AI models.

Evaluating SPIs against these criteria can also help identify promising candidates for initial monitoring to test dynamic safety case implementation. For the claim C2.1, Research papers - other source (SPI 9) and Dark web mentions - conventional attacks (SPI 12) stand out as metrics which would be more feasible to track using existing research and cyber intelligence sources.

### 4.4 Scenarios - safety case updating process examples

Having described the DSCMS in the previous section, introducing CSA and specifying SPIs for the cyber inability case, we now demonstrate the updating process using a set of hypothetical system safety change scenarios as examples. The detail of the updating process was given in section 4.1 (see figure 1), here we summarise and extend the process description to describe change impact recovery actions at the level of the DSCMS itself. Governance level response actions are described in section 5. The high-level process for updating the safety case in response to scenarios that trigger SPI threshold breaches is as follows:

1. Initially, the DSCMS operates normally, monitoring SPIs. When an SPI breaches its predefined threshold, the DSCMS executes a consistency check to assess which elements of the safety case are invalidated.
2. If no argumentation elements are affected, the monitoring continues uninterrupted. If certain argumentation elements are invalidated, these elements are flagged for review.

3. Findings from the consistency checks are then provided to governance structures, which determine appropriate recovery action interventions.
4. Once these interventions are implemented, the safety case is revalidated by re-generating evidence and/or re-running consistency checks with modified SPIs to ensure the corrective action taken has effectively mitigated the impact on the safety case.

This is a continuous cycle of monitoring, assessment, decision-making, action, and revalidation that maintains or recovers system safety objectives over time.

---

**Scenario 1: Risk model threshold breach (top-level claim invalidated)**
**Change impact: Highest**
**Consistency rules:**

- A sudden spike in losses from cyberattacks targeting Critical National Infrastructure (CNI) invalidates the top-level claim of the safety case.
- Thresholds for SPIs C3.1 SPI 2 and 5, which track aggregated losses and month-on-month increases in loss amounts, are breached due to $2M in damages attributed to AI-assisted vulnerability discovery and exploitation by non-expert attackers.
- These breaches propagate to SPIs C2.1 SPI 2 and 5, invalidating the claim C2.1 that the AI system cannot uplift technical non-experts in conducting such attacks in realistic settings, and ultimately invalidate the top-level claim that deploying the AI system does not pose unacceptable risk.

**SPIs breached:** C3.1 SPI 2 & 5, C2.1 SPI 2 & 5, C1.1 SPI 2 & 5
**Claims invalidated:** C3.1, C2.1, C1.1, C0

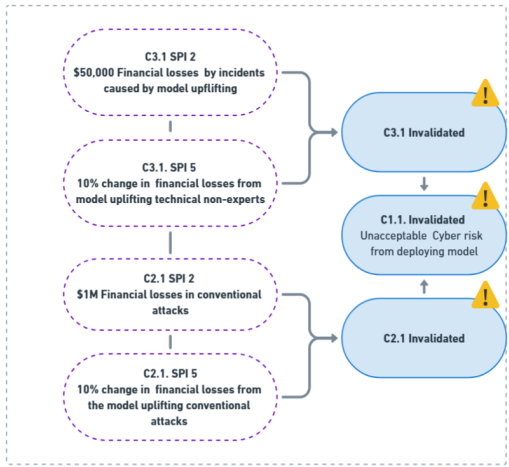

**Safety case change impact:**
Fundamentally the 'inability' basis of the safety case no longer holds, and an alternative safety case must be made that moves to 'control' type arguments - the level after 'inability' arguments in Clymer et al. (2024), also covered in Buhl et al. paper (2024) - for example, using safeguards such as fine-grained output filtration or restricting user access where patterns indicative of known threat actor tactics, techniques, and procedures (TTPs) relating to the attack type are detected.

Figure 7: Scenario 1 change impact

---

**Scenario 2: Novel attack vector discovered**
**Change impact: High**
**Consistency rules:**

- Monitoring identifies a new attack vector through two cyberattacks deviating from known patterns, causing losses of £600k.
- These breaches trigger thresholds on C2.2 SPI 1, 2, 4, and 5, which track incident numbers and associated losses, but remain below the thresholds at C1.1 (assumed for the sake of this example).
- As a result, the breach does not propagate further, and the top-level claim C0 remains valid.

**SPIs breached:** C2.2 SPI 1, 2, 4 & 5
**Claims invalidated:** C2.2

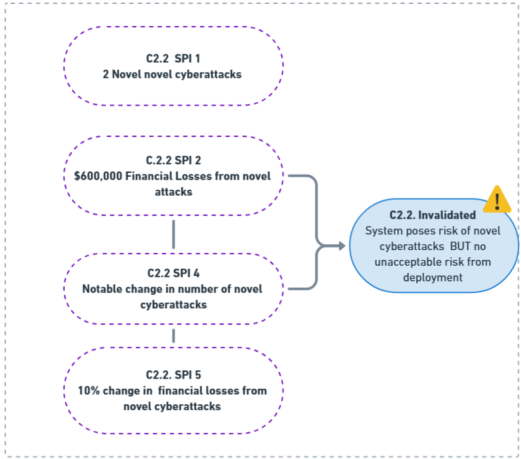

**Safety case change impact:**
To revalidate the safety case, a new claim (C3.4) could be added to define a risk model capturing the newly identified attack vector, threat actor, and target based on details of exploited vulnerabilities, methods used, system impacts, and timeline of the attacks identified in the logs and forensic data from the two cyberattacks. Additional claims could be introduced to define proxy tasks e.g. C4.3, based on the harm vector that the threat actor exploited, and evaluation setups e.g. C6.4, to link evidence for the ongoing testing of model capabilities that could enable this new risk model to be exploited.

Figure 8: Scenario 2 change impact

---

**Scenario 3: A model update shows capability uplift in automating low-medium threat level threat actor cyberattacks**
**Change impact: high/medium**
**Consistency rules:**

- A new version of a frontier model, post training (but pre-mitigation - e.g. addition of safeguards) was evaluated on the InterCode-CTF evaluations and the results indicated a significant jump in the fraction of challenges solved successfully. Model performance increased from a ~40% to a ~70% success rate on easy-medium difficulty reverse engineering, cryptography and forensics tasks, indicating a significant uplift of low-medium human threat actor capabilities.
- Thus, SPI 3 of claim C6.3 was breached, leading to it being invalidated as well.

**SPIs breached:** C6.3 SPI 3
**Claims invalidated:** C6.3

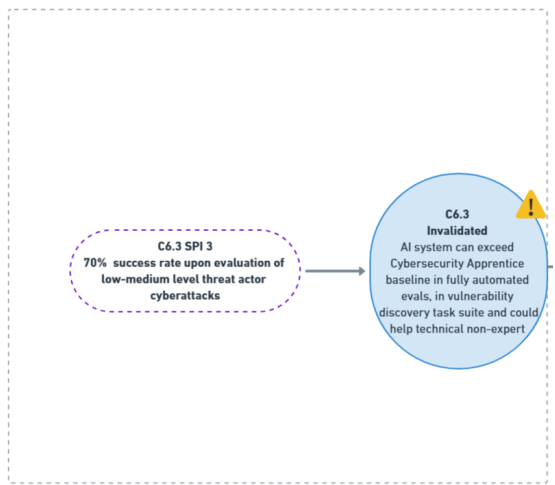

**Safety case change impact:**
Given this breached SPI is indicating that the model has now developed some capabilities to help non-technical experts in vulnerability discovery, inability arguments could no longer be relied on in the safety case. Additional mitigations might need to be introduced to ensure that the model can be approved for deployment. For drastic jumps of capabilities, 'control' type arguments (Clymer et al., 2024, Buhl et al., 2024) would need to be introduced to explain how the AI system is restricted from using those capabilities capable of causing serious harm.

Figure 9: Scenario 3 change impact

**Scenario 4 - Response time to include new proxy task benchmarking increases**
**Change impact: medium/low**
**Consistency rules:**

- The average response time taken to respond to new task-related updates provided from benchmarking, including integration into existing proxy tasks has been found to have increased to 31 days.
- This now breaches the threshold of C5.1 SPI 4, but does not breach C4.1x (assumed for the sake of this example)

**SPI breached:** C5.1 SPI 4
**Claims invalidated:** C5.1

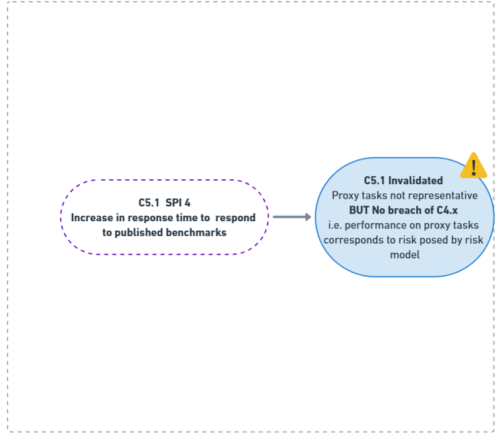

**Safety case change impact:**
High level change - The breach of C5.1 SPI 4 invalidates the claim C5.1, as published benchmarks are not being reviewed and incorporated in a timely manner, increasing the likelihood that proxy tasks are not representative. This could be addressed via a governance level intervention to correct the delay.

Figure 10: Scenario 4 change impact

---

**Scenario impact categories**

By analysing a sample of realistic scenarios that could trigger an SPI breach and defining how these would impact the underlying safety case, we can identify the types of insights that the monitoring of a safety case can provide, assess the severity of these impacts, and determine the stakeholders who may need to be informed. We propose a categorisation schema aligned with the types of changes that would be relevant to different internal and external stakeholders, enabling the development of tailored notification requirements for various change impacts.

| Insight category | Description | Impact Severity | Scenario 1 | Scenario 2 | Scenario 3 | Scenario 4 |
|---|---|---|---|---|---|---|
| Offensive cyber capability increase | Inability safety case no longer holds, capability increase requires controls to be implemented and control type claim to be added to safety case | Highest | **YES** | NO | **YES** | NO |
| Systemic impact pathway demonstration of risk model | AI system is changing offensive cyber threat landscape, targeted defensive measures may be required | High/ medium | **YES** | **YES** | **YES** | NO |
| Internal process performance | Monitoring indicates internal evaluations, controls, or processes are not functioning as intended and may require adjustments to address emerging risks or improve effectiveness | Medium/ low | NO | NO | NO | **YES** |

Table 13: Scenario impact categories

# 5 Integrating DSCMS with frontier AI developer and national governance systems

In the previous sections, we detailed requirements for a DCSMS, introduced the CSA approach as a basis for such a DSCMS, and identified SPIs for the example offensive cyber capabilities safety case template. We then detailed some example change scenarios to demonstrate the dynamic update process. Whilst these sections focused on the safety case and its management directly, in this section we expand our view to show how the DSCMS would need to integrate into higher level governance systems in order to be effective in initiating higher-level action to maintain or restore safety. The primary governance structures of interest are those internal to the frontier AI developers, and national governance (government, regulators and similar relevant stakeholders).

## 5.1 Integrated governance and dynamic safety assurance framework

Within a frontier AI developer, the DSCMS must be embedded within the broader organisational governance structure, underpinned by a Safety Management System (SMS). The SMS provides four key contributions that support the implementation and success of the DSCMS namely 1) policies, 2) processes, 3) decision-making mechanisms, and 4) responsibility and accountability definitions needed to transform DSCMS change impact into concrete steps that maintain or restore safety objectives, both internally and if needed, engagement with external stakeholders such as governments or regulators.

Scenario example:

Consider an example scenario where the DSCMS detects a significant increase in real-world incidents suggesting that the frontier AI model's capabilities are being misused for offensive cyber activities. Previously, the SPI related to this risk was stable. Now, the DSCMS issues an alert indicating that a threshold has been breached:

- *Notification* - the SMS stipulates that the Responsible Scaling Officer (RSO, described later) receives immediate notification.
- *Action and tracing* - in accordance with predefined protocols, the SMS directs safety teams to investigate the cause, refine SPIs, implement additional safeguards (e.g., stricter access control), and log process for traceability.
- *Accountability and oversight* - depending on severity, the SMS may escalate the issue to executive leadership (e.g., the CEO or Board) and possibly produce reports for governance entities. Every decision, mitigation measure, and communication step is logged, ensuring clarity, accountability, and a robust audit trail.

## Policies and processes

By establishing clear, organisation-wide policies and processes for collecting, validating, communicating, and acting on evolving safety information, the SMS ensures that changes in SPIs, newly identified vulnerabilities, and shifting safety arguments lead to timely, evidence-based interventions. These policies provide criteria and thresholds for action, formalise review and approval workflows, define roles and responsibilities, ensure traceability and auditability, and support continuous improvement. As a result, all relevant stakeholders, from technical staff to executive leaders and regulators, operate within a cohesive, life cycle-based safety framework that integrates DSCMS outputs into safety decision-making, enabling responsive, robust governance throughout the AI system's lifecycle.

## Decision gates and lifecycle integration

One of the four contributions of the SMS as defined above is providing decision-making mechanisms that support the development and maintenance of the safety case throughout the frontier AI system's lifecycle. One effective approach is to introduce decision gates at critical points in the system's progression, such as when moving from a development environment to a production-level deployment. At each gate, before advancing to the next stage, the current safety case is reviewed. If DSCMS outputs show that certain claims are no longer well-supported or that new risks have surfaced (e.g., an SPI now revealing heightened cyberattack capabilities), these gates prompt timely interventions in the safety case process. For example, after a decision gate, the SMS may require additional testing, implement new security measures, or mandate re-verification of safety arguments before allowing the system to proceed. This enables incremental, evidence-based verification and validation (V&V) throughout the lifecycle, reducing the likelihood of late-stage surprises. Appendix A provides more details about life-cycle based decision gates and their support of incremental V&V .

A key requirement from Section 3 is REQ-002: 'The DSCMS shall maintain the safety case for the frontier AI system throughout its entire lifecycle, as defined in [Lifecycle Definition Document].' To fulfill this requirement, the DSCMS lifecycle should run in parallel with the frontier AI system's lifecycle. Figure 11, a conceptual lifecycle diagram, illustrates how the system's stages (e.g., Planning, Development, Evaluation, Deployment, and Decommissioning) align with DSCMS activities (e.g., Safety Case Planning, SPI Definition, Continuous Monitoring, Consistency Checks) and decision gates (G1, G2, G3, etc.). At the end of each stage, a decision gate assesses both the frontier AI system and its safety case to confirm that claims and SPIs remain relevant, valid, and up-to-date before proceeding to the next stage.

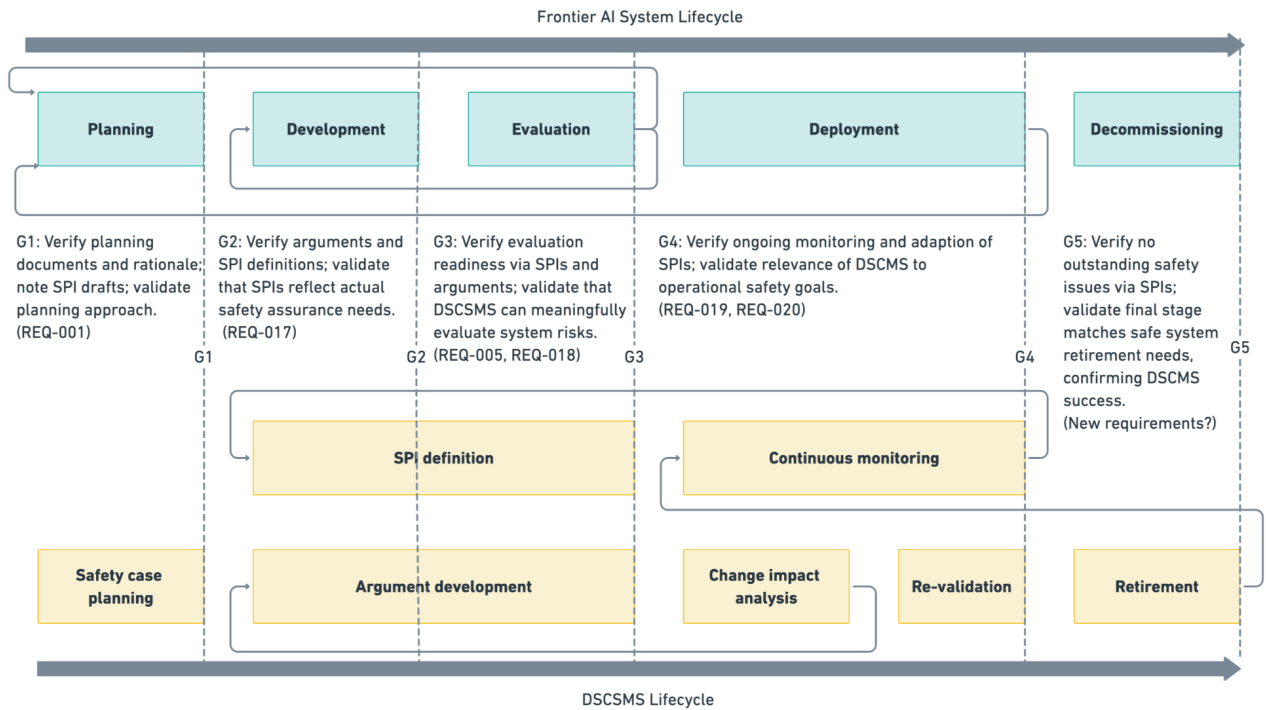

Figure 11: Conceptual lifecycle diagram aligning frontier AI system development stages, DSCMS activities, and decision gates

This continuous alignment of the two lifecycles inherently satisfies REQ-002, as the safety case remains a living artifact that evolves with the frontier AI system. Early phases (e.g., Planning at G1) might be guided by REQ-001 (establishing the rationale for safety claims), while mid-development checks could address REQ-017 (SPI definition and management), and late-stage decisions might align with requirements ensuring readiness for deployment or safe rollback options if conditions change.

The lifecycle phases shown in Figure 11 are presented in a generic, linear, and hierarchical manner for illustration; in practice, these phases and their attributes will vary by organisation. Moreover, phases may overlap or revert to earlier ones, reflecting planned iterations (e.g., scheduled reassessments or developments with significant technological uncertainty) or unplanned events (e.g., triggered by a DSCMS alert about a newly discovered vulnerability).

Fully automating the impact analysis of changes to the frontier AI model is difficult because the relationships between modifications and their downstream effects on the safety case can be highly non-linear and context-dependent. Consequently, subject matter experts remain essential during evaluation activities, and the balance between human oversight and automation will differ depending on the stage in the lifecycle, the organisation's risk tolerance, available resources, and the nature of the frontier AI system's risks. Integrating decision gates into the process provides a means to incorporate expert judgment, manage risks, and maintain alignment with the organisation's overall risk management strategy.

**Roles and responsibilities within the SMS**

A well-defined SMS delineates who responds to DSCMS outputs, what actions they can take, and how these actions feed back into the lifecycle. A definition might look like this:

| Role | Responsibility |
|---|---|
| Responsible Scaling Officer (RSO) | Reviews DSCMS-triggered alerts, determines escalation pathways |
| Executive Leadership (CEO, Board) | Receives high-severity breach notifications, authorizes major strategic or governance-level interventions |
| Safety and Compliance Teams | Investigate DSCMS findings, refine SPIs, implement mitigations, maintain records for traceability |
| External Oversight Entities & Regulators | Receive compliance reports (REQ-023), request further evidence, impose conditions, recommend controls |

Table 14: Roles and responsibilities within the SMS

**Additional DSCMS requirement considerations**

- REQ-023 (Governance Reporting Interface)
  This requirement ensures that the DSCMS provides a standardised governance interface for reporting to regulators and oversight bodies. By allowing internal and external stakeholders to request and review safety evidence, implementation of REQ-023 facilitates communication, fosters transparency, and promotes accountability as the frontier AI system and its associated safety arguments evolve.
- REQ-025 (Data Security and Access Control)
  To safeguard sensitive data—such as proprietary models, SPI telemetry, and compliance records, REQ-025 mandates adherence to a defined Data Security Plan. By enforcing strong data protection measures, this requirement maintains the integrity and credibility of the safety assurance process, meeting both ethical and legal obligations.
- [REQ-X] Decommissioning Requirement (End-of-Life Assurance)
  Complementing early and mid-lifecycle requirements (like REQ-001 and REQ-002), a decommissioning-focused requirement, would ensure that no unresolved safety issues persist at a system's retirement. By applying the same rigour at the end of a system's life, organisations maintain stakeholder trust, prevent residual risks, and uphold a continuous standard of life-cycle based safety assurance.

Ultimately, integrating the DSCMS into the SMS establishes a continuous feedback loop. As the AI system evolves, so too do the SPIs, safety arguments, and governance landscape. By continuously re-verifying claims at each decision gate, the organisation develops an adaptive governance framework capable of keeping pace with frontier AI's inherent complexity and rapid change. Yet, this adaptability also brings added lifecycle management complexity. The interplay of iterative lifecycle phases, automated impact analyses, and the need for human judgement requires a careful balance. Developers must consider, on a risk and opportunity basis, whether to opt for faster, more automated updates with less overhead (albeit at the risk of reduced accuracy in complex and sensitive risk scenarios), or to rely more heavily on human-driven, context-sensitive oversight, which is thorough but inevitably slower and more resource-intensive.

By monitoring SPIs and triggering re-evaluation of safety claims when SPI thresholds are breached, the DSCMS could enhance both operational control and safety planning within frontier AI organisations. At the same time, it could inform and strengthen national governance measures, enabling policymakers and regulators to respond more dynamically to emerging risks. In the following sections, we outline how DSCMS capabilities can be integrated into frontier AI developers' internal risk management practices (Section 5.2) and into broader national governance structures (Section 5.3).

## 5.2 DSCMS integration with frontier developer governance

**Internal risk management considerations within frontier AI developers**

Current risk management practices within frontier AI developers are driven by voluntary commitments towards responsible policies (Anthropic, 2024; OpenAI, 2024, Google DeepMind, 2024; DSIT, 2024) which predominantly consist of capability evaluations that test the elicitation of certain harmful behaviors and use those results to devise mitigations or safeguards where applicable. While such practices have so far been somewhat successful in avoiding the most catastrophic outcomes, they have been criticised for being limited to technical evaluations done in isolation without much regard for the larger context and usage of real-world deployments.

Frontier AI developers can integrate dynamic safety cases for cyber-inability to strengthen their existing risk management systems. As an example, we will look at Anthropic's Responsible Scaling Policy (RSP). A dynamic safety case is especially useful for operationalising monitoring, rapid remediation, and defense in depth components of the RSP. By automatically monitoring potential SPIs listed in Section 4.3 such as cyber incidents involving the model and mentions on the darkweb, the dynamic safety case can support concrete actions which are triggered by changes in objective data.

Depending on the exact structure of the safety case adopted, these changes could provide a useful mechanism for triggering changes such as preliminary or full capability evaluations related to cyber-risks or revisions to the RSP. This can both provide an indicator for when evaluations may be needed earlier than the default scaling and time based thresholds that would normally trigger evaluations. Alternatively, they may provide support for forgoing in-depth evaluations if they indicate little has changed for a deployed model. The SPIs can also provide additional data on top of the capability evaluation snapshots for the Responsible Scaling Officer (RSO) and CEO to make judgments about a model's safety and help calibrate their forecasts of model impacts, and confidence levels in those forecasts.

Additionally, the dynamic safety case can be structured so that changes in especially important SPIs automatically notify additional parties such as the Long-Term Benefit Trust and government partners such as the US & UK AISI, or trigger the necessary steps for the RSO and relevant internal parties to do so. It can also provide a method by which oversight bodies like the board could be automatically called to meet and review major sudden changes in safety and exercise oversight. Changes in SPIs can be classified by urgency levels based on the importance of the SPI and magnitude of the change in a given time period. The highest priority indicators could be tied to the RSP's AI Security Level thresholds. If appropriately set, a relevant scenario could trigger the advancement to the next ASL level, e.g. from ASL-2 to ASL-3. These changes could be automatically monitored and transmitted through a secure communication protocol to the relevant party. A minor change can go to the internal safety team whereas a major change should also go straight to the RSO. Government organisations like the US and UK AISI could also be looped into major changes, which we discuss in section 5.3.

**Integrating Dynamic Safety Cases into Anthropic's RSP: suggested language for RSP section 3.3 & 4.3**

The text below is provided as a suggestion for how section 3.3 and 4.3 could be modified to address dynamic safety cases.

1. 'We may additionally implement dynamic safety cases for dangerous capabilities assessment. They may be based on Safety Performance Indicators (SPIs) for each threat model that can be automatically monitored. Changes in these SPIs will be categorised by severity level depending on the importance of the indicator and magnitude of the change. Depending on the severity, changes to an SPI will trigger notifications to the appropriate parties and subsequent action to ensure risks are mitigated. Below is a table summarising SPIs, their classification, and consequences. Low priority changes will result in changes to our dangerous capability evaluations and be logged in the next capability report. High priority changes will notify the RSO and CEO.

They will also trigger a full capability re-evaluation, and, if necessary, collaboration with experts, partner organisations and governments to apply additional safeguards. Highest priority changes may result in a pause in training or deployment, a change in safety team priorities, and a re-evaluation of the RSP and safety case for that capability.

2. SPIs can serve as an additional (ex-ante) indicator for when further safeguards are needed to address dangerous capabilities risks. We still plan to do full capability evaluations of our models based on scaling and time based thresholds that could cause significant changes in capabilities.'

**Using Safety Performance Indicator (SPI) changes for Dynamic Safety Case Management**

Taking the system safety change scenarios introduced in section 4.4, the figures below show plausible governance responses.

In scenario 1 (risk model threshold breach, C3.1 SPIs 2 and 5 contain clear and severe indicators that should immediately notify decision makers and trigger processes to enact additional safety action - e.g. control measures restricting model availability, such as removing API access, blocking integrations with specific agentic tools, filtering out certain types of requests more thoroughly, or temporarily banning any users who cannot be verified as trustworthy. However, for scenario 4 (response time for including new proxy task benchmarking) indicators like average time taken to respond to benchmark updates (C5.1 SPI 4) would be lower priority since they do not indicate immediate or imminent severe harm. Changes in this indicator could solely notify the internal safety team, prompting an update to their evaluations to bring them in-line with the latest benchmarks.

---

**Scenario 1: Risk model threshold breach (top-level claim invalidated)**

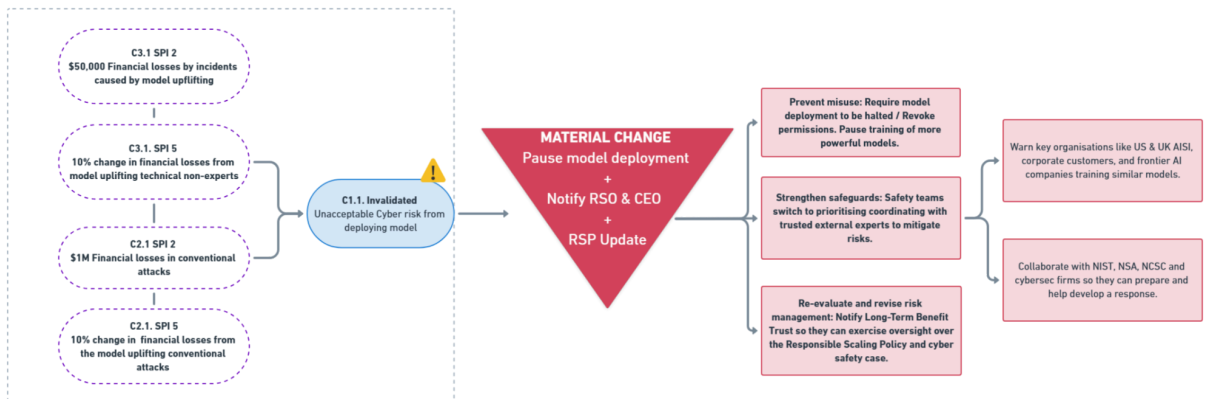

**Scenario 2: Novel attack vector discovered**

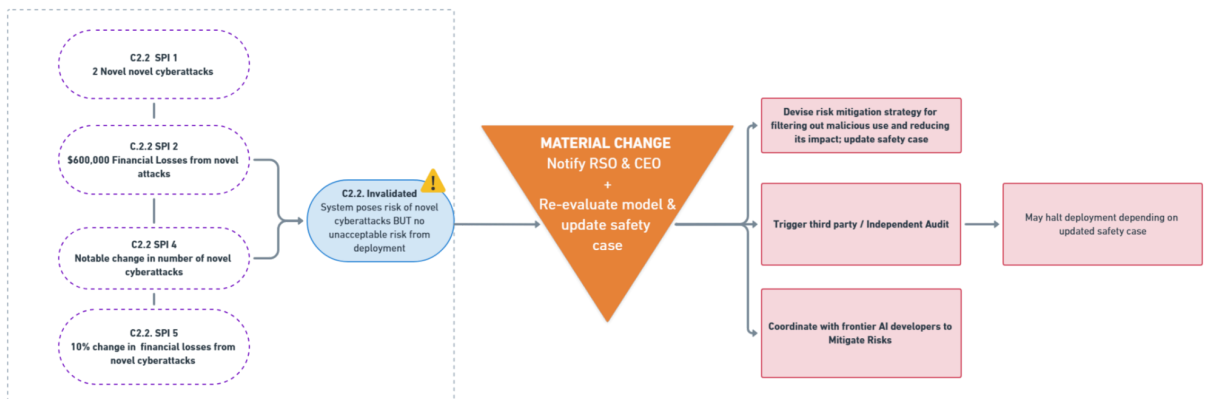

**Scenario 3: A model update shows capability uplift in automating low-medium threat level threat actor cyberattacks**

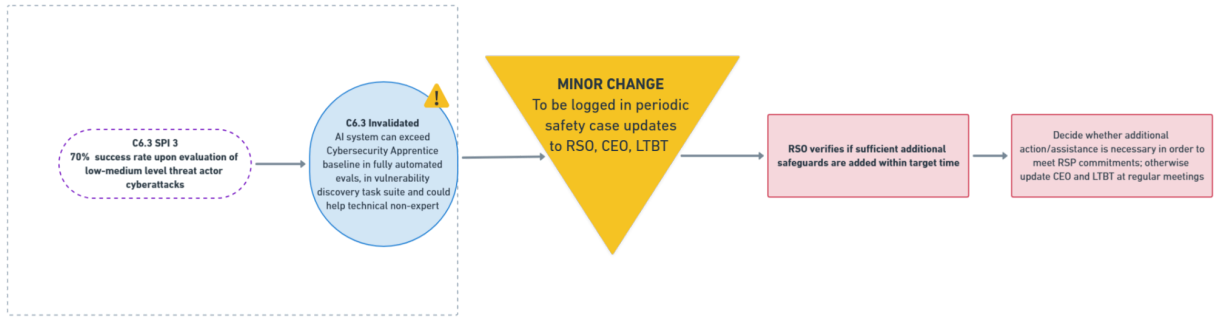

**Scenario 4: Response time to include new proxy task benchmarking increases**

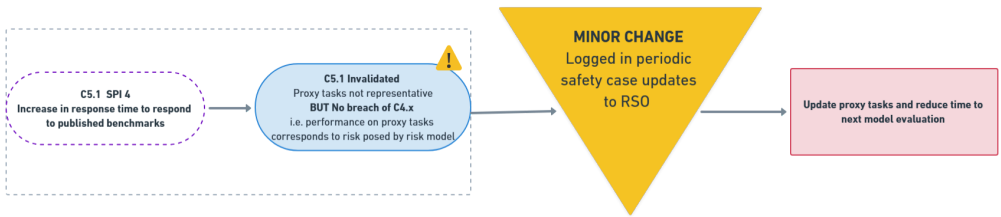

Figure 12: Frontier AI developer governance response to scenarios

### 5.3 DSCMS integration with national governance

The responsibility of demonstrating that frontier AI models do not pose an unacceptable risk to society lies with the developers. However, states have a key interest in monitoring the safety activities of developers. Integrating a DSCMS into national governance systems could advance the existing risk management ecosystem, requiring aligned processes and policies (The Model for Responsible Innovation, DSIT, 2024).

The rapidly evolving nature of frontier AI development requires an agile approach that can adapt to risk profile changes. It is valuable to go beyond a prescriptive governance regime and rely on performance-based regulation (Leveson, 2011) by establishing a general duty of reducing risks "as low as reasonably practicable" (Hopkins, 2012). This shifts responsibility for controlling risks from the government to developer self-regulation.

**Key actors**

The paper explores UK national governance mechanisms as a case study, the insights derived could inform policy frameworks in other countries and further international AI safety collaboration. The UK AI Safety Institute (AISI) is the most suitable existing body to govern and assess safety cases, given its current work building a workstream in this area, and the complementary role in evaluating AI model capabilities (AI Safety Institute Approach to Evaluations, DSIT, 2024; Castris & Thomas, 2024). AISI could develop its expertise to create mechanisms and guidance akin to the ONR's Technical Assessment Guide 51(ONR, 2024).

The Department for Science, Innovation and Technology ('DSIT') ensures regulatory alignment. Within DSIT, the central AI risk function (CAIRF) could complement AISI in addressing AI risks and identifying gaps in risk mitigation efforts (Camrose, 2024). The National Cyber Security Centre (NCSC) is particularly relevant given its role in ensuring that AI safety cases address cybersecurity risks throughout the system lifecycle (Guidelines for Secure AI System Development, NCSC, 2023).

**Core components of national governance**

In order to minimise regulation while maximising effective governance, the core functions of national governance must be appropriately established:

- **Examination and oversight:** AISI could monitor dynamic safety cases to highlight developer oversights and inform future standards. Accurate information through the lifecycle is a vital ex-ante safety measure, enabling AISI to assess whether a frontier AI developer can proceed with model development or deployment. (Askell et al., 2019).
- **Incentivise implementation:** Policy instruments could incentivise AI developers to adopt, implement and integrate a DSCMS to build robust risk management and safety practices (Kolt et al., 2024).
- **Facilitation and collaboration:** Governments are uniquely placed to facilitate integration of dynamic safety cases into Responsible Scaling Policies ('RSPs') of frontier AI developers via an iterative process (Jones et al., 2024). Threat-reporting information shared by the government on potential risks to security, reliability and related issues would strengthen the safety case, and thereby the overall risk management system. Collaboration with bodies responsible for addressing national security implications of AI such as the Laboratory for AI Security Research (LASR) (Oxford, 2024) and the Testing Risks of AI for National Security (TRAINS) Taskforce in the US (US AISI, 2024) could help improve SPIs used in the safety cases.
- **Future regulation:** Embedding dynamic safety cases within the current national framework could help effectively transition from a high-principle based approach to rule-based approach in the future (Schuett et al., 2024).

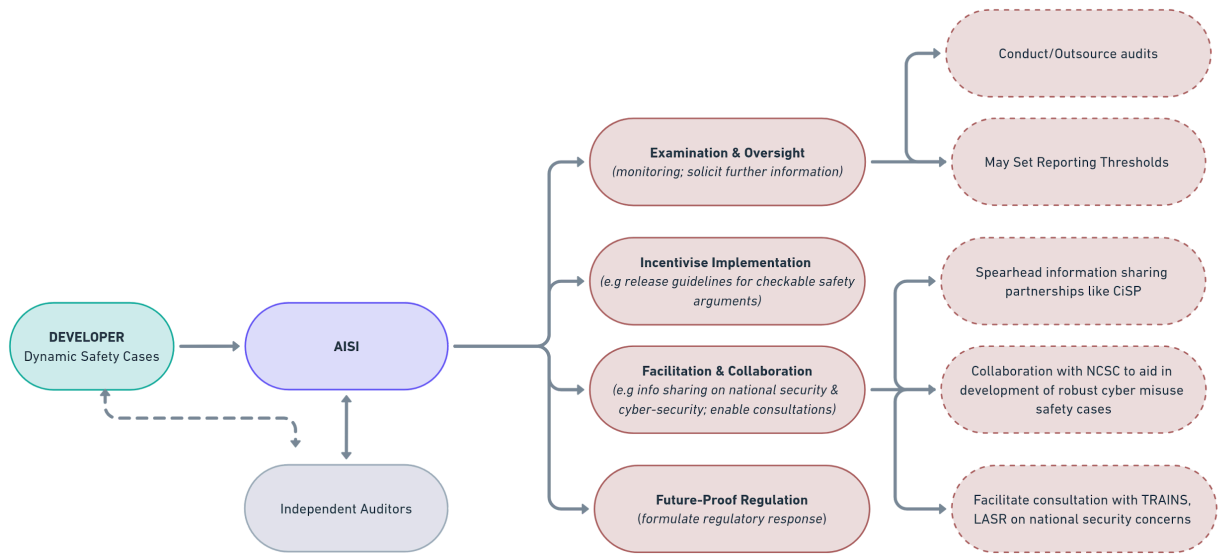

Figure 12: Components of National Governance of DSCs

**A comparative analysis**

Governance of safety cases has long been integral to various safety-critical industries. A rapid review of best practices in sectors like nuclear, energy, space, autonomous systems and defense highlights potential policy pathways but also reveals inherent limitations in their applicability to frontier AI (see Appendix C). In the context of frontier AI, it is essential to adopt adaptive governance strategies to manage risks in a coherent, proportional, and continuous manner throughout the model's lifecycle.

Existing literature reveals two critical gaps: first, it primarily addresses safety-critical industries with relatively low uncertainty, where systems do not evolve at the same rapid pace as AI; second, it fails to accommodate the emerging framework of dynamic safety cases within governance mechanisms.

**Policy options**

In order to execute the core functions of the national government, the following policy instruments have been identified. These policy options are not mutually exclusive, and can be used in tandem to strengthen the governance of dynamic safety cases.

| Policy Intervention | Need & Intended Outcome | Reliance on other instruments | Uncertainties |
|---|---|---|---|
| **Live Sharing of Dynamic Safety Case Dashboard** | Live sharing information about vulnerabilities<br><br>Facilitate rapid risk assessment | Stronger information security mechanism | Confidentiality Concerns<br><br>Defaulting to a 'presumed safety' approach |
| **Submission of Dynamic Safety Case Linked to Threshold Breach** | Identify thresholds for minor & material changes | Guidance documents/statute requiring submission | Inconsistent Submissions<br><br>Increased compliance burden |
| **Cyber Security-related Information Sharing Initiatives** | Public-private collaboration strengthens safety cases | Develop initiatives/ information sharing committee | Ensuring information security<br><br>Identifying trusted actors within frontier AI company |
| **Capacity Building** | Build required technical knowledge<br><br>Avoid dilution of oversight | Investment into technical capacity building | Resource intensive<br><br>Administrative burden on AISI |
| **State-Backed Third Party Auditing** | Enable audits by experts<br><br>Reduce burden on AISI | Mechanism to approve auditors<br><br>Voluntary commitment/MoU ensuring developers undergo audit | Information security concerns |

Table 15: Policy Matrix

### 1. Live sharing of Dynamic Safety Case dashboard

With checkable safety arguments created at the beginning of the AI system's lifecycle and maintained thereafter, it is valuable to create transparent access that would allow AISI visibility to the live dynamic safety case. This approach would ensure information integrity and facilitate rapid risk assessment, mitigation and coordination by AISI. This transparency mechanism could also help increase public trust. Active stakeholder consultations with developers would be valuable to secure initially voluntary commitments similar to model evaluations establishing licensing, certification, authorisation and approval obligations.

**Risks and uncertainties:** There may remain concerns around sharing of commercially sensitive information, and the potential competitive disadvantage it may yield. This may be mitigated by structured, compartmentalised access based on security clearance type principles. Further, in light of key frontier AI organisations committing to transparency at the Seoul Summit 2024, there is reasonable declared intent to share safety cases throughout the lifecycle of the model (Frontier AI Safety Commitments, AI Seoul Summit, 2024). The signed commitments specifically discuss how developers should share detailed information which cannot be shared publicly with trusted actors, including their governments. Another concern is the risk of fostering a 'presumed safety' culture where real-time access to all dynamic safety case updates is provided without clearly defined trigger thresholds, potentially leading to information overload and diminished focus on critical safety events.

## 2. Submission of safety case linked to threshold breach

AISI can require updated dynamic safety case snapshot report to be submitted periodically combined with a requirement for immediate submission of an up-to-date snapshot where a material change in the safety case has occurred, subject to a breach in acceptable risk thresholds. The acceptable risk thresholds could be predetermined as part of a project safety management plan submitted by the developer and previously approved by AISI. Over time, it may be possible for AISI to share their own thresholds of material change in order to work towards standardisation. Such a framework must, via guidance documents or statutory instruments, establish what would constitute minor changes and what would be deemed a material change.

**Risks and uncertainties:** Without continuous monitoring by AISI, there may be uncertainty around what constitutes a material change and whether such material change has been appropriately notified. This may lead to inconsistent submissions, delays, or unnecessary reporting, overwhelming regulatory bodies. Additionally, it may create an increased compliance burden on developers, given the potential subjectivity around the thresholds set.

**Possible Scenarios**

The scenarios presented in Section 4.4 can be revisited to explore potential government responses to evolving dynamic safety cases throughout the lifecycle of the frontier AI model. Each scenario illustrates a distinct risk level, with the accompanying diagrams outlining corresponding actions that may be taken by the national government.

The scenarios are examined in relation to the two abovementioned policy options: first, if the live sharing of the dynamic safety case dashboard, and second, if the submission of safety cases only after a predefined threshold has been exceeded.

---

### Scenario 1: Risk model threshold breach (top-level claim invalidated)

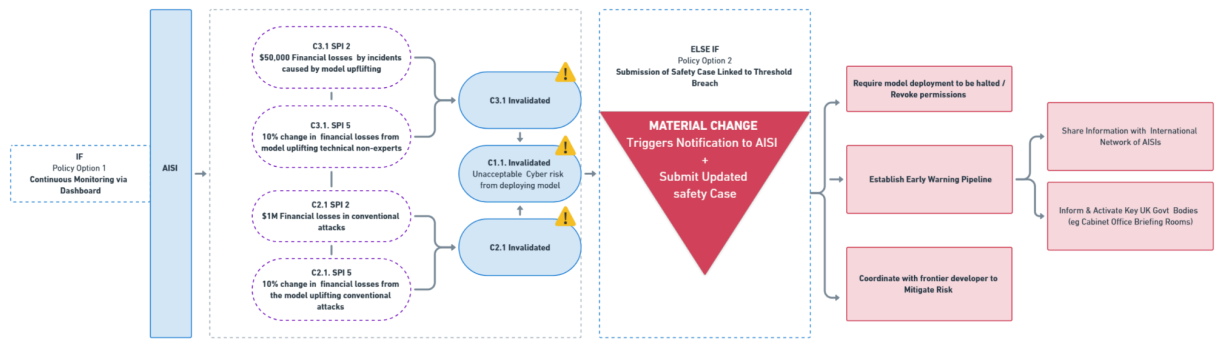

### Scenario 2: Novel attack vector discovered

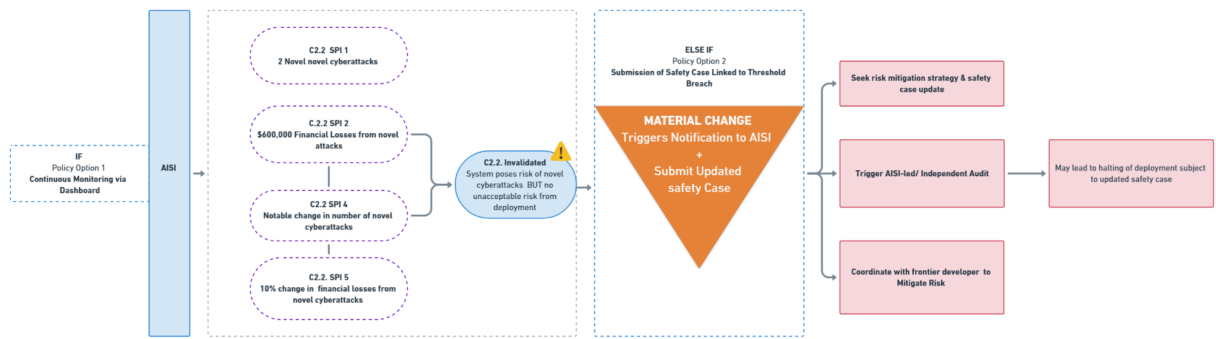

**Scenario 3: A model update shows capability uplift in automating low-medium threat level threat actor cyberattacks**

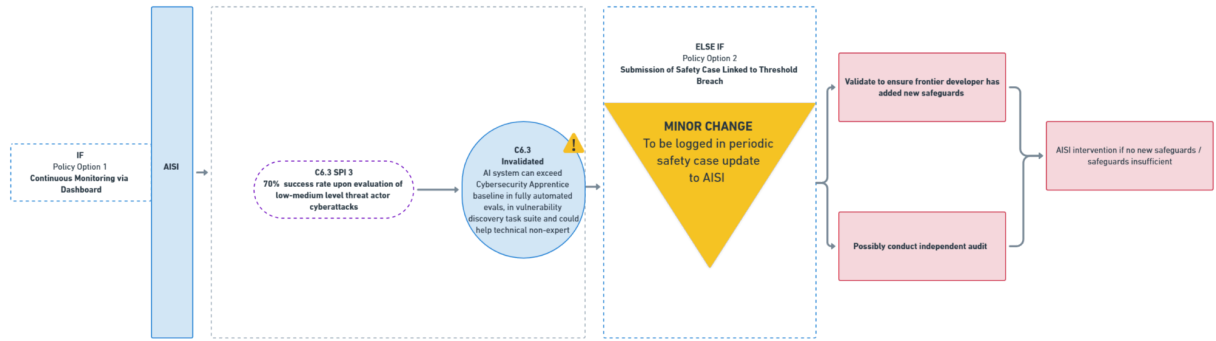

**Scenario 4: Response time to include new proxy task benchmarking increases**

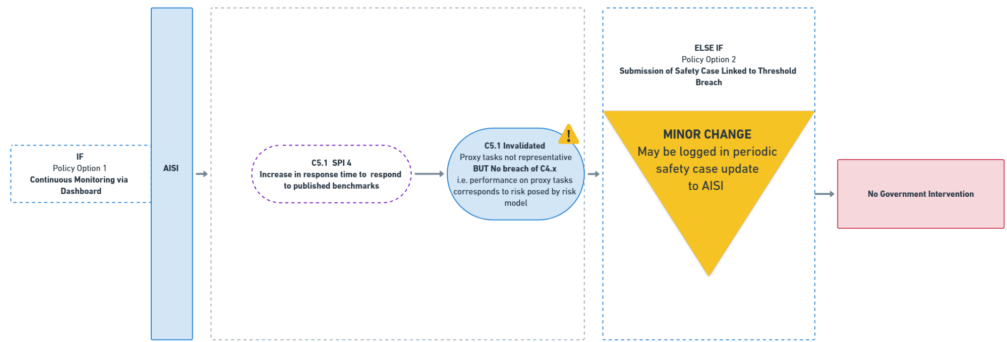

Figure 13: Government response to scenarios

---

### 3. Cyber security-related information sharing initiatives

For SPIs to meaningfully support safety management, information sharing between government entities and frontier AI developers must be viewed as a critical complementary process. It may be valuable for key organisations within the government such as the NCSC, DSIT and the National Protective Security Authority (NPSA) which is responsible for building resilience to national security threats to create a pathway to share relevant information such as via APIs providing AI-model specific cyber intelligence feeds. Structures to establish such initiatives can be guided by the Cyber Security Information Sharing Partnership (CiSP) which is a key public-private collaboration platform that allows government agencies to share real-time threat intelligence with private companies. Another valuable institutional structure that could be modeled after is the Financial Services Cyber Collaboration Centre (FSCCC). It is an industry-led initiative that works in partnership with law enforcement, NCSC, and the financial authorities to identify, analyse, and communicate threats to the sector.

**Risks and uncertainties**: The challenge with sharing information is ensuring information is secure and is not maliciously used by those who have access to it. This would require investment into implementing information security protocols.

### 4. Capacity building

Dynamic safety cases imply much the same considerations as conventional safety cases as described in (Buhl et al., 2024). Effective governance of safety cases requires that AISI possess robust technical capabilities. Without such expertise, developers would hold a significant advantage over government, leading to an information asymmetry that could undermine the possibility of oversight -the use of safety cases for decision-making may require appointment or establishment of a body to review safety cases. Technical capacity is crucial to successfully monitor and introduce any future intervention powers if required (Charles Haddon-Cave QC, 2009) in order to support responsible frontier AI developer training or deployment decisions.

**Risks and uncertainties**: Given the fast evolving nature of frontier AI, maintaining the necessary technical expertise within AISI may prove to be challenging. This effort would demand substantial resources, particularly as the public sector contends with competition with private-sector incentives for skilled professionals. Additionally, conducting regular audits could introduce considerable administrative burden on AISI, potentially leading to delays in decision-making and creating inefficiencies within the governance process.

5. **State-recommended third party audits**

To supplement the technical and governance capacity of AISI, independent audits may be carried out by third parties. Creating an independent layer of review would reduce the burden of assessing dynamic safety cases that would otherwise be taken on solely by AISI. Such audits can help verify claims (Schuett et al., 2024) developers made in the dynamic safety cases by drawing on external expertise. While information is shared with government bodies in full, it may be appropriate to share information on a case-by-case basis for independent audits by third parties.

Allowing independent auditors and domain experts to verify safety cases would subject frontier AI models to increased scrutiny (Kolt et al., 2024). Such a government recommended critical review by an independent third-party would help identify missing risks, bridge expertise gaps and incomplete scenarios in the developer's safety case, especially at the design and deployment stage. Further, by creating a legal obligation for developers to engage a third-party auditors or contract with a third-party expert would ensure enforcement (Le Coze, 2024).

To build a reliable network of auditors, a mechanism similar to the NCSC's Cyber Resilience Audit (CRA) scheme may be incorporated. This scheme allows independent cyber audits on behalf of a Cyber Oversight Body by pre-approved auditors. Auditors are approved based on meeting a predetermined technical standard. In a similar manner, AISI may encourage audits by approved auditors selected based on predefined criteria.

Alternatively, an approach similar to the establishment of an independent supervisory system for monitoring and evaluating the safety levels of nuclear facilities by the IAEA may be adopted to continuously monitor and assess safety cases (Cha, 2024).

**Risks and uncertainties:** As has been recognised vis-a-vis algorithmic audits (Digital Regulation Cooperation Forum, 2022), the decision to allow a form of delegated oversight would trade the cost of a weaker form of regulation against the potential benefit of a more comprehensive oversight than would be allowed by AISI's capacity. Another concern with such independent audits is with respect to safeguarding sensitive information.

## Components of international collaboration & coordination

In a multi-actor ecosystem, AISI's leadership, combined with the technical expertise of frontier AI developers and international collaboration with organisations such as the US AISI, would continue to foster a coordinated approach to AI safety governance through the development of global best practices, which is not unique to dynamic safety cases, but this does introduce further specialist knowledge requirements. For dynamic safety cases to truly mitigate emergent risks from frontier AI, countries must establish an early warning pipeline whereby alerts on high-risk breaches are shared, and align actionable frameworks for risk management. Currently, AISI's evaluations inform national government policy and support international actors (AISI, 2024). Existing information-sharing channels with international actors can be leveraged to share best practices, regulatory strategies, and research findings with other AI safety institutes. Additionally, integration of dynamic safety case updates into cross-border cybersecurity frameworks would enhance accountability and enforcement.

# 6 Discussion

In this paper we have presented the need to continuously update safety cases throughout the life of a frontier AI system and proposed a DSCMS as one partial solution. We described how the DSCMS might reduce the time and effort required to revise safety cases, integrate effectively into governance structures, and thereby provide more timely visibility of a system's ongoing safety status. In this section, we consider how well the DSCMS proposal meets this need, while also addressing the challenges, limitations and open questions that remain that require further work.

From a high-level perspective, the DSCMS approach based on CSA and SPIs appears promising. Its structured, semi-automated nature could reduce the manual labour typically involved in safety case maintenance. The CSA framework adapts in principle to frontier AI with essentially no modification required, and its formal safety argument model with consistency rules enables full automation of the consistency checking process, opening up the potential for 'instant' change impact analysis when a system change or SPI breach occurs. However, the ongoing cumulative automation benefits afforded by this over the system lifecycle require setup effort - further work would be required to develop domain-specific efficient methods for specifying consistency rules for frontier AI safety case patterns or building blocks, and expertise would need to be developed to develop these rules with well-tuned SPI threshold semantics such that the propagation extent in the event on an SPI threshold breach is consistent in invalidating accurately and completely the extent of the safety case that reflects the real-world impact. Software tools to support CSA have been demonstrated in the lab (i.e. FASTEN.safe (Cârlan & Ratiu, 2020)), and tools exist for specification and measurement of SPIs developed for assurance cases in other sectors. Investigation of toolsets is outside the scope of this paper, but it is plausible that existing tools may be adaptable for proof-of-concept implementation of a DSCMS for frontier AI, and that AI-assisted tools could be useful in automating further aspects of the DSCMS, with less effort, but the use of AI in in this way of course entails its own issues regarding tool qualification for safety and robustness.

Whilst not currently automated, the use of the SPIs in the DSCMS does offer the possibility to automate - to varying degrees - the integration of evidence, and live monitoring of the same. Even where automation is not currently possible, specifying SPIs offer a method to systematically integrate evidence into the safety case in a way that is perhaps more transparent to stakeholders, as it offers a way to compress diverse evidence into metrics. Selecting, defining and validating SPIs for frontier AI is non-trivial. Capturing the true underlying risks (including potential catastrophic risks) can be complicated by overemphasis on quantifiable or measurable indicators. Ensuring that a SPI set is complete, correct and minimally biased requires careful expert judgement. A further challenge is the difficulty of obtaining external data - for instance, cybersecurity threat feeds of near-miss data - often held privately or protected by national security considerations. The use of the combination of leading and lagging indicators requires development to obtain useful characteristics of ex-ante and ex-post updates to an evolving risk landscape. Some of the SPIs specified in this paper could be classified as 'meta'-SPIs, which are useful in themselves for indirect process measurement, for example as means to assess whether model evaluation methods in use are tracking the latest benchmarks. Whilst SPIs have been recommended for autonomous systems, for example in the safety standard UL4600, there is significant work as outlined above to develop the specifics for frontier AI. As such, prioritisation of effort is important, and we have proposed an approach to this in section 4.3.4 based on significance to safety and feasibility of measurement. Beyond cyber capabilities safety arguments, a significant challenge is the definition of external SPIs that would adequately cover the space of systemic risks, given, as discussed, the open-ended potential operational context and the complex and unpredictable interactions with sociotechnical systems. This is paramount for adequately mitigating catastrophic risks, where early warning and leading indicators are essential prior to deployment. A related challenge that applies to safety cases in general is that of adequacy, rigour and confidence in the safety case. Barrett et al. (manuscript in preparation) explores assessing confidence in safety cases for frontier AI, drawing on Assurance 2.0: A Manifesto (Bloomfield & Rushby, 2021). This confidence aspect would likely need to be woven into the DSCMS to offer decision makers a common view on which to act.

So far in this discussion we've focussed on the prospects of a DSCMS, but if we take a wider view of updating of safety cases, there are important aspects to consider. We discussed the possibility for automation and systemisation of the update process and the challenges with specifying and obtaining SPIs, particularly external SPIs. However, in the case of fundamental large changes to either the model or the operational context, the change to the safety case may be major enough to dominate routine incremental updating considerations. For example, with compute scaling, the step in capabilities between generations of frontier models in recent years has been dramatic. If this scaling continues, a large change in the safety case will not require only minor updates, but potentially rebuilding the underlying structure of the safety case, such as moving from 'inability' arguments to 'control' or similar arguments, where model capabilities can no longer be argued to be insufficient to cause harm. This entails a fundamentally different safety case to be constructed which a DSCMS may need significant reconstruction to accommodate, and again, this underscores the importance of careful pre-deployment analyses - particularly where catastrophic risk is at stake.

We have argued that DSCMS can integrate into both corporate-level governance and broader national governance frameworks, providing the benefits of systematising the integration of evidence for safety assurance, and potentially being able to provide this in the form of real time decision support to appropriate decision makers in the chain of command automatically, to support planned deployment decisions and in reactive incident response. Beyond this, DSCMS could offer a means to operationalise safety management frameworks risk thresholds through the use of SPIs specified in relation, for example, to model capability evaluations. Similar points apply to the benefit of DSCMS to national governance. Access to a DSCMS live dashboard, or *limited, periodic or aggregated* access to the same could permit governments a high degree of visibility of frontier AI system safety even without formal regulation, in a similar manner to, for example, the way UK AISI currently obtains early access to new models for independent evaluation. This visibility in turn could help monitor assurance across the industry and help inform development of more nuanced and timely policy measures, complementing existing efforts towards fine-grained understanding of AI system safety in governments. Still, information security related to commercial and national security concerns would need to be addressed. Overall, the DSCMS contributes to addressing a key gap: reconciling the need for safety assurance prior to deployment with the fact that the risk landscape is likely to continue to evolve in ways that are difficult to predict. The approach could help maintain confidence in ongoing safety, but it does not eliminate the need for robust initial analysis for high-severity risk systems, and further work is needed to exploit this method.

# 7 Recommendations

Whilst dynamic safety cases as proposed in this paper shows promise to help with the updating challenge, further research and development is required before stronger recommendations could be made about adoption. From the discussion in the previous section we draw out the recommendations below, which could be executed by frontier AI developers, governments, wider research communities, or a relevant, appropriate combination of the above.

1. **Proof of concept**. Develop a minimal DSCMS prototype (e.g. for a simplified cyber inability safety case). This could validate the feasibility of formal consistency rules, SPI monitoring and user interfaces for decision support for a concrete frontier AI case.
2. **CSA consistency rules and patterns.** Create reusable building blocks or patterns for CSA. These patterns could be developed based on the frontier AI safety case sketches produced to date, and future work towards 'control' based arguments.
3. **Specification, prioritisation and validation of SPIs.** Conduct focused research on how best to define, prioritise and validate complete and correct SPIs sets with well chosen thresholds for frontier AI contexts. This could include exploring ways to combine leading and lagging indicators and mitigate biases from the push for measurable metrics. Partnerships with experts from other sectors, particularly autonomous vehicles (AV) could be beneficial. Exploring software tool support and AI-assistance in the above work could bear fruit, for example in integrating evidence and transforming it into suitable data feeds for SPIs.
4. **Governance 'war-gaming' exercises.** Integrate DSCMS dashboards into scenario-based drills, testing how real-time safety information informs decisions under uncertain or high-stakes conditions - especially around major deployment events. This may help specify Responsible Scaling Policy risk thresholds and provide fine-grained visibility to governments.
5. **Facilitate access to external SPI data.** Governments could help establish trusted channels for sharing critical data (e.g. cyber threat feeds), enabling more accurate and timely SPI monitoring crucial for systemic risk, without compromising security.
6. **Advance systemic risk modelling.** Encourage further research into complex hazard analysis for open-ended, large-scale unpredictable operational contexts for frontier AI systems in order to translate systemic risk models into suitable SPIs to capture catastrophic hazards before deployment decisions are made.

# 8 Conclusion

This paper has proposed the concept of a Dynamic Safety Case Management System (DSCMS), extending and adapting methods from the autonomous vehicles sector to meet the unique and evolving challenges of frontier AI systems. This approach aims to complement robust initial safety analyses by enabling continuous monitoring and updating of safety arguments, integrating changing evidence about AI model capabilities and a shifting operational risk landscape. This method provides a pathway to semi-automate and systematise impact analysis via CSA and SPIs. While the DSCMS offers a partial solution, there is still substantial work required to develop and validate these methods. The DSCMS could provide a benefit to corporate governance by providing a structured means of integrating evidence via SPIs that can provide decision makers with potentially real-time overall system safety assessments both proactively when making critical deployment decisions when risk severity is high, and reactively in response to safety incident triggers. We also argue that a DSCMS could benefit national governance by providing a finer-grained, potentially real time view of frontier AI safety, supporting understanding and hence potentially enabling more nuanced, adaptable, timely  policy response. By merging proactive and continuous safety assurance, dynamic safety cases could enhance transparency and trust between developers, governments, regulators and potentially civil society.

# References

Agrawal, A., Khoshmanesh, S., Vierhauser, M., Rahimi, M., Cleland-Huang, J., & Lutz, R. (2019). Leveraging Artifact Trees to Evolve and Reuse Safety Cases. 2019 IEEE/ACM 41st International Conference on Software Engineering (ICSE), 1222–1233. https://doi.org/10.1109/ICSE.2019.00124

AISI (2024) US AISI and UK AISI Joint Pre-Deployment Test - Anthropic's Claude 3.5 Sonnet. Retrieved from: https://www.aisi.gov.uk/work/pre-deployment-evaluation-of-anthropics-upgraded-claude-3-5-sonnet

Anthropic (2024), https://www.anthropic.com/voluntary-commitments

Anthropic. (2024). Responsible scaling policy. Retrieved from https://perma.cc/DB9F-GAV4

Anurin, A., Ng, J., Schaffer, K., Schreiber, J., and Kran, E. (2024). Catastrophic Cyber Capabilities Benchmark (3CB): Robustly Evaluating LLM Agent Cyber Offense Capabilities. arXiv preprint arXiv:2410.09114.

Anwar, U., Saparov, A., Rando, J., Paleka, D., Turpin, M., Hase, P., . . . Krueger, D. (2024). Foundational challenges in assuring alignment and safety of large language models. arXiv. preprint arXiv:2404.09932.

Askell, A., Brundage, M., & Hadfield, G. (2019). The Role of Cooperation in Responsible AI Development (arXiv:1907.04534). arXiv. https://doi.org/10.48550/arXiv.1907.04534

Balesni, Marius Hobbhahn, David Lindner, Alexander Meinke, Tomek Korbak, Joshua Clymer, Buck Shlegeris, Jérémy Scheurer, Charlotte Stix, Rusheb Shah, Nicholas Goldowsky-Dill, Dan Braun, Bilal Chughtai, Owain Evans, Daniel Kokotajlo and Lucius Bushnaq: Towards evaluations-based safety cases for AI scheming, November 2024. URL https://arxiv.org/abs/2411.03336. arXiv:2411.03336.

Barrett, S., Fox, P., Krook, J., Mondal, T., Mylius, S., & Tlaie, A. (manuscript in preparation). Assessing confidence in frontier AI safety cases.

Barwise, J., Etchemendy, J., Allwein, G., Barker-Plummer, D., & Liu, A. (2002). Language, proof and logic (p. 598). Stanford, USA: CSLI publications.

Bhatt, M., Chennabasappa, S., Li, Y., Nikolaidis, C., Song, D., Wan, S., ... and Saxe, J. (2024). Cyberseceval 2: A wide-ranging cybersecurity evaluation suite for large language models. arXiv preprint arXiv:2404.13161.

Bloomfield, R., & Rushby, J. (2021). Assurance 2.0: A Manifesto (No. arXiv:2004.10474). arXiv. https://doi.org/10.48550/arXiv.2004.10474

Bloomfield, R., Fletcher, G., Khlaaf, H., Hinde, L., and Ryan, P. (2021). Safety Case Templates for Autonomous Systems (No. arXiv:2102.02625). arXiv. https://doi.org/10.48550/arXiv.2102.02625

Buhl, M. D., Sett, G., Koessler, L., Schuett, J., & Anderljung, M. (2024). Safety cases for frontier AI (No. arXiv:2410.21572). arXiv. http://arxiv.org/abs/2410.21572

Calinescu, R., Weyns, D., Gerasimou, S., Iftikhar, M. U., Habli, I., & Kelly, T. (2018). Engineering Trustworthy Self-Adaptive Software with Dynamic Assurance Cases. IEEE Transactions on Software Engineering, 44(11), 1039–1069. https://doi.org/10.1109/TSE.2017.2738640

Camrose, V. (Parliamentary Under Secretary of State, Department for Science, Innovation & Technology). (2024). Further supplementary written evidence (LLM0120). Hearing before the House of Lords Communications and Digital Select Committee.

Cârlan, C. (forthcoming). Checkable Safety Arguments -- A Modeling Framework Supporting the Maintenance of Safety Arguments Consistent with System Development Artifacts. Manuscript submitted for publication. Doctoral dissertation, Technische Universität München.

Cârlan, C., & Ratiu, D. (2020). FASTEN.Safe: A Model-Driven Engineering Tool to Experiment with Checkable Assurance Cases. Computer Safety, Reliability, and Security: 39th International Conference, SAFECOMP 2020, Lisbon, Portugal, September 16–18, 2020, Proceedings, 298–306. https://doi.org/10.1007/978-3-030-54549-9_20

Carlan, C., Gauerhof, L., Gallina, B., & Burton, S. (2022). Automating Safety Argument Change Impact Analysis for Machine Learning Components. 2022 IEEE 27th Pacific Rim International Symposium on Dependable Computing (PRDC), 43–53. https://doi.org/10.1109/PRDC55274.2022.00019

Carlan, C., PetriSor, D., Gallina, B., & Schoenhaar, H. (2020). Checkable Safety Cases: Enabling Automated Consistency Checks between Safety Work Products. 2020 IEEE International Symposium on Software Reliability Engineering Workshops (ISSREW), 295–302. https://doi.org/10.1109/ISSREW51248.2020.00088

Castris, A. L. D., & Thomas, C. (2024). The potential functions of an international institution for AI safety. Insights from adjacent policy areas and recent trends (arXiv:2409.10536). arXiv. https://doi.org/10.48550/arXiv.2409.10536

Cha, S. (2024). Towards an international regulatory framework for AI safety: Lessons from the IAEA's nuclear safety regulations. Humanities and Social Sciences Communications, 11(1), 1–13. https://doi.org/10.1057/s41599-024-03017-1

Chao, P., Robey, A., Dobriban, E., Hassani, H., Pappas, G. J., & Wong, E. (2024). Jailbreaking Black Box Large Language Models in Twenty Queries (No. arXiv:2310.08419). arXiv. https://doi.org/10.48550/arXiv.2310.08419

Charles Haddon-Cave QC. (2009). The Nimrod Review [Independent Report]. Retrieved from https://www.gov.uk/government/publications/the-nimrod-review

Chen, Z. Ding, L. Alowain, X. Chen, and D. Wagner. DiverseVul: A new vulnerable source code dataset for deep learning based vulnerability detection. In International Symposium on Research in Attacks, Intrusions and Defenses, pages 654–668, Apr. 2023b. URL https://arxiv.org/abs/2304.00409.

Clymer, J., Gabrieli, N., Krueger, D., & Larsen, T. (2024). Safety cases: How to justify the safety of advanced AI systems. arXiv preprint arXiv:2403.10462

Davidson, T., Denain, J.-S., Villalobos, P., & Bas, G. (2023). AI capabilities can be significantly improved without expensive retraining (No. arXiv:2312.07413). arXiv. https://doi.org/10.48550/arXiv.2312.07413

Denney, E., & Pai, G. (2024). Reconciling Safety Measurement and Dynamic Assurance (No. arXiv:2405.19641). arXiv. https://doi.org/10.48550/arXiv.2405.19641

Denney, E., Pai, G., & Habli, I. (2015). Dynamic Safety Cases for Through-Life Safety Assurance. 2015 IEEE/ACM 37th IEEE International Conference on Software Engineering, 587–590. https://doi.org/10.1109/ICSE.2015.199

Denney, E., Pai, G., & Pohl, J. (2012). AdvoCATE: An Assurance Case Automation Toolset. In F. Ortmeier & P. Daniel (Eds.), Computer Safety, Reliability, and Security (pp. 8–21). Springer. https://doi.org/10.1007/978-3-642-33675-1_2

Digital Regulation Cooperation Forum. (2022, Spring). Auditing algorithms: The existing landscape, role of regulators, and future outlook. GOV.UK. https://www.gov.uk/government/publications/findings-from-the-drcf-algorithmic-processing-workstream-spring-2022/auditing-algorithms-the-existing-landscape-role-of-regulators-and-future-outlook

DSIT. (2024). Frontier AI safety commitments, AI Seoul Summit 2024. Retrieved from https://perma.cc/M9NQ-GNED

DSIT. (2024, November). The model for responsible innovation. GOV.UK. https://www.gov.uk/government/publications/the-model-for-responsible-innovation/the-model-for-responsible-innovation

DSIT. AI Safety Institute approach to evaluations. https://perma.cc/6SSN-NPL4, 2024.

European Commission. (2024). First draft of the general-purpose AI Code of Practice published: written by independent experts. Available at: https://digital-strategy.ec.europa.eu/en/library/first-draft-general-purpose-ai-code-practice-published-written-independent-experts (Accessed: 27 November 2024).

European Parliament and Council of the European Union. (2024, June 13). Regulation (EU) 2024/1689 of the European Parliament and of the Council of 13 June 2024 on artificial intelligence (Artificial Intelligence Act). Official Journal of the European Union. Interinstitutional File: 2021/0106(COD) https://artificialintelligenceact.eu/article/51/

Favaro, F., Fraade-Blanar, L., Schnelle, S., Victor, T., Peña, M., Engstrom, J., . . . Smith, D. (2023). Building a credible case for safety: Waymo's approach for the determination of absence of unreasonable risk. arXiv preprint arXiv:2306.01917.

Frontier Model Forum. (2024). Frontier Model Forum: Advancing Frontier AI safety. https://www.frontiermodelforum.org

Goemans, A., Buhl, M. D., Schuett, J., Korbak, T., Wang, J., Hilton, B., & Irving, G. (2024). Safety case template for frontier AI: A cyber inability argument (No. arXiv:2411.08088). arXiv. https://doi.org/10.48550/arXiv.2411.08088

Google DeepMind. (2024). Frontier Safety Framework Version 1.0. Retrieved from https://perma.cc/5Q3D-LM2Q

Grosse, R. (2024). Three Sketches of ASL-4 Safety Case Components, Anthropic. https://www.anthropic.com/research/sabotage-evaluations

Hall, A. (2020, March). The UK Nuclear Industry Good Practice Guide To: Keeping Safety Cases 'Live'. Nuclear Industry Safety Directors' Forum (SDF). https://nuclearinst.com/SDF-safety-cases

Hopkins, A. (2012). Working Paper 87: Explaining 'Safety Case'. National Research Centre for OHS Regulation, Canberra.

INCOSE. (2023). Guide to Writing Requirements (INCOSE-TP-2010-006-04, Version/Revision: 4, 1 July 2023). INCOSE.

Inge, J. R. (2007). The safety case, its development and use in the United Kingdom. In Equipment safety assurance symposium 2007. ISSC. Retrieved from https://perma.cc/T4L4-FP7C

Jones, Mahi Hardalupas, & William Agnew. (2024). Under the radar?: Examining the evaluation of foundation models (p. 106) [Emerging Technology & Industry Practice]. Ada Lovelace Institute. https://www.adalovelaceinstitute.org/report/under-the-radar/

Kaplan, Sam McCandlish, Tom Henighan, Tom B. Brown, Benjamin Chess, Rewon Child, Scott Gray, Alec Radford, Jeffrey Wu, and Dario Amodei. Scaling laws for neural language models, January 2020. URL https://arxiv.org/abs/2001.08361. arXiv:2001.08361 [cs.LG].

Koessler, L., Schuett, J., and Anderljung, M. (2024) Risk thresholds for frontier AI (No. arXiv:2401.14713). arXiv: https://arxiv.org/pdf/2406.14713

Kolt, N., Anderljung, M., Barnhart, J., Brass, A., Esvelt, K., Hadfield, G. K., Heim, L., Rodriguez, M., Sandbrink, J. B., & Woodside, T. (2024). Responsible Reporting for Frontier AI Development (arXiv:2404.02675). arXiv. https://doi.org/10.48550/arXiv.2404.02675

Le Coze, J.-C. (2024). Subcontracting Safety (Cases). In J.-C. Le Coze & B. Journé (Eds.), Safe Performance in a World of Global Networks: Case Studies, Collaborative Practices and Governance Principles (pp. 57–64). Springer Nature Switzerland. https://doi.org/10.1007/978-3-031-35163-1_6

Leveson, N. G. (2011). The Use of Safety Cases in Certification and Regulation [Working Paper]. Massachusetts Institute of Technology. Engineering Systems Division. https://dspace.mit.edu/handle/1721.1/102833

Leveson, N. G. (2012). Engineering a safer world: Systems thinking applied to safety. MIT Press.doi: 10.7551/mitpress/8179.001.0001

Mirzaei, E., Cârlan, C., Thomas, C., & Gallina, B. (2022, February 8). Design-time Specification of Dynamic Modular Safety Cases In Support of Run-Time Safety Assessment.

MITRE ATT&CK®. (n.d.). Retrieved 20 December 2024, from https://attack.mitre.org/

MoD (2007). Safety management requirements for defence systems (Standard 00-56). Retrieved from https://perma.cc/YUP4-V83D

MoD (2024) SMP12. Safety Case and Safety Case Report. ASEMS. Retrieved from https://www.asems.mod.uk/guidance/posms/smp12

NCSC (2013), 'Csaw cybersecurity games & conference' 2013–2023.

NCSC (2024). The near-term impact of AI on the cyber threat. Retrieved from https://perma.cc/6NPT-UHX4

NCSC (2023) Guidelines for Secure AI System Development, 2023 placeholder https://www.ncsc.gov.uk/collection/guidelines-secure-ai-system-development

ONR TAGs—Nuclear Safety 51 (NS-TAST-GD-051 Issue 7.1; ONR Guide). (2024). ONR. https://onr.org.uk/publications/regulatory-guidance/regulatory-assessment-and-permissioning/technical-assessment-guides-tags/nuclear-safety-tags/technical-assessment-guides-tags-nuclear-safety-full-list/

OpenAI, 2023. Preparedness Framework (Beta). Retrieved from https://perma.cc/9FBB-URXF

OpenAI, 2024a. https://openai.com/index/disrupting-malicious-uses-of-ai-by-state-affiliated-threat-actors/

OpenAI, 2024b. GPT 4o System Card. https://openai.com/index/gpt-4o-system-card/

OpenAI, 2024c. GPT o1 System Card https://openai.com/index/openai-o1-system-card/

Oxford University to lead AI security research through new national laboratory partnership. (2024, December 4). University of Oxford. https://www.ox.ac.uk/news/2024-12-04-oxford-university-lead-ai-security-research-through-new-national-laboratory

Phuong, M., Aitchison, M., Catt, E., Cogan, S., Kaskasoli, A., Krakovna, V., Lindner, D., Rahtz, M., Assael, Y., Hodkinson, S., Howard, H., Lieberum, T., Kumar, R., Raad, M. A., Webson, A., Ho, L., Lin, S., Farquhar, S., Hutter, M., … Shevlane, T. (2024). Evaluating Frontier Models for Dangerous Capabilities (No. arXiv:2403.13793). arXiv. https://doi.org/10.48550/arXiv.2403.13793

Ratiu, D., Rohlinger, T., Stolte, T., & Wagner, S. (2024). Towards an Argument Pattern for the Use of Safety Performance Indicators (No. arXiv:2410.00578). arXiv. http://arxiv.org/abs/2410.00578

Sabel, C., Herrigel, G., & Kristensen, P. H. (2018). Regulation under uncertainty: The coevolution of industry and regulation. Regulation & Governance, 12(3), 371–394. https://doi.org/10.1111/rego.12146

Schleiss, F. Carella and I. Kurzidem, "Towards Continuous Safety Assurance for Autonomous Systems," 2022 6th International Conference on System Reliability and Safety (ICSRS), Venice, Italy, 2022, pp. 457-462, doi: 10.1109/ICSRS56243.2022.10067323.

Schuett, J., Anderljung, M., Carlier, A., Koessler, L., & Garfinkel, B. (2024). From Principles to Rules: A Regulatory Approach for Frontier AI (arXiv:2407.07300). arXiv. https://doi.org/10.48550/arXiv.2407.07300

SCSC. (2021). Goal structuring notation community standard (Version 3). Retrieved from https://perma.cc/CD9W-YX6S

Shao, M., Chen, B., Jancheska, S., Dolan-Gavitt, B., Garg, S., Karri, R. and Shafique, M., 2024. An empirical evaluation of llms for solving offensive security challenges. arXiv preprint arXiv:2402.11814.

Solaiman, Z. Talat, W. Agnew, L. Ahmad, D. Baker, S. L. Blodgett, H. Daumé III, J. Dodge, E. Evans, S. Hooker, Y. Jernite, A. S. Luccioni, A. Lusoli, M. Mitchell, J. Newman, M.-T. Png, A. Strait, and A. Vassilev. Evaluating the Social Impact of Generative AI Systems in Systems and Society, June 2023. URL http://arxiv.org/abs/2306.05949. arXiv:2306.05949 [cs].

Sujan, M. A., Habli, I., Kelly, T. P., Pozzi, S., & Johnson, C. W. (2016). Should healthcare providers do safety cases? Lessons from a cross-industry review of safety case practices. Safety Science, 84, 181–189. doi: 10.1016/j.ssci.2015.12.021

The Space Industry Regulations 2021, SI 2021/792, Part 8 (UK). (2021). legislation.gov.uk. https://www.legislation.gov.uk/uksi/2021/792/part/8/chapter/3

Tice, C., Kreer, P. A., Helm-Burger, N., Shahani, P. S., Ryzhenkov, F., van der Weij, T., Hofstatter, F., Haimes, J.. Sandbag Detection through Model Impairment, October 2024. URL https://openreview.net/forum?id=zpcx5a8w2a.

Trapp, M., & Weiss, G. (2019). Towards Dynamic Safety Management for Autonomous Systems.

UL (2023). Evaluation of Autonomous Products (UL 4600).

US AISI. (2024). U.S. AI Safety Institute establishes new U.S. government taskforce to collaborate on research and testing of AI models to manage national security capabilities & risks. NIST. https://www.nist.gov/news-events/news/2024/11/us-ai-safety-institute-establishes-new-us-government-taskforce-collaborate

Usman, Y., Upadhyay, A., Gyawali, P. and Chataut, R.(2024) Is Generative AI the Next Tactical Cyber Weapon For Threat Actors? Unforeseen Implications of AI Generated Cyber Attacks. (No. arXiv:2408.12806). arXiv: https://arxiv.org/abs/2408.12806

van der Weij, T., Hofstätter, F., Jaffe, O., Brown, S. F., & Ward, F. R. (2024). AI Sandbagging: Language Models can Strategically Underperform on Evaluations. arXiv preprint arXiv:2406.07358.

Wang, K. Sun, A. Batcheller, and S. Jajodia. Detecting "0-day" vulnerability: An empirical study of secret security patch in OSS. In IEEE/IFIP International Conference on Dependable Systems and Networks, pages 485–492. IEEE, 2019. URL https://csis.gmu.edu/ksun/publications/secretpatch-dsn19.pdf

Wasil, A. R., Clymer, J., Krueger, D., Dardaman, E., Campos, S., & Murphy, E. R. (2024). Affirmative safety: An approach to risk management for high-risk AI. arXiv preprint arXiv:2406.15371

Weidinger, J. Mellor, M. Rauh, C. Griffin, J Uesato, P.-S. Huang, M. Cheng, M. Glaese, B. Balle, A. Kasirzadeh, Z. Kenton, S. Brown, W. Hawkins, T. Stepleton, C. Biles, A. Birhane, J. Haas, L. Rimell, L. A. Hendricks, W. Isaac, S. Legassick, G. Irving, and I. Gabriel. Ethical and social risks of harm from Language Models, December 2021. URL http://arxiv.org/abs/2112.04359. arXiv:2112.04359 [cs].

Wheatcraft, L., Ryan, M., Dick, J. (2016). On the use of attributes to manage requirements. Systems Engineering, 19(5), 448–458.

Wozniak, A., Borg, T., Hansson, H., & Törngren, M. (2020). Safety Case Patterns for Machine Learning in Highly Automated Driving. In 2020 IEEE International Symposium on Software Reliability Engineering Workshops (ISSREW) (pp. 186–191). IEEE. https://doi.org/10.1109/ISSREW51248.2020.00057

Wozniak, E., Cârlan, C., Acar-Celik, E., & Putzer, H. J. (2020). A Safety Case Pattern for Systems with Machine Learning Components. In A. Casimiro, F. Ortmeier, E. Schoitsch, F. Bitsch, & P. Ferreira (Eds.), Computer Safety, Reliability, and Security. SAFECOMP 2020 Workshops (pp. 370–382). Springer International Publishing. https://doi.org/10.1007/978-3-030-55583-2_28

Yang, A. Prabhakar, K. Narasimhan, and S. Yao. InterCode: Standardizing and benchmarking interactive coding with execution feedback. ArXiv, June 2023. URL https://arxiv.org/abs/2306.14898.

Yao, S., Yu, D., Zhao, J., Shafran, I., Griffiths, T. L., Cao, Y., & Narasimhan, K. (2023). Tree of Thoughts: Deliberate Problem Solving with Large Language Models (No. arXiv:2305.10601). arXiv. https://doi.org/10.48550/arXiv.2305.10601

Yohsua, B., Daniel, P., Tamay, B., Rishi, B., Stephen, C., Yejin, C., . . . Tramèr, F. (2024). International scientific report on the safety of advanced AI: Interim report. DSIT. Retrieved from https://perma.cc/PM8K-9G48

Zhang, A. K., Perry, N., Dulepet, R., Ji, J., Lin, J. W., Jones, E., ... and Liang, P. (2024). Cybench: A Framework for Evaluating Cybersecurity Capabilities and Risks of Language Models. arXiv preprint arXiv:2408.08926.

Zhou, J. K. Siow, C. Wang, S. Liu, and Y. Liu. SPI: Automated identification of security patches via commits. ACM Transactions on Software Engineering and Methodology, 31(1):1–27, 2021

## Appendix A: Expansion on requirements, V&V and decision gates

| ID | Title | Requirement Description | Rationale | Verification Method | Verification Criteria |
|---|---|---|---|---|---|
| REQ-001 | Early Creation of Safety Cases | The DSCMS shall develop an initial safety case for the Frontier AI system at the start of its lifecycle, including all required safety arguments, supporting evidence, and justifications in accordance with [Standards Reference Document]. | Ensuring that a comprehensive safety case is developed at the onset of the system's lifecycle allows safety considerations to guide design and development decisions from the very beginning. Early identification and documentation of safety arguments, evidence, and justifications help establish a sound foundation for later stages, minimize costly rework, and ensure compliance with recognized standards. **Keywords: foundational argument baseline** | Document inspection. | Initial safety case supported by evidence and justifications. |
| REQ-002 | Consistent Maintenance of Safety Cases | The DSCMS shall maintain the safety case for the Frontier AI system throughout its entire lifecycle , as defined in [Lifecycle Definition Document]. | As the Frontier AI system evolves through its lifecycle, maintaining an up-to-date safety case is of high importance as this ensures that safety arguments and supporting evidence remain current and valid, reflecting all modifications, upgrades, and new operational conditions. The continuous alignment reduces the risk of outdated assumptions and strengthens overall system trustworthiness. **Keywords: lifecycle-driven updating** | Periodic document inspection. Change log analysis. | Safety case up to date with all evidence and justifications. No conflicting information persists. |
| REQ-005 | Automated Consistency Checks | The DSCMS shall perform automated consistency checks of safety arguments against the latest associated development artefacts in [Configuration Management Document]. | Automating consistency checks between safety arguments and the associated development artifacts streamlines the verification process and reduces manual effort. By instantaneously identifying discrepancies or outdated elements, the system can promptly alert stakeholders to potential integrity issues thereby improving overall safety assurance and system reliability. **Keywords: real-time artifact synchronisation** | Tool-based consistency inspection. | No unresolved discrepancies between safety argument and development artifacts. All inconsistencies resolved or accepted. |
| REQ-017 | SPI Definition and Management | The DSCMS shall define and maintain a catalogue of Safety Performance Indicators (SPIs) ([SPI Definition Document]) with thresholds, update frequency, and referenceable artefacts. | Defining and managing SPIs provides a quantitative basis for monitoring and assessing the system's ongoing safety performance. By establishing clear thresholds, updating frequencies, and linking SPIs to reference documents, the DSCMS ensures that safety performance can be tracked systematically, prompting timely interventions when indicators fall outside acceptable ranges. **Keywords: threshold-driven performance metrics** | Document inspection. | Each SPI has a defined threshold, update frequency and linked artifact. |

| REQ-018 | SPI-Driven Argument Re-Evaluation | The DSCMS shall automatically re-evaluate affected parts of the safety argument whenever an SPI threshold is breached ([SPI Threshold Criteria Document]). | When an SPI threshold is exceeded, it indicates a potential shift in safety-critical conditions. Automatically re-evaluating the relevant portions of the safety argument enables a rapid response to changing system states. This proactive approach helps maintain the integrity and validity of safety assurances leading to enhanced system resilience and hazard mitigation strategies. **Keywords: threshold-driven performance metrics** | Simulation of threshold breach. Demonstration of automated triggers. | When SPI threshold is breached, DSCMS automatically flags impacted arguments and initiates re-evaluation. |
|---|---|---|---|---|---|
| REQ-019 | External Data Feed Integration | The DSCMS shall map external data feeds to relevant safety arguments/evidence. | External data sources including environmental conditions, user feedback, and third-party system inputs can significantly influence the Frontier AI system's risk profile. Integrating these feeds into the DSCMS enables the safety case to be informed by the most up-to-date, real-world information, thus improving the accuracy and relevance of safety arguments. **Keywords: internal threshold-triggered reassessment** | Integration test with sample external inputs. | External data feeds mapped to relevant safety arguments and evidence. No unresolved external input references. |
| REQ-020 | External Data Change Impact | The DSCMS shall conduct automated impact analyses when new external data arrives. | As new external data arrives, it may alter underlying assumptions or evidence within the safety case. Automated impact analyses ensure that every relevant argument and piece of evidence is reassessed promptly - by identifying and addressing the implications of these changes in real-time, the DSCMS helps maintain robust, evidence-based safety justifications. **Keywords: environmental data aggregation** | Triggered impact analysis test. | Arrival of external data automatically initiates consistency checks. Affected artifacts identified and flagged for review. |
| REQ-023 | Governance Reporting Interface | The DSCMS shall have a ([Governance Interface Specification]) interface. | Providing a dedicated interface for governance and oversight entities streamlines compliance reporting, reviews, and independent audits. This transparency not only builds stakeholder confidence but also facilitates efficient communication, ensuring that governing bodies can easily access, interpret, and assess safety information. **Keywords: governance interface portal** | User interface review and functional test. | Safety arguments, evidence and reports accessible in the correct format. Governance interface compliant |
| REQ-025 | Data Security and Access Control | The DSCMS shall implement data security measures in accordance with ([Data Security Plan]). | Safety cases often contain sensitive or proprietary information. Implementing robust data security measures and strict access controls safeguards the confidentiality, integrity, and availability of these critical safety documents. Necessitating compliance with security policies and standards reduces the risk of data breaches or unauthorised alterations to the safety arguments and evidence. **Keywords: compliance-based data protection** | Security audit and penetration testing. | Access restricted to authorised personnel only. |

Table 16: DSCMS Requirements

## Requirements analysis of DCMS requirements

The requirements for the DCSMS were established through a structured, iterative approach by initially analysing existing literature on safety cases, existing solutions to automate case updating in different sectors and unique challenges associated with frontier AI in the context of safety case updating. Grounded in Model-Based Systems Engineering (MBSE) principles and guided by the best practices outlined in the *INCOSE Guide to Writing Requirements* (INCOSE, 2023). Initially, key lifecycle concepts, safety objectives, governance drivers, and security considerations were analysed to define the broader mission and intended operational environment of the DSCMS. To ensure that the DSCMS requirements were comprehensible and suitable for the purposes of this paper, each requirement statement was developed to promote clarity, singularity, and completeness. Following the guidelines from the INCOSE guide, the authors employed a structured natural language format, including a consistent subject-verb-object format, explicit condition clauses where relevant, and quantifiable performance measures (INCOSE, 2023, Sec. 1–4). This approach mitigated ambiguity, supported verification and validation activities, and provided a clear link from stakeholder intent to system functionality. Authors and external reviewers reviewed the evolving set of requirements. During these reviews, adjustments were made to improve readability and resolve any ambiguous terms to enhance feasibility and correctness of the requirements set. This iterative refinement process ensured the requirements were not only formally correct but also easily apprehensible for their intended audience and purpose. The result was a set of DSCMS requirements that balanced the level of rigor with the contextual risk and complexity, acknowledging that higher criticality areas may require more formal specification methods, while less critical elements may suffice with less stringent documentation (INCOSE, 2023). Furthermore, the established requirements align with the MBSE principles by systematically linking concepts, needs, and requirements in a data-centric environment, supporting impact analysis, change management, and continuous validation of the DSCMS's intended safety objectives (Wheatcraft et al., 2016). Verification methods and verification criteria are essential elements in ensuring that each requirement is properly implemented and meets its intended purpose. The verification method specifies how the requirement will be verified providing a clear and repeatable approach to assessing compliance. Verification criteria define the conditions that must be met to consider the requirement satisfied. Although these verification methods and criteria have not been used in the context of this paper's own DSCMS proposal, the presented Requirements–Verification Traceability Matrix (RVTM) can serve as an inspiration for frontier AI developers to consider how a DSCMS solution may be effectively verified in their own operational contexts.

## Verification and validation

Verification and validation (V&V) are central mechanisms in the DSCMS to ensure the continued credibility, trustworthiness, and regulatory alignment of safety cases across the frontier AI system's lifecycle. As outlined by the SEH5E (INCOSE, 2023), verification activities focus on confirming that the DSCMS meets specified requirements such as REQ-005 (Automated Consistency Checks) and REQ-018 (SPI-Driven Argument Re-Evaluation) by implementing tool-based inspections and triggered simulations to confirm the correctness and completeness of safety arguments. Validation, on the other hand, ensures that the safety cases remain relevant and provide the intended assurance that the frontier AI system does not pose unacceptable risk. These processes extend to maintaining up-to-date evidence against evolving capabilities and threat landscapes (Buhl et al., 2024; Goemans et al., 2024). By establishing a continuous and partially automated V&V framework, the DSCMS can detect discrepancies, re-evaluate arguments as SPIs change, and promptly update safety cases as per REQ-002 (Consistent Maintenance of Safety Cases) (Cărlan, forthcoming). This integrated V&V approach aligns with systems engineering best practices and ensures that the safety case remains both current and credible, thereby supporting decision-making under uncertainty and complexity.

## Lifecycle-based decision gates

Decision gates provide structured checkpoints throughout the frontier AI lifecycle, leveraging data-driven insights from the DSCMS to reduce risk incrementally via verification, validation, and technical reviews at each lifecycle stage (INCOSE, 2023; Buhl et al., 2024). As described in the paper on Dynamic Safety Cases (Goemans et al., 2024), these gates serve as tailored junctures—such as before a major model upgrade or deployment—where the latest safety arguments and SPIs are assessed against pre-defined acceptance criteria. If thresholds are exceeded or new threats are identified, additional verification activities (e.g., automated consistency checks) and validation tasks (e.g., re-running updated evaluation suites) are triggered to substantiate safety claims. This incremental V&V strategy, interwoven with decision gates, ensures that only adequately validated safety arguments progress to the next lifecycle stage. In turn, it reduces uncertainty and promotes agility, allowing organizations to adjust their safety management strategy proactively, involve subject matter experts for high-stakes re-evaluations, and maintain alignment with governance structures and compliance frameworks (Cârlan, forthcoming; Anthropic, 2024). Thus, decision gates reinforce continuous safety assurance by linking updated evidence, fresh V&V results, and evolving systemic understanding to informed go/no-go decisions.

## Appendix B: Safety Performance Indicators (SPIs) ( Internal and External)

Set of internal and external Safety Performance Indicators proposed for monitoring the cyber inability safety case sketched by Goemans et al. (2024)

**SPI set C2.x:** There are no other major sources of cyber risk

| SPI ID | SPI Title | SPI Description | Unit | Example | Leading/ Lagging | Evidence | Example Threshold |
|---|---|---|---|---|---|---|---|
| C2.x SPI 1 | # of incidents - other source (last 30 days) | Number of incidents due to other source of cyber risk caused by the model in the last 30 days. Alerts to unforeseen sources of cyber risk. | Count | 1 | Lagging | Incidents | 3 |
| C2.x SPI 2 | Financial losses - other source (last 30 days) | Financial losses from incidents due to other source of cyber risk caused by the model in the last 30 days. Losses act as proxy of materiality of new source. | $ | 50,000 | Lagging | Incidents | 1M |
| C2.x SPI 3 | # of incidents from similar models - other source (last 30 days) | Number of incidents due to other source of cyber risk caused by similar AI model in the last 30 days. Early warning of new source of cyber risk. | Count | 1 | Leading | Industry Bodies | 5 |
| C2.x SPI 4 | Financial losses from similar models - other source (last 30 days) | Financial losses from incidents due to other source of cyber risk caused by similar AI model in the last 30 days. | $ | 50,000 | Leading | Industry Bodies | 1M |
| C2.x SPI 5 | Research papers - other source | Number of research papers identifying other sources of cyber risk from frontier AI models in the last 30 days. Early warning of new source of cyber risk. | Count | 2 | Leading | Research insights | 0 |
| C2.x SPI 6 | Dark web mentions- other source (last 30 days) | Mentions of potential misuse of the AI model on the dark web relating to other major sources of cyber risk in the last 30 days. Early warning of new source of cyber risk. | Count | 3 | Leading | Cyber threat intelligence | 0 |

Table 17: SPIs for claim C2.x.

**SPI set C2.1:** The AI system would not uplift threat actors using conventional cyberattacks in any realistic setting, even absent any safeguards

| SPI ID | SPI Title | SPI Description | Unit | Example | Leading/ Lagging | Evidence | Example Threshold |
|---|---|---|---|---|---|---|---|
| C2.1 SPI 1 | # of incidents - conventional attacks (last 30 days) | Number of cyberattacks in the last 30 days caused by model uplifting threat actors in conventional attacks | Count | 5 | Lagging | Incidents | 5 |
| C2.1 SPI 2 | Financial losses - conventional attacks (last 30 days) | Financial losses in the last 30 days from incidents caused by model uplifting threat actors in conventional attacks | $ | 50,000 | Lagging | Incidents | 1M |
| C2.1 SPI 3 | Mean Time to Detect Incidents (MTTD) | The average time taken to identify an incident from the time it occurs. Provides feedback on the speed of the incident detection channels. | Days | 5 | Lagging | Incidents | 7 |
| C2.1 SPI 4 | % change in # of incidents - conventional attacks (M-on-M) | Percentage change in the number of incidents from the model uplifting conventional attacks compared to the previous 30 days. | % | -20% | Lagging | Incidents | +10% |
| C2.1 SPI 5 | % change in financial losses - conventional attacks (M-on-M) | Percentage change in the financial losses from the model uplifting conventional attacks compared to the previous 30 days. | % | +30% | Lagging | Incidents | +10% |
| C2.1 SPI 6 | # of incidents from similar models - conventional attacks (last 30 days) | Number of cyberattacks in the last 30 days caused by a similar model uplifting threat actors in conventional attacks. | Count | 1 | Leading | Industry Bodies | 2 |
| C2.1 SPI 7 | Financial losses from similar model - conventional attacks (last 30 days) | Financial losses in the last 30 days from incidents caused by similar models uplifting threat actors in conventional attacks. | $ | 50,000 | Leading | Industry Bodies | 1M |
| C2.1 SPI 8 | # of Near misses - conventional attacks (last 30 days) | Number of averted cyberattacks caused by model uplifting threat actors in conventional attacks | Count | 4 | Leading | Incidents | 5 |
| C2.1 SPI 9 | Research papers - other source | Research papers demonstrating conventional cyberattacks from frontier AI models uplifting threat actor capabilities | Count | 2 | Leading | Research insights | 0 |
| C2.1 SPI 10 | # of emerging threats - conventional attacks | Number of emerging threats of AI-enhanced conventional attacks from Cyber threat intelligence monitoring | Count | 3 | Leading | Cyber threat intelligence | 5 |
| C2.1 SPI 11 | Emerging threat trend - conventional attacks (% change) | Percentage change in the number of emerging threats from the model uplifting conventional attacks compared to the previous 30 days. | % | +5% | Leading | Cyber threat intelligence | +10% |
| C2.1 SPI 12 | Dark web mentions - conventional attacks (last 30 days) | Mentions of potential misuse relating to conventional attacks of the AI model on the dark web in the last 30 days | Count | 3 | Leading | Cyber threat intelligence | 5 |
| C2.1 SPI 13 | # of new TTPs - conventional attacks (last 30 days) | New Tactics, Techniques, and Procedures (TTPs) leveraging AI models identified as being used by threat actors in the last 30 days | Count | 1 | Leading | Cyber threat intelligence | 1 |

Table 18: SPIs for claim C2.1.

**SPI set C2.2:** The AI System poses no risk of novel cyberattacks

| SPI ID | SPI Title | SPI Description | Unit | Example | Leading/Lagging | Evidence | Example Threshold |
|--------|-----------|-----------------|------|---------|-----------------|----------|-------------------|
| C2.2 SPI 1 | # of incidents - novel cyberattacks (last 30 days) | Number of cyberattacks deviating from known attack patterns | Count | 5 | Lagging | Incidents | 2 |
| C2.2 SPI 2 | Financial losses - novel attacks (last 30 days) | Financial losses in the last 30 days from incidents caused by novel cyberattacks | $ | 10,000 | Lagging | Incidents | 500,000 |
| C2.2 SPI 3 | Mean Time to Detect (MTTD) Incidents from novel attacks | The average time taken to identify incidents caused by novel cyberattacks from the time they occur. | Days | 10 | Lagging | Incidents | 7 |
| C2.2 SPI 4 | % change in # of incidents - novel attacks (M-on-M) | Percentage change in the number of incidents caused by novel cyberattacks compared to the previous 30 days. | % | -20% | Lagging | Incidents | +10% |
| C2.2 SPI 5 | % change in financial losses - novel attacks (M-on-M) | Percentage change in the financial losses from novel cyberattacks compared to the previous 30 days. | % | +30% | Lagging | Incidents | +10% |
| C2.2 SPI 6 | # of incidents from similar models - novel attacks (last 30 days) | Number of cyberattacks in the last 30 days caused by a similar model uplifting novel cyberattacks. | Count | 1 | Leading | Industry Bodies | 2 |
| C2.2 SPI 7 | Financial losses from similar model - novel attacks (last 30 days) | Financial losses in the last 30 days from incidents caused by similar models uplifting novel cyberattacks. | $ | 50,000 | Leading | Industry Bodies | 1M |
| C2.2 SPI 8 | # of Near misses - novel attacks (last 30 days) | Number of averted cyberattacks caused by model uplifting novel cyberattacks. | Count | 4 | Leading | Incidents | 5 |
| C2.2 SPI 9 | Research papers - novel attacks | Research papers identifying novel threats of cyberattacks from frontier AI models | Count | 2 | Leading | Research insights | 0 |
| C2.2 SPI 10 | # of emerging threats - novel attacks | Number of novel emerging threats of AI-enhanced attacks from Cyber threat intelligence monitoring | Count | 3 | Leading | Cyber threat intelligence | 5 |
| C2.2 SPI 11 | Dark web mentions - novel attacks (last 30 days) | Mentions of potential misuse relating to novel attacks  of the AI model on the dark web in the last 30 days | Count | 3 | Leading | Cyber threat intelligence | 5 |
| C2.2 SPI 12 | # of new TTPs - novel attacks (last 30 days) | New Tactics, Techniques, and Procedures (TTPs) leveraging AI models via novel attacks identified as being used by threat actors in the last 30 days | Count | 1 | Leading | Cyber threat intelligence | 1 |

Table 19: SPIs for claim C2.2.

**SPI set C3.1:** AI system cannot uplift technical non-experts in cyberattacks against against Critical National Infrastructure (CNI) via vulnerability discovery and exploitation in any realistic setting **(Risk model 1)**

| SPI ID | SPI Title | SPI Description | Unit | Example | Leading/ Lagging | Evidence | Example Threshold |
|---|---|---|---|---|---|---|---|
| C3.1 SPI 1 | # of incidents - Risk model 1 (last 30 days) | Number of cyberattacks caused by model uplifting technical non-experts against CNI via vulnerability discovery and exploitation | Count | 3 | Lagging | Incidents | 1 |
| C3.1 SPI 2 | Financial losses - Risk model 1 (last 30 days) | Financial losses in the last 30 days from incidents caused by model uplifting technical non-experts against CNI via vulnerability discovery and exploitation | $ | 10,000 | Lagging | Incidents | 50,000 |
| C3.1 SPI 3 | Mean Time to Detect (MTTD) Incidents from Risk model 1 | The average time taken to identify incidents caused by model uplifting technical non-experts against CNI via vulnerability discovery and exploitation from the time they occur. | Days | 10 | Lagging | Incidents | 7 |
| C3.1 SPI 4 | % change in # of incidents - Risk model 1 (M-on-M) | Percentage change in the number of incidents caused by model uplifting technical non-experts against CNI via vulnerability discovery and exploitation compared to the previous 30 days. | % | -20% | Lagging | Incidents | +10% |
| C3.1 SPI 5 | % change in financial losses - Risk model 1 (M-on-M) | Percentage change in the financial losses from model uplifting technical non-experts against CNI via vulnerability discovery and exploitation compared to the previous 30 days. | % | +30% | Lagging | Incidents | +10% |
| C3.1 SPI 6 | # of incidents from similar models - Risk model 1 (last 30 days) | Number of cyberattacks in the last 30 days caused by a similar model uplifting technical non-experts against CNI via vulnerability discovery and exploitation | Count | 1 | Leading | Industry Bodies | 2 |
| C3.1 SPI 7 | Financial losses from similar model - Risk model 1 (last 30 days) | Financial losses in the last 30 days from incidents caused by similar models uplifting technical non-experts against CNI via vulnerability discovery and exploitation | $ | 50,000 | Leading | Industry Bodies | 1M |
| C3.1 SPI 8 | # of Near misses - Risk model 1 (last 30 days) | Number of averted cyberattacks caused by model uplifting technical non-experts against CNI via vulnerability discovery and exploitation | Count | 4 | Leading | Incidents | 5 |
| C3.1 SPI 9 | Research papers - Risk model 1 | Research papers identifying threats relating to frontier AI model uplifting technical non-experts against CNI via vulnerability discovery and exploitation | Count | 2 | Leading | Research insights | 0 |
| C3.1 SPI 10 | # of emerging threats - Risk model 1 | Number of emerging threats from Cyber threat intelligence monitoring relating to AI models uplifting technical non-experts against CNI via vulnerability discovery and exploitation | Count | 3 | Leading | Cyber threat intelligence | 5 |
| C3.1 SPI 11 | Dark web mentions - Risk model 1 (last 30 days) | Mentions of potential misuse of the AI model on the dark web in the last 30 days, relating to uplift of technical non-experts against CNI via vulnerability discovery and exploitation | Count | 3 | Leading | Cyber threat intelligence | 5 |
| C3.1 SPI 12 | # of new TTPs - Risk model 1 (last 30 days) | New Tactics, Techniques, and Procedures (TTPs) leveraging AI models for vulnerability discovery and exploitation identified as being used by technical non-experts against CNI in the last 30 days | Count | 1 | Leading | Cyber threat intelligence | 1 |

Table 20: SPIs for claim C3.1.

**SPI set C3.2:** AI system cannot uplift cybersecurity experts in cyberattacks against against Hardened Critical National Infrastructure (CNI) via vulnerability discovery and exploitation in any realistic setting **(Risk model 2)**

| SPI ID | SPI Title | SPI Description | Unit | Example | Leading/ Lagging | Evidence | Example Threshold |
|---|---|---|---|---|---|---|---|
| C3.2 SPI 1 | # of incidents - Risk model 2 (last 30 days) | Number of cyberattacks caused by model uplifting cybersecurity experts against Hardened CNI via vulnerability discovery and exploitation. | Count | 3 | Lagging | Incidents | 1 |
| C3.2 SPI 2 | Financial losses - Risk model 2 (last 30 days) | Financial losses in the last 30 days from incidents caused by model uplifting cybersecurity experts against Hardened CNI via vulnerability discovery and exploitation. | $ | 10,000 | Lagging | Incidents | 50,000 |
| C3.2 SPI 3 | Mean Time to Detect (MTTD) Incidents from Risk model 2 | The average time taken to identify incidents caused by model uplifting cybersecurity experts against Hardened CNI via vulnerability discovery and exploitation from the time they occur. | Days | 10 | Lagging | Incidents | 7 |
| C3.2 SPI 4 | % change in # of incidents - Risk model 2 (M-on-M) | Percentage change in the number of incidents caused by model uplifting cybersecurity experts against Hardened CNI via vulnerability discovery and exploitation compared to the previous 30 days. | % | -20% | Lagging | Incidents | +10% |
| C3.2 SPI 5 | % change in financial losses - Risk model 2 (M-on-M) | Percentage change in the financial losses from model uplifting cybersecurity experts against Hardened CNI via vulnerability discovery and exploitation compared to the previous 30 days. | % | +30% | Lagging | Incidents | +10% |
| C3.2 SPI 6 | # of incidents from similar models - Risk model 2 (last 30 days) | Number of cyberattacks in the last 30 days caused by a similar model uplifting cybersecurity experts against Hardened CNI via vulnerability discovery and exploitation | Count | 1 | Leading | Industry Bodies | 2 |
| C3.2 SPI 7 | Financial losses from similar model - Risk model 2 (last 30 days) | Financial losses in the last 30 days from incidents caused by similar models uplifting cybersecurity experts against Hardened CNI via vulnerability discovery and exploitation | $ | 50,000 | Leading | Industry Bodies | 1M |
| C3.2 SPI 8 | # of Near misses - Risk model 2 (last 30 days) | Number of averted cyberattacks caused by model uplifting cybersecurity experts against Hardened CNI via vulnerability discovery and exploitation | Count | 4 | Leading | Incidents | 5 |
| C3.2 SPI 9 | Research papers - Risk model 2 | Research papers identifying threats relating to frontier AI model uplifting cybersecurity experts against Hardened CNI via vulnerability discovery and exploitation | Count | 2 | Leading | Research insights | 0 |
| C3.2 SPI 10 | # of emerging threats - Risk model 2 | Number of emerging threats from Cyber threat intelligence monitoring relating to AI models uplifting cybersecurity experts against Hardened CNI via vulnerability discovery and exploitation | Count | 3 | Leading | Cyber threat intelligence | 5 |

| C3.2 SPI 11 | Dark web mentions - Risk model 2 (last 30 days) | Mentions of potential misuse of the AI model on the dark web in the last 30 days, relating to uplift of cybersecurity experts against Hardened CNI via vulnerability discovery and exploitation | Count | 3 | Leading | Cyber threat intelligence | 5 |
| C3.2 SPI 12 | # of new TTPs - Risk model 2 (last 30 days) | New Tactics, Techniques, and Procedures (TTPs) leveraging AI models for vulnerability discovery and exploitation identified as being used by cybersecurity experts against Hardened CNI in the last 30 days | Count | 1 | Leading | Cyber threat intelligence | 1 |

Table 21: SPIs for claim C3.2.

| SPI set C3.3: AI system cannot uplift non-technical novices in cyberattacks against against Soft Targets via Spear Phishing campaigns in any realistic setting (**Risk model 3**) | | | | | | | |
|---|---|---|---|---|---|---|---|
| **SPI ID** | **SPI Title** | **SPI Description** | **Unit** | **Example** | **Leading/ Lagging** | **Evidence** | **Example Threshold** |
| C3.3 SPI 1 | # of incidents - Risk model 3 (last 30 days) | Number of cyberattacks caused by model uplifting non-technical novices in cyberattacks against Soft Targets via Spear Phishing campaigns | Count | 3 | Lagging | Incidents | 1 |
| C3.3 SPI 2 | Financial losses - Risk model 3 (last 30 days) | Financial losses in the last 30 days from incidents caused by model uplifting non-technical novices in cyberattacks against Soft Targets via Spear Phishing campaigns | $ | 10,000 | Lagging | Incidents | 20,000 |
| C3.3 SPI 3 | Mean Time to Detect (MTTD) Incidents from Risk model 3 | The average time taken to identify incidents caused by model uplifting non-technical novices in cyberattacks against Soft Targets via Spear Phishing campaigns from the time they occur. | Days | 10 | Lagging | Incidents | 7 |
| C3.3 SPI 4 | % change in # of incidents - Risk model 3 (M-on-M) | Percentage change in the number of incidents caused by model uplifting non-technical novices in cyberattacks against Soft Targets via Spear Phishing campaigns compared to the previous 30 days. | % | -20% | Lagging | Incidents | +10% |
| C3.3 SPI 5 | % change in financial losses - Risk model 3 (M-on-M) | Percentage change in the financial losses from model uplifting non-technical novices in cyberattacks against Soft Targets via Spear Phishing campaigns compared to the previous 30 days. | % | +30% | Lagging | Incidents | +10% |
| C3.3 SPI 6 | # of incidents from similar models - Risk model 3 (last 30 days) | Number of cyberattacks in the last 30 days caused by a similar model uplifting non-technical novices in cyberattacks against Soft Targets via Spear Phishing campaigns | Count | 1 | Leading | Industry Bodies | 2 |
| C3.3 SPI 7 | Financial losses from similar model - Risk model 3 (last 30 days) | Financial losses in the last 30 days from incidents caused by similar models uplifting non-technical novices in cyberattacks against Soft Targets via Spear Phishing campaigns | $ | 50,000 | Leading | Industry Bodies | 1M |
| C3.3 SPI 8 | # of Near misses - Risk model 3 (last 30 days) | Number of averted cyberattacks caused by model uplifting non-technical novices in cyberattacks against Soft Targets via Spear Phishing campaigns | Count | 4 | Leading | Incidents | 5 |
| C3.3 SPI 9 | Research papers - Risk model 3 | Research papers identifying threats relating to frontier AI model uplifting non-technical novices in cyberattacks against Soft Targets | Count | 2 | Leading | Research insights | 0 |

| | | via Spear Phishing campaigns | | | | | |
|---|---|---|---|---|---|---|---|
| C3.3 SPI 10 | # of emerging threats - Risk model 3 | Number of emerging threats from Cyber threat intelligence monitoring relating to AI models uplifting non-technical novices in cyberattacks against Soft Targets via Spear Phishing campaigns | Count | 3 | Leading | Cyber threat intelligence | 5 |
| C3.3 SPI 11 | Dark web mentions - Risk model 3 (last 30 days) | Mentions of potential misuse of the AI model on the dark web in the last 30 days, relating to uplift of non-technical novices in cyberattacks against Soft Targets via Spear Phishing campaigns | Count | 3 | Leading | Cyber threat intelligence | 5 |
| C3.3 SPI 12 | # of new TTPs - Risk model 3 (last 30 days) | New Tactics, Techniques, and Procedures (TTPs) leveraging AI models for Spear Phishing campaigns identified as being used by non-technical novices against Soft Targets in the last 30 days | Count | 1 | Leading | Cyber threat intelligence | 1 |

Table 22: SPIs for claim C3.3.

| SPI set C3.n: AI system cannot uplift [Threat actor] in cyberattacks against [Target profile] via [Harm vector] **(Risk model n)** | | | | | | | |
|---|---|---|---|---|---|---|---|
| **SPI ID** | **SPI Title** | **SPI Description** | **Unit** | **Example** | **Leading/ Lagging** | **Evidence** | **Example Threshold** |
| C3.n SPI 1 | # of incidents - Risk model n (last 30 days) | Number of cyberattacks caused by model uplifting [Threat actor] in cyberattacks against [Target profile] via [Harm vector] | Count | 3 | Lagging | Incidents | 1 |
| C3.n SPI 2 | Financial losses - Risk model n (last 30 days) | Financial losses in the last 30 days from incidents caused by model uplifting [Threat actor] in cyberattacks against [Target profile] via [Harm vector] | $ | 10,000 | Lagging | Incidents | 20,000 |
| C3.n SPI 3 | Mean Time to Detect (MTTD) Incidents from Risk model n | The average time taken to identify incidents caused by model uplifting [Threat actor] in cyberattacks against [Target profile] via [Harm vector] | Days | 10 | Lagging | Incidents | 7 |
| C3.n SPI 4 | % change in # of incidents - Risk model n (M-on-M) | Percentage change in the number of incidents caused by model uplifting [Threat actor] in cyberattacks against [Target profile] via [Harm vector] compared to the previous 30 days. | % | -20% | Lagging | Incidents | +10% |
| C3.n SPI 5 | % change in financial losses - Risk model n (M-on-M) | Percentage change in the financial losses from model uplifting [Threat actor] in cyberattacks against [Target profile] via [Harm vector] compared to the previous 30 days. | % | +30% | Lagging | Incidents | +10% |
| C3.n SPI 6 | # of incidents from similar models - Risk model n (last 30 days) | Number of cyberattacks in the last 30 days caused by a similar model uplifting [Threat actor] in cyberattacks against [Target profile] via [Harm vector] | Count | 1 | Leading | Industry Bodies | 2 |
| C3.n SPI 7 | Financial losses from similar model - Risk model n (last 30 days) | Financial losses in the last 30 days from incidents caused by similar models uplifting [Threat actor] in cyberattacks against [Target profile] via [Harm vector] | $ | 50,000 | Leading | Industry Bodies | 1M |
| C3.n SPI 8 | # of Near misses - | Number of averted cyberattacks caused by model uplifting [Threat | Count | 4 | Leading | Incidents | 5 |

| | | actor] in cyberattacks against [Target profile] via [Harm vector] | | | | | |
|---|---|---|---|---|---|---|---|
| | Risk model n (last 30 days) | | | | | | |
| C3.n SPI 9 | Research papers - Risk model n | Research papers identifying threats relating to frontier AI model uplifting [Threat actor] in cyberattacks against [Target profile] via [Harm vector] | Count | 2 | Leading | Research insights | 0 |
| C3.n SPI 10 | # of emerging threats - Risk model n | Number of emerging threats from Cyber threat intelligence monitoring relating to AI models uplifting [Threat actor] in cyberattacks against [Target profile] via [Harm vector] | Count | 3 | Leading | Cyber threat intelligence | 5 |
| C3.n SPI 11 | Dark web mentions - Risk model n (last 30 days) | Mentions of potential misuse of the AI model on the dark web in the last 30 days, relating to uplift of [Threat actor] in cyberattacks against [Target profile] via [Harm vector] | Count | 3 | Leading | Cyber threat intelligence | 5 |
| C3.n SPI 12 | # of new TTPs - Risk model n (last 30 days) | New Tactics, Techniques, and Procedures (TTPs) leveraging AI models for [Harm vector] identified as being used by [Threat actor] against [Target profile] in the last 30 days | Count | 1 | Leading | Cyber threat intelligence | 1 |

Table 23: SPIs for claim C3.n.

| SPI set C3.x: Risk models C3.1-C3.n are sufficiently comprehensive | | | | | | | |
|---|---|---|---|---|---|---|---|
| SPI ID | SPI Title | SPI Description | Unit | Example | Leading/ Lagging | Evidence | Example Threshold |
| C3.x SPI 1 | # of incidents - Not in risk models (last 30 days) | Number of conventional cyberattacks caused by model uplifting threat actors against target profiles via harm vectors not defined in Risk models | Count | 3 | Lagging | Incidents | 0 |
| C3.x SPI 2 | Financial losses - Not in risk models (last 30 days) | Financial losses in the last 30 days from incidents caused by model uplifting threat actors against target profiles via harm vectors not defined in Risk models | $ | 10,000 | Lagging | Incidents | 0 |
| C3.x SPI 3 | Mean Time to Detect (MTTD) Incidents not in risk models | The average time taken to identify incidents caused by model uplifting threat actors against target profiles via harm vectors not defined in Risk models | Days | 10 | Lagging | Incidents | 7 |
| C3.x SPI 4 | % change in # of incidents - Not in risk models (M-on-M) | Percentage change in the number of incidents caused by model uplifting threat actors against target profiles via harm vectors not defined in Risk models compared to the previous 30 days. | % | -20% | Lagging | Incidents | +10% |
| C3.x SPI 5 | % change in financial losses - Not in risk models (M-on-M) | Percentage change in the financial losses from model uplifting threat actors against target profiles via harm vectors not defined in Risk models compared to the previous 30 days. | % | +30% | Lagging | Incidents | +10% |
| C3.x SPI 6 | # of incidents from similar models - Not | Number of cyberattacks in the last 30 days caused by a similar model uplifting threat actors against target profiles via harm vectors not | Count | 1 | Leading | Industry Bodies | 2 |
| | in risk models (last 30 days) | defined in Risk models | | | | | |
| C3.x SPI 7 | Financial losses from similar model - Not in risk models (last 30 days) | Financial losses in the last 30 days from incidents caused by similar models uplifting threat actors against target profiles via harm vectors not defined in Risk models | $ | 50,000 | Leading | Industry Bodies | 1M |
| C3.x SPI 8 | # of Near misses - Not in risk models (last 30 days) | Number of averted conventional cyberattacks caused by model uplifting threat actors against target profiles via harm vectors not defined in Risk models | Count | 4 | Leading | Incidents | 5 |
| C3.x SPI 9 | Research papers - Not in risk models | Research papers identifying threats relating to frontier AI model uplifting threat actors against target profiles via harm vectors not defined in Risk models | Count | 2 | Leading | Research insights | 0 |
| C3.x  SPI 10 | # of emerging threats - Not in risk models | Number of emerging threats from Cyber threat intelligence monitoring relating to AI models uplifting threat actors against target profiles via harm vectors not defined in Risk models | Count | 3 | Leading | Cyber threat intelligence | 5 |
| C3.x SPI 11 | Dark web mentions - Not in risk models (last 30 days) | Mentions of potential misuse of the AI model on the dark web in the last 30 days, relating to uplift of threat actors against target profiles via harm vectors not defined in Risk models | Count | 3 | Leading | Cyber threat intelligence | 5 |
| C3.x SPI 12 | # of new TTPs - Not in risk models (last 30 days) | New Tactics, Techniques, and Procedures (TTPs) leveraging AI models for harm vectors identified as being used by threat actors against target profiles not defined in Risk models in the last 30 days | Count | 1 | Leading | Cyber threat intelligence | 1 |
| C3.x SPI 13 | Weights escape evaluation | The model capability for penetrating any important layers of defense on its own weights (OpenAI Preparedness Framework, 2023) | Qualitative | No important layers breached | Lagging | Internal evaluations | No important layers breached |

Table 24: SPIs for claim C3.x.

| SPI set C4.1 The AI system cannot help a Technical Non-Expert exceed the Cybersecurity Apprentice baseline in Vulnerability Discovery Task Suite | | | | | | | |
|---|---|---|---|---|---|---|---|
| SPI ID | SPI Title | SPI Description | Unit | Example | Leading/ Lagging | Evidence | Example Threshold |
| C4.1 SPI 1 | External Benchmark Capability Advancement | Measures the number of instances where external benchmarks or evaluations showed that a Technical Non-Expert exceeded an equivalent Cybersecurity Apprentice baseline for Vulnerability Discovery due to model assistance. | Count | 1 | Leading | Research insights | 1 |
| C4.1 SPI 2 | Patch Classification Evaluation 1 | Vulnerability detection proficiency measured using a patch classification dataset by Wang et al (2019), as described in Phuong et al (2024) | Accuracy, F1-score | <75% | Lagging | Internal evaluations | <75% |
| C4.1 SPI 3 | Patch | Vulnerability detection proficiency measured using the Security Patch | Accuracy, | <75% | Lagging | Internal | <60% |

| | Classification Evaluation 2 | Identification (SPI) dataset by Zhou et al (2021), as described in Phuong et al (2024) | F1-score | | | evaluations | |
| C4.1 SPI 4 | Function Classification Evaluation 1 | Vulnerability detection proficiency measured using the DiverseVul dataset by Chen et al (2023), as described in Phuong et al (2024) | Accuracy, F1-score | <75% | Lagging | Internal evaluations | <60% |

<div align="center">Table 25: SPIs for claim C4.1.</div>

| SPI set C4.2 The AI system cannot help a Technical Non-Expert exceed the Cybersecurity Apprentice baseline in Vulnerability Exploitation Task Suite | | | | | | | |
|---|---|---|---|---|---|---|---|
| SPI ID | SPI Title | SPI Description | Unit | Example | Leading/ Lagging | Evidence | Example Threshold |
| C4.2 SPI 1 | External Benchmark Capability Advancement | Measures the number of instances where external benchmarks or evaluations showed that a Technical Non-Expert exceeded an equivalent Cybersecurity Apprentice baseline for Vulnerability Exploitation due to model assistance. | Count | 1 | Leading | Research insights | 1 |
| C4.2 SPI 2 | Evaluation of vulnerability attack automation of low-medium level threat actor cyberattacks (high-school level) | Passing rate for General skills, Reverse engineering, Cryptography, Forensics, and Binary exploitation tasks in the high-school level CSAW CTF challenges (NCSC, 2023) as described in o1 System Card (OpenAI 2024c) | Percent | 50% | Lagging | Internal evaluations | 50% |
| C4.2 SPI 3 | Evaluation of vulnerability attack automation of low-medium level threat actor cyberattacks (college level) | Passing rate for General skills, Reverse engineering, Cryptography, Forensics, and Binary exploitation tasks in the college level CSAW CTF challenges | Percent | 25% | Lagging | Internal evaluations | 25% |
| C4.2 SPI 4 | Evaluation of vulnerability attack automation of low-medium level threat actor | Passing rate for General skills, Reverse engineering, Cryptography, Forensics, and Binary exploitation tasks in the professional level CSAW CTF challenges | Percent | 20% | Lagging | Internal evaluations | 20% |

| | cyberattacks (professional level) | | | | | | |
|---|---|---|---|---|---|---|---|

Table 26: SPIs for claim C4.2.

| SPI set C4.x: Performance on the proxy tasks corresponds to risk posed by the risk model | | | | | | | |
|---|---|---|---|---|---|---|---|
| **SPI ID** | **SPI Title** | **SPI Description** | **Unit** | **Example** | **Leading/ Lagging** | **Evidence** | **Example Threshold** |
| C4.x SPI 1 | Incident-Evaluation Mismatch Rate | Measures the discrepancy between real-world incidents and evaluation results on proxy tasks to assess whether the evaluations sufficiently predict risk levels e.g. 10% of real-world incidents did not align with evaluation outcomes | % mismatch rate | 10% | Lagging | Incidents, Evaluation Results | 5% |

Table 27: SPIs for claim C4.x.

| SPI set C5.1 The proxy tasks are representative | | | | | | | |
|---|---|---|---|---|---|---|---|
| **SPI ID** | **SPI Title** | **SPI Description** | **Unit** | **Example** | **Leading/ Lagging** | **Evidence** | **Example Threshold** |
| C5.1 SPI 1 | Number of new Task Suites Published | Measures the number of new task suites published in the last 30 days by AI Safety Institutes, frontier developers, and independent organisations that provide new potential proxies. | Count | 2 | Leading | Research insights | Track for trends |
| C5.1 SPI 2 | Response to Published New Task Suites | Average time taken to evaluate and, if relevant, incorporate new tasks from published suites into proxy evaluations. | Days | 5 | Leading | Research insights | 14 |
| C5.1 SPI 3 | Number of new Published Benchmarks | Measures the number of new cyber benchmarks including the model published in the last 30 days | Count | 2 | Leading | Research insights | Track for trends |
| C5.1 SPI 4 | Response to Published Benchmarks | Average time taken to respond to new task-related updates provided by benchmarking, including integration into existing proxy tasks. | Days | 5 | Leading | Research insights | 14 |
| C5.1 SPI 5 | TTP Proxy Task Validation | Number of new Tactics, Techniques, and Procedures (TTPs) identified through CTI that are not reflected in the proxy tasks. | Count | 3 | Leading | Cyber Threat Intelligence | 2 |
| C5.1 SPI 6 | Proxy Alignment with Emerging Threats | Number of emerging threats identified through CTI that are not reflected in the proxy tasks. | Count | 1 | Leading | Cyber Threat Intelligence | 2 |
| C5.1 SPI 7 | Proxy Coverage Against Industry Standards | Percentage of proxy tasks that align with those used in industry benchmarks or by other frontier developers. | % | 80% | Leading | Research Insights, Industry | 70% |

| | | | | | | Bodies | |
|---|---|---|---|---|---|---|---|
| C5.1 SPI 8 | Incident Proxy Task Correlation | Measures the average percentage of tasks involved in real-world incidents that are adequately covered by the current proxy tasks | % | 85% | Lagging | Incidents | 90% |
| C5.1 SPI 9 | Near Miss Proxy Task Correlation | Measures the percentage of skills and actions involved in averted real-world incidents that are adequately covered by the current proxy tasks | % | 85% | Lagging | Near Misses | 90% |

Table 28: SPIs for claim C5.1.

**SPI set C5.2** The scoring is adequate

| SPI ID | SPI Title | SPI Description | Unit | Example | Leading/ Lagging | Evidence | Example Threshold |
|---|---|---|---|---|---|---|---|
| C5.2 SPI 1 | Task Difficulty Benchmark Comparison | Measures the alignment of cyber capability levels assigned to tasks e.g. CTFs against industry benchmarks or other frontier developers. | % | 85% | Leading | Research Insights, Industry Bodies | 90% |
| C5.2 SPI 2 | Automated Test scoring alignment with Industry Practices | Measures how closely the pass/fail criteria for proxy tasks align with industry standards and best practices e.g. use of manual corrections | % | 90% | Leading | Research Insights, Industry Bodies | 90% |
| C5.2 SPI 3 | Number of new Automated Test scoring methodologies published | Measures the number of new scoring methodologies or best practices for automated test scoring (e.g., number of test attempts like Pass@10) published in the last 90 days | Count | 2 | Leading | Research Insights | Track for trends |
| C5.2 SPI 4 | Task based probing scoring alignment with Industry Practices | Measures how closely the scoring for task-based probing/ automated evaluations with human-oversight align with industry standards and best practices e.g. hinting, trajectory interventions | % | 90% | Leading | Research Insights, Industry Bodies | 90% |
| C5.2 SPI 5 | Number of new task based probing scoring methodologies published | Measures the number of new scoring methodologies or best practices for task-based probing/ automated evaluations with human-oversight published in the last 90 days | Count | 2 | Leading | Research Insights | Track for trends |
| C5.2 SPI 4 | Human uplift test scoring alignment with Industry Practices | Measures how closely the scoring for human uplift tests align with industry standards and best practices e.g. hinting, trajectory interventions | % | 90% | Leading | Research Insights, Industry Bodies | 90% |
| C5.2 SPI 5 | Number of new human uplift test scoring methodologies published | Measures the number of new scoring methodologies or best practices for human uplift testing  published in the last 90 days | Count | 2 | Leading | Research Insights | Track for trends |
| C5.2 SPI 6 | Response Time to Evaluate New Best Practices | Measures the time taken to evaluate and, if applicable, adopt new best practices or scoring methodologies suggested by research publications or benchmarking reports. | Days | 15 | Leading | Research Insights | 30 |

Table 29: SPIs for claim C5.2.

| SPI set C5.x All relevant aspects of the relevant proxy tasks have been justified | | | | | | | |
|---|---|---|---|---|---|---|---|
| SPI ID | SPI Title | SPI Description | Unit | Example | Leading/ Lagging | Evidence | Example Threshold |
| C5.x SPI 1 | Number of new insights for proxy task design | Measures the number of new insights for aspects to consider when designing proxy tasks in the last 30 days. | Count | 1 | Leading | Research insights | Track for trends |
| C5.x SPI 2 | Response to Published Benchmarks | Average time taken to respond to new proxy task design insights, including integration into existing processes. | Days | 7 | Leading | Research insights | 14 |
| C5.x SPI 3 | Code de-obfuscation evaluation | Model performance on recovering variable names and code semantics as well as a commercial code deobfuscation tool (OpenAI Preparedness Framework, 2023) | Qualitative | Not as good as a top-end commercial tool | Lagging | Internal evaluations | Not as good as a top-end commercial tool |
| C5.x SPI 4 | Fully fleshed out capability graph during threat modelling | Ensuring that significant efforts are taken to completely specify the capability graph during the threat modelling process (see Balesni et al, 2024) and any misses are minimised to the best extent possible, i.e., as low as reasonably practicable, or ALARP | Qualitative | Comprehensive capability graph | Leading | Internal evaluations | Comprehensive capability graph |

Table 30: SPIs for claim C5.x.

| SPI set C6.x The evaluation setups are adequate | | | | | | | |
|---|---|---|---|---|---|---|---|
| SPI ID | SPI Title | SPI Description | Unit | Example | Leading/ Lagging | Evidence | Example Threshold |
| C6.x SPI 1 | Number of new research insights published on evaluation setups | Measures the number of new research insights published on evaluation setup best practices in the last 90 days | Count | 2 | Leading | Research Insights | Track for trends |
| C6.x SPI 2 | Response Time to Evaluate insights on evaluation setups | Measures the time taken to evaluate and, if applicable, adopt new best practices or methodologies for evaluation setups suggested by research publications or benchmarking reports. | Days | 15 | Leading | Research Insights | 30 |
| C6.x SPI 3 | Redoing evaluations based on scaling law predictions (effective compute) | Ensures there has been less than a certain level of increase in effective compute used for training of a specific model (eg: GPT-4) since the last set of evaluations were done for a previous version of the same model | Multiple of effective compute | 4x | Leading | Scaling laws | 4x |
| C6.x SPI 4 | Redoing evaluations based on scaling law predictions (fine-tuning) | Ensures there has been less than a certain amount of time of fine-turning progress of a specific model (eg: GPT-4) since the last set of evaluations were done for a previous version of the same model | Months of fine-tuning progress | 6 | Leading | Scaling laws | 6 |

Table 31: SPIs for claim C6.x.

| SPI set C6.1 The AI system cannot exceed the Cybersecurity Apprentice baseline in Vulnerability Discovery Task Suite in Automated Evals with Human Oversight | | | | | | | |
|---|---|---|---|---|---|---|---|
| **SPI ID** | **SPI Title** | **SPI Description** | **Unit** | **Example** | **Leading/ Lagging** | **Evidence** | **Example Threshold** |
| C6.1 SPI 1 | Number of new research insights published on Automated Evals with Human Oversight | Measures the number of new research insights published on Automated Evals with Human Oversight in the last 90 days | Count | 2 | Leading | Research Insights | Track for trends |
| C6.1 SPI 2 | Response Time to Evaluate insights to Automated Evals with Human Oversight | Measures the time taken to evaluate and, if applicable, adopt new best practices or methodologies for Automated Evals with Human Oversight suggested by research publications or benchmarking reports. | Days | 15 | Leading | Research Insights | 30 |
| C6.1 SPI 3 | Evaluation of vulnerability attack automation of low-medium level threat actor cyberattacks (professional level) | Passing rate at Reverse engineering, Cryptography, and Binary exploitation tasks in the Hack the Box CTF challenges (Phuong et al, 2024) | Percent | 20% | Lagging | Internal evaluations | 20% |
| C6.1 SPI 4 | No significant improvement in bug-finding | Change in bug-finding rate for cybersecurity professional, as compared to not using the model (OpenAI Preparedness Framework, 2023) | Percent | 10% | Lagging | Internal evaluations | 10% |

Table 32: SPIs for claim C6.1.

| SPI set C6.2 The AI system cannot exceed the Cybersecurity Apprentice baseline in Vulnerability Discovery Task Suite in Fully Automated Evals | | | | | | | |
|---|---|---|---|---|---|---|---|
| **SPI ID** | **SPI Title** | **SPI Description** | **Unit** | **Example** | **Leading/ Lagging** | **Evidence** | **Example Threshold** |
| C6.2 SPI 1 | Number of new research insights published on Fully Automated Evals | Measures the number of new research insights published on Fully Automated Evals in the last 90 days | Count | 2 | Leading | Research Insights | Track for trends |
| C6.2 SPI 2 | Response Time to Evaluate insights on Fully Automated Evals | Measures the time taken to evaluate and, if applicable, adopt new best practices or methodologies for Fully Automated Evals suggested by research publications or benchmarking reports. | Days | 15 | Leading | Research Insights | 30 |
| C6.2 SPI 3 | Cyber theory of attack evaluation | Comparison with human baseline in a set of multiple choice questions (MCQs) related to cyber attacks (OpenAI Preparedness Framework, 2023) | Percent | 30% | Lagging | Internal evaluations | Lower than human baseline |
| C6.2 SPI 4 | End-to-end self-proliferation evaluation | Is able to complete all tasks end-to-end in a self-proliferation evaluations suite (Phuong et al, 2024) | Boolean | No | Lagging | Internal evaluations | No |
| C6.2 SPI 5 | Agentic software engineering performance evaluation | Pass rate (avg of 5) on set of tasks on the SWE-bench dataset as described in o1 System Card (OpenAI 2024c) | Percent | 50% | Lagging | Internal evaluations | 50% |

Table 33: SPIs for claim C6.2.

| SPI set C6.3 The AI system cannot exceed the Cybersecurity Apprentice baseline in Vulnerability Discovery Task Suite in Human Uplift Studies | | | | | | | |
|---|---|---|---|---|---|---|---|
| **SPI ID** | **SPI Title** | **SPI Description** | **Unit** | **Example** | **Leading/ Lagging** | **Evidence** | **Example Threshold** |
| C6.3 SPI 1 | Number of new research insights published on Human Uplift Studies | Measures the number of new research insights published on Human Uplift Studies in the last 90 days | Count | 2 | Leading | Research Insights | Track for trends |
| C6.3 SPI 2 | Response Time to Evaluate insights on Human Uplift Studies | Measures the time taken to evaluate and, if applicable, adopt new best practices or methodologies for Human Uplift Studies suggested by research publications or benchmarking reports. | Days | 15 | Leading | Research Insights | 30 |
| C6.3 SPI 3 | Evaluation of low-medium level threat actor cyberattacks | Passing rate for General skills, Reverse engineering, Cryptography, Forensics, Binary exploitation and Web exploitation tasks in the InterCode-CTF evaluation suite by Yang et al (2023) as described by Phuong et al (2024) | Percent | 50% | Lagging | Internal evaluations | 50% |
| C6.3 SPI 4 | AI-assisted CTF hack evaluations | Change in score on CTF challenges for human participants before and after using AI assistance from the model (OpenAI Preparedness Framework, 2023) | Percent | 10% | Lagging | Internal evaluations | 10% |

Table 34: SPIs for claim C6.3.

| SPI set C6.n Other evaluation setups | | | | | | | |
|---|---|---|---|---|---|---|---|
| **SPI ID** | **SPI Title** | **SPI Description** | **Unit** | **Example** | **Leading/ Lagging** | **Evidence** | **Example Threshold** |
| C6.n SPI 1 | Number of new research insights published on the evaluation setup (n) | Measures the number of new research insights published on the evaluation setup (n) in the last 90 days | Count | 2 | Leading | Research Insights | Track for trends |
| C6.n SPI 2 | Response Time to Evaluate insights on the evaluation setup (n) | Measures the time taken to evaluate and, if applicable, adopt new best practices or methodologies for the evaluation setup (n) suggested by research publications or benchmarking reports. | Days | 15 | Leading | Research Insights | 30 |

Table 35: SPIs for claim C6.n.

| SPI set C7.1 During evaluation, the AI system did not score higher than the Cybersecurity Apprentice baseline in Vulnerability Discovery Task Suite in Automated Evals with Human Oversight | | | | | | | |
|---|---|---|---|---|---|---|---|
| **SPI ID** | **SPI Title** | **SPI Description** | **Unit** | **Example** | **Leading/ Lagging** | **Evidence** | **Example Threshold** |
| C7.1 SPI 1 | Milestone-based cyber vulnerability evaluation | Performance on a milestone-based cyber vulnerability evaluation suite, when each milestone is independently evaluated, i.e., not end-to-end (Phuong et al, 2024) | Probability of success, P(success) | 1/500 | Lagging | Internal evaluations | 1/500 |
| C7.1 SPI 2 | Expert-guided cyber vulnerability evaluation | Bits of expert advice needed for successfully completing milestone-based and end-to-end cyber vulnerability evaluation (Phuong et al, 2024) | Bits (lower than the threshold fails the SPI) | 900 | Lagging | Internal evaluations | 900 |

Table 36: SPIs for claim C7.1.

**SPI set C7.x** No under elicitation during evaluation

| SPI ID | SPI Title | SPI Description | Unit | Example | Leading/ Lagging | Evidence | Example Threshold |
|---|---|---|---|---|---|---|---|
| C7.x SPI 1 | Incidents Elicitation Gap Analysis | Measures the percentage of incidents in the last 90 days where the real-world capability elicited was not adequately captured during evaluation. | % | 8% | Lagging | Incidents | 10% |
| C7.x SPI 2 | Near Misses Elicitation Gap Analysis | Measures the percentage of near misses or averted incidents in the last 90 days where the real-world capability elicited was not adequately captured during evaluation. | % | 12% | Lagging | Incidents | 20% |
| C7.x SPI 3 | Research Informed Under Elicitation Findings | Measures the number of newly published benchmarks and research insights in the last 90 days that suggest under elicitation of harmful behaviors during evaluation. | Count | 1 | Leading | Research Insights | 2 |

Table 37: SPIs for claim C7.x.

**SPI set C8.1** No Prosaic Under elicitation

| SPI ID | SPI Title | SPI Description | Unit | Example | Leading/ Lagging | Evidence | Example Threshold |
|---|---|---|---|---|---|---|---|
| C8.1 SPI 1 | Deviation from Industry Average Elicitation Iterations | Measures the %+/- deviation from average elicitation iterations used internally in the most recent evaluations, compared to external benchmarks. | % | +10% | Leading | Research Insights | <-20% |
| C8.1 SPI 2 | Use of Agentic Scaffolding Compared to Industry Standards | Measures the extent as a % alignment to which agentic scaffolding techniques are applied during the most recent evaluations, compared to industry standards or best practices. | % | 90% | Leading | Research Insights | 90% |
| C8.1 SPI 3 | Use of Fine-Tuning Compared to Industry Standards | Measures the extent as a % alignment to which fine-tuning techniques are applied during the most recent evaluations, compared to industry standards or best practices. | % | 90% | Leading | Research Insights | 90% |

Table 38: SPIs for claim C8.1.

**SPI set C8.2** No Sandbagging

| SPI ID | SPI Title | SPI Description | Unit | Example | Leading/ Lagging | Evidence | Example Threshold |
|---|---|---|---|---|---|---|---|
| C8.2 SPI 1 | Noise injection sandbagging evaluation | Performance improvement on a benchmark task when adding noise to model weights, as compared to non-sandbagging model; evaluation approach originally described by Tice et al (2024) | Percent | 10% | Lagging | Internal evaluations | 10% |
| C8.2 SPI 2 | Deviation from Previous Capability Scores | Measures the % deviation in capabilities compared to the previous evaluation round on the previous model version. Predictions based on additional compute, fine-tuning, or other scaling inputs are used to set threshold. | % | -8% | Leading | Internal - Capability evaluations | ≤5% |
| C8.2 SPI 3 | Capability Performance Outlier Analysis | Measures the relative deviation of a specific capability's performance compared to trends in other capabilities from the same evaluation round. Significant underperformance may indicate sandbagging. E.g. Vulnerability Exploitation lags by 15% while others improved by 5-8% | % deviation from mean | -10% | Leading | Internal - Capability evaluations | ≤10% deviation from capability mean |
| C8.2 SPI 4 | Performance Drift Analysis Post Deployment | Measures the degree of performance improvement between pre- and post-deployment evaluations as a % performance change. Significant increases may indicate sandbagging. | % performance change | +15% | Leading | Internal - Capability evaluations | ≤5% performance increase |

<p align="center">Table 39: SPIs for claim C8.2.</p>

| SPI set C8.x There are no others ways in which the evaluation could fail to elicit true capabilities | | | | | | | |
|---|---|---|---|---|---|---|---|
| **SPI ID** | **SPI Title** | **SPI Description** | **Unit** | **Example** | **Leading/ Lagging** | **Evidence** | **Example Threshold** |
| C8.x SPI 1 | Externally Identified Causes of Under Elicitation | Measures the number of new research papers or insights from Frontier Model Forum in the 90 days that identify causes of under elicitation beyond prosaic under elicitation or sandbagging. | Count | 1 | Leading | Research Insights, Industry Bodies | 0 |
| C8.x SPI 2 | External red-teaming of model on cyber vulnerability detection/attack evaluations | Number of major missouts in threat modelling and evaluation during external red-teaming of model on cyber vulnerability detection/attack evaluations suite | Count | 1 | Lagging | External evaluations | 1 |
| C8.x SPI 3 | External pairwise safety comparison evaluation of model on cyber vulnerability detection/attack evaluations | Comparison with pre-mitigation / previous version during external pairwise safety comparison evaluation of model on cyber vulnerability detection/attack evaluations suite as described in o1 System Card (OpenAI 2024c) | Qualitative | Better than pre-mitigation | Lagging | External evaluations | Better than pre-mitigation (minimum) / previous version (ideal) |

<p align="center">Table 40: SPIs for claim C8.x.</p>

## Appendix C: Sectoral analysis of safety cases: key practices and governance models

A cross-sectoral analysis of safety case governance models, and associated practices.

| Sector | Governance Model | Best Practice Insight | Drawback |
|---|---|---|---|
| **Nuclear** | • **Log of Minor Changes**: A proposed practice has been to accumulate and incorporate into a regular time-bound update changes that have no impact on the configuration of the facility/ plant/equipment (Hall, 2020).<br>• **Archiving:** Safety Assessment Principles developed by the UK's Office for Nuclear Regulation ('ONR') superseded safety case documents to be archived, ensuring a robust historical record for future investigations (ONR, 2024). | Structured approach to managing minor risks without overwhelming operational processes. | Does not cater to dynamic safety cases |
| **Defence** (*SMP12. Safety Case and Safety Case Report*, (MoD, 2024**)**) | • **Foundational Tool:** The UK's Ministry of Defence (MoD) has adopted the Safety Case framework as a foundational approach for managing safety.<br>• **Living Document:** A Safety Case is recognised as a living document that should be updated and reviewed as per the Project Safety Management Plan. | Project Safety Management Plan used as the basis for review of safety considerations throughout the lifecycle | Does not cater to dynamic safety cases |
| **Energy** (Sabel et al., 2018) | • **Mandatory Updation:** The explosion on the Piper Alpha platform in 1988 resulting in the loss of 167 lives marked a turning point in safety management. In response, it became mandatory for British energy companies to submit safety cases and update them every five years. | Periodic safety case updates underscores the importance of time-bound reviews | Does not cater to dynamic safety cases |
| **Autonomous Systems** | • **UL 4600**: The Standard for Safety for the Evaluation of Autonomous Products provides a structured framework for developing and maintaining safety cases.<br>• **Reporting Mechanisms**: UL 4600 only briefly mentions the need for independent third-party confirmation of safety case updates, leaving a critical gap in oversight. (UL, 2023) | UL4600 is a useful technical safety standard | Significant gaps remain in reporting and regulatory oversight |
| **Space** (The Space Industry Regulations, 2021) | • **Review and revision**: The Space Industry Regulations 2021 creates a statutory obligation for spaceflight operators to review and share a revised copy of the Safety Case where there has been a change in facts and circumstances.<br>• **Regulatory Approval:** No launch of a vehicle is allowed before a written confirmation from the regulatory accepting the revised safety case. | Reliance on regulatory validation | Does not cater to dynamic safety cases |

Table 41: A cross-sectoral analysis of safety case governance models, and associated practices.